\documentclass{article}

\usepackage{xcolor}
\usepackage{soul}
\usepackage[sc]{mathpazo}
\usepackage{amsmath} 
\usepackage{bbm} 
\usepackage{amsfonts} 
\usepackage{xparse} 
\usepackage[margin=1.5in]{geometry} 
\usepackage{enumerate}
\usepackage{verbatim}
\usepackage{hyperref}

\usepackage[shortcuts]{extdash}
\usepackage{amsthm}
\theoremstyle{definition}

\theoremstyle{plain}

\newtheorem{theorem}{Theorem}

\newtheorem{conjecture}{Conjecture}

\usepackage{multirow}
\usepackage[pdftex]{graphicx}

\def\complex{\mathbb{C}}
\def\real{\mathbb{R}}

\def\Tr{\text{Tr}}
\def\T{\text{T}}

\newcommand\I{\mathbbm{1}}
\newcommand{\poly}[2]{\textnormal{P}(#1,#2)}

\def\Texp{\mathcal{T}\exp}

\def\F{\mathcal{F}}

\def\R{\mathbb{R}}
\def\Utot{U_{\textnormal{total}}}

\def\Gv{G_{\textnormal{v}}}

\def\tGv{\widetilde{G}_{\textnormal{v}}}
\def\Gv{G_{\textnormal{v}}}
\def\Gvo{G_{\textnormal{v1}}}
\def\Gvt{G_{\textnormal{v2}}}
\def\tGv{\widetilde{G}_{\textnormal{v}}}
\def\tGvo{\widetilde{G}_{\textnormal{v1}}}
\def\tGvt{\widetilde{G}_{\textnormal{v2}}}
\def\Utog{U_{\textnormal{tog}}}
\def\Inte{\textnormal{Int}}
\def\Dyson{\mathcal{D}}
\def\SM{\textnormal{SM}}
\def\Ex{\textnormal{Ex}}
\newcommand{\mathematica}{\texttt{Mathematica}}
\def\M{\textnormal{M}}

\usepackage[utf8]{inputenc}
\DeclareUnicodeCharacter{2009}{\,} 

\newcommand{\vertiii}[1]{{\left\vert\kern-0.25ex\left\vert\kern-0.25ex\left\vert #1 
    \right\vert\kern-0.25ex\right\vert\kern-0.25ex\right\vert}}

\NewDocumentCommand\linear{mo}{
							\IfNoValueTF{#2}
							{\text{L}(\mathcal{#1})}
							{\text{L}(\mathcal{#1}, \mathcal{#2})}
							}

\usepackage{authblk}

\title{Engineering Effective Hamiltonians}

\author[1,6,*]{Holger Haas}
\author[2,3,*]{Daniel Puzzuoli}
\author[4]{Feihao Zhang}
\author[5,6,7,8]{David G. Cory}

\def\affilsize{\normalsize}

\affil[1]{\affilsize Department of Physics, University of Waterloo, Canada}
\affil[2]{\affilsize Department of Mathematics and Statistics, University of Ottawa, Canada}
\affil[3]{\affilsize School of Mathematics and Statistics, Carleton University, Canada}
\affil[4]{\affilsize State Key Laboratory of Low-Dimensional Quantum Physics and Department of Physics, Tsinghua University, China}
\affil[5]{\affilsize Department of Chemistry, University of Waterloo, Canada}
\affil[6]{\affilsize Institute for Quantum Computing, University of Waterloo, Canada}
\affil[7]{\affilsize Perimeter Institute for Theoretical Physics, Canada}
\affil[8]{\affilsize Canadian Institute for Advanced Research, Canada}
\affil[*]{\affilsize These authors contributed equally to this work.}

\begin{document}
\maketitle

\begin{abstract}
In the field of quantum control, effective Hamiltonian engineering is a powerful tool that utilises perturbation theory to mitigate or enhance the effect that a variation in the Hamiltonian has on the evolution of the system. Here, we provide a general framework for computing arbitrary time-dependent perturbation theory terms, as well as their gradients with respect to control variations, enabling the use of gradient methods for optimizing these terms. In particular, we show that effective Hamiltonian engineering is an instance of a bilinear control problem -- the same general problem class as that of standard unitary design -- and hence the same optimization algorithms apply. We demonstrate this method in various examples, including decoupling, recoupling, and robustness to control errors and stochastic errors. We also present a control engineering example that was used in experiment, demonstrating the practical feasibility of this approach. 
\end{abstract}

\section{\label{sec:introduction}Introduction}
Efficient tools for engineering control sequences that drive a quantum system to undertake desired evolution are critical for quantum computing, sensing, and spectroscopy. In the case of quantum computing \cite{Vandersypen_2005, Krantz_2019,Bruzewicz_2019}, it is imperative that the effective evolution corresponds, as closely as possible, to that of the experimenter's best characterization of the system Hamiltonian, as only this can reliably lead to high fidelity unitary operations. In realistic settings this requires successful suppression of numerous unwanted, yet unavoidable, physical effects: couplings to uncharted or unaccountable external degrees of freedom \cite{viola_1998, viola_1999, viola_robust_2003}, leakage out of the computational subspace \cite{lucero_reduced_2010,motzoi_simple_2009,chen_measuring_2016}, as well as uncertainties and stochastic variations in the system's internal and control Hamiltonians \cite{Gamliel_1995,kabytayev_robustness_2014,soare_experimental_2014}. In the case of sensing and spectroscopy \cite{haeberlen_1968, mehring_1983, Degen_2017,casola_probing_2018,zhou_quantum_2019}, the experimenter is interested in letting the system evolve under some Hamiltonian that is not fully characterized, while removing the effects of other, potentially unknown and potentially much stronger Hamiltonian terms that interfere with the effects of the Hamiltonian of interest, limiting sensing ability or spectroscopic resolution. In most cases, the unwanted and wanted effects both arise from Hamiltonian terms that are either not fully characterized or cannot be fully accounted for, and hence methodologies for suppressing undesired effects -- while potentially retaining detectability of others -- have to rely on perturbation theory analysis.

To formalize the above, we say that our quantum system is controlled over a period $0 \le t \le T$, and denote the unitary evolution over this time period, as is generated by the experimenter's best characterization of the system's internal and control Hamiltonians, by $U(0 \le t \le T)$. Successful engineering of the desired effective evolution boils down to ensuring that $U(T)$ and variations of it with respect to particular Hamiltonian variations, take a desired form. Such variations can generally be expressed as time-dependent perturbation theory expressions of the following form:
\begin{equation}
  \label{eq:nestedIntegrals}
  \int_0^T dt_1 \int_0^{t_1} dt_2 ... \int_0^{t_{n-1}} dt_n~ f(t_1,t_2,...,t_n)~ U^{-1}(t_1) A(t_1) U(t_1) ~...~ U^{-1}(t_n) A(t_n) U(t_n) ,
\end{equation}
where $f(t_1,t_2,...,t_n)$ is a scalar function and $\lbrace A(t_i) \rbrace$ is a set of, possibly time-dependent, operators. These integrals arise in a variety of existing treatments \cite{haeberlen_1968, viola_1998, viola_1999, kubo_1963, green_2012} and are further discussed in the upcoming paragraphs. Control design for quantum computing implementations often requires ensuring that some list of such nested integrals are minimized or, better yet, equal to zero. This demand can occasionally be fulfilled somewhat incidentally; by ensuring that the control fields are as strong as possible the experimenter tends to minimize the control period and thereby the effect of some perturbations. Conversely, sensing and spectroscopy applications typically need control sequences that minimize some set of the nested integrals above while maximizing others, hence, the fastest control approach does not suffice.

Analytical perturbative tools for engineering effective Hamiltonians were introduced by Haeberlen and Waugh \cite{haeberlen_1968} with their average Hamiltonian theory (AHT). AHT prescribed a systematic approach for setting perturbation theory integrals of the kind in Equation~\eqref{eq:nestedIntegrals} with $f=1$ to some desired values. AHT immediately proved an indispensable tool for the development of a vast number of magnetic resonance control sequences, e.g., dipolar sequences \cite{waugh_1968, mansfield_1971, rhim_1973, mehring_1973, takegoshi_1985, cory_1991}, composite pulses for control and internal Hamiltonian variations \cite{levitt_1986}, imaging sequences \cite{cory_1990} and many more. Another $f=1$ analytical treatment was given by dynamical decoupling (DD) \cite{viola_1998, viola_1999, khodjasteh_2005} and dynamically corrected gates \cite{khodjasteh_2009, khodjasteh_2010} introduced in the context of quantum computing. Perturbation theory terms with $f \neq 1$ in Equation~\eqref{eq:nestedIntegrals} appear when solving for the ensemble averaged evolution of a quantum system under stochastic operators, as in stochastic Liouville theory \cite{kubo_1963}. In such cases, $f(t_1,t_2,...,t_n)$ will be composed of correlation functions that characterize the stochastic operators. Analytic control design seeking to minimize nested integrals of that kind was performed in \cite{cappellaro_2006}. 

In addition to the above considerations, achieving the most efficient and accurate control of any quantum system -- or an ensemble of quantum systems -- requires tailoring of control sequences for the particular experimental setup and physical system at hand. When it comes to flexible tailored control design, numerical control optimization has a number of advantages over analytical control design: (i) it can easily deal with simultaneous control of an ensemble \cite{borneman_2010, li_2011}, (ii) it is not specific to any Hilbert space dimension, (iii) it can accommodate any experimental constraints present for the specific hardware configuration, e.g., amplitude and bandwidth constraints for the control waveform \cite{borneman_2012}, (iv) it can account for deterministic control distortions due to control hardware \cite{hincks_2015}, and (v) it stands a better chance of yielding control sequences that are closer to being time optimal given (iii) and (iv). Furthermore, recent technical advances such as the use of graphics processing units and automatic differentiation \cite{Leung_2017} hold promise of significantly improving efficiency, and streamlining the implementation of numerical control engineering routines.

Given the benefits of perturbative tools and numerical control design, there has been increasing interest in numerical optimization of control sequences that implement effective Hamiltonians. A filter function formalism for mitigating the effect of stochastic noise in quantum control was introduced by Green~\emph{et~al} \cite{green_2012, green_2013}, and has been combined with gradient free numerical optimization, leading to experimental advancements \cite{soare_2014}. The filter function approach was fully generalized to be applicable to general classical and quantum noise in \cite{pazsilva_2014}, and furthermore, a set of \emph{fundamental filter functions}, out of which all other filter functions can be constructed, was identified. Although there are individual, problem specific, numerical approaches that have previously been taken \cite{pasini_2009, grace_2012}, a complete framework for numerical control optimization that would yield a desired value for $U(T)$ simultaneously with values for an arbitrary set of perturbation terms has so far been lacking.

With this manuscript, we provide a general method for the numerical evaluation of $U(T)$ simultaneously with the evaluation, or arbitrarily close approximation, of any number of nested integrals of the kind in Equation~\eqref{eq:nestedIntegrals}. Furthermore, this method also enables straightforward computation of gradients of these integrals, which is crucial for efficiently searching large control landscapes. We accomplish this by generalizing the work of Van~Loan \cite{vanloan_1978}, Carbonell~\emph{et al} \cite{carbonell_2008} and, more recently, Goodwin and Kuprov \cite{goodwin_2015}, who showed that certain nested integrals involving matrix exponentials can be evaluated via exponentiation of a single block matrix. This method has found application in unitary engineering \cite{goodwin_2016}, as it provides an accurate and efficient tool for evaluating partial derivatives of $U(T)$. We also note that the first order version the method outlined here has been observed in \cite{machnes_2018, kirchoff_2018} for evaluating functional derivatives of $U(T)$ with respect to control amplitudes. 

Our method is aimed at complementing the existing quantum control tools dealing with open quantum systems and systems interacting with non-Markovian environments which have been reviewed in \cite{Glaser_2015} and \cite{Brif_2010}. We generalize the pre-existing work by extending the block-matrix methods to compute the perturbation theory terms in Equation~\eqref{eq:nestedIntegrals} to arbitrary order, and also develop tools for approximating nested integrals involving arbitrary scalar functions $f(t_1,t_2,...,t_n)$. This is done by showing that the perturbation theory terms may themselves be written as parts of solutions to first order matrix differential equations, which we call the \emph{Van~Loan equations}. The Van~Loan equations have the same form as the Schr{\"o}dinger equation, and in particular depend on the control amplitudes in the same way. The immediate benefit of this formulation is that control and optimization of the perturbation theory terms is now a computational problem of the same kind as standard unitary design, and as such the same optimization methods, including those that use gradient information, can be employed. Most of this manuscript is devoted to attempting to clearly demonstrate how to exploit the differential equation formulation for the purpose of numerical control searches that involve various perturbation expressions.

Some of the authors of this manuscript have successfully employed the methods presented for nanoscale magnetic resonance imaging experiments \cite{rose_2018}. These experiments posed a very challenging control setting -- we were dealing with an ensemble of strongly dipolar coupled proton spins that experienced a vast Rabi (control) field strength ($|a(t)|$) variation of $0.9~\text{MHz}\le |a(t)| \le 1.7~\text{MHz}$., while the phase coherence time ($T_2$) of the coupled spins was 11~$\mu$s. Our numerical tools helped us to find control sequences that yielded a $\pi/2$ unitary rotation that was insensitive to first order perturbations due to dipolar and chemical shift Hamiltonians for the entire spin ensemble simultaneously. Even though the rotation took 7.5~$\mu$s to implement, it enabled an increase of spin $T_2$ by a factor of $\sim 500$. We strongly believe that such coherence time enhancements would not have been possible without the numerical tools developed here.

In this manuscript, we first give some background for matrix differential equations and effective Hamiltonians in Section~\ref{sec:aht}. We then specify our general approach for tackling control problems in Section~\ref{sec:controlSetup}. With Section~\ref{sec:diff_comp_methods}, we present a solution for a general time dependent upper triangular block matrix differential equation and highlight how it can be used for calculating nested integrals in Equation~\eqref{eq:nestedIntegrals}. Subsequently, we exemplify the construction of Van Loan block matrix differential equations and numerical control optimizations with five examples in Section~\ref{sec:examples}, which include the control sequence we engineered for the aforementioned nanoscale magnetic resonance experiments that was optimized to be implemented in the presence of a non-trivial transfer function for the control hardware.

\section{\label{sec:aht}Effective Hamiltonians}
We denote the set of $n \times n$ complex matrices by $\M_n$. The starting point for effective Hamiltonian analysis is an initial value problem (IVP) of the form:
\begin{equation}
	\dot{U}(t) = G(t) U(t),
\end{equation}
where $G,U : [0,T] \rightarrow \M_n$ are matrix valued functions, the initial value is $U(0) = \I_n$, and $\dot{U}$ denotes the time derivative of $U$. In the context of quantum control, we will typically have $G(t) = -iH(t)$, for $H(t)$ a time dependent Hamiltonian, but $G(t)$ could also represent the generator for a master equation, and in any case it is notationally convenient to consider a general $G(t)$. We call $G(t)$ the \emph{generator} of the above IVP, and $U(t)$ the \emph{propagator}. Under assumptions on the generator $G(t)$ (which we will not explicitly state or worry about) this IVP has a unique solution \cite{hale_2009}, which we will write using the \emph{time-ordered exponential} notation:
\begin{equation}
	U(t) = \Texp\left( \int_0^t dt_1 G(t_1) \right).
\end{equation}
In this manuscript we will not work with the ``time-ordering'' operator, we simply regard the above expression as a choice of notation for the solution of the above IVP. 

The goal of any effective Hamiltonian treatment such as AHT, DD or filter function formalism is to analyze the effect that a variation in generator has on the propagator. Formally, for two functions $G(t),\Gv(t) : [0,T] \rightarrow \M_n$ we want to analyze how the evolution of a system with generator $G(t) + \Gv(t)$ is different from a system with generator $G(t)$, where we are viewing $\Gv(t)$ as a \emph{variation} of the generator $G(t)$.

The \emph{toggling frame} provides a way of writing the propagator of a system evolving under $G(t) + \Gv(t)$ in a way that clearly separates out the deviation caused by $\Gv(t)$.\footnote{The toggling frame concept was utilized in \cite{evans_timedependent_1967}, though our presentation more closely follows that of \cite{haeberlen_1968}. Neither of these references use the terminology of \emph{toggling frame}, which appeared later (see, e.g., the presentation of \cite{mehring_1983}).} We denote the propagator under $G(t)$ alone:
\begin{equation}
U(t) =  \mathcal{T}\exp\left( \int_0^t dt_1 G(t_1)\right),
\end{equation}
the propagator under both:
\begin{equation}
\Utot(t) = \mathcal{T}\exp\left( \int_0^t dt_1 [G(t_1) + \Gv(t_1)]\right),
\end{equation}
and the \emph{toggling frame propagator}, defined as:
\begin{align}
	\Utog(t) =  \mathcal{T}\exp\left( \int_0^t dt_1 \tGv(t_1)\right),
\end{align}
where $\tGv(t) = U^{-1}(t)\Gv(t)U(t)$. With these definitions, it holds that
\begin{equation}
	\Utot(t) = U(t)\Utog(t),
\end{equation}
which may be verified by differentiating both sides of the equation and verifying that they are solutions to the same IVP. 

The decomposition $\Utot(t) = U(t)\Utog(t)$ packages all variation of $\Utot(t)$ as an effect of $\Gv(t)$ into $\Utog(t)$, and hence, the deviation of $\Utot(t)$ from $U(t)$ caused by $\Gv(t)$ may be analyzed by studying $\Utog(t)$. Operating in the perturbative limit, effective Hamiltonian schemes analyze $\Utog(t)$ via series expansion, either through the \emph{Dyson series} \cite{dyson_1949}:
\begin{equation}
\Utog(t) = \I_n + \int_0^t dt_1 \tilde{G}_v(t_1) + \int_0^t dt_1\int_0^{t_1} dt_2 \tilde{G}_v(t_1)\tilde{G}_v(t_2) + \dots
\end{equation}
or via the \emph{Magnus expansion} \cite{magnus_1954,blanes_2009}, which under certain conditions gives $\Utog(t) = \exp(\Omega(t))$ for
\begin{equation}
	\Omega(t) = \int_0^t dt_1 \tilde{G}_v(t_1) + \int_0^t dt_1 \int_0^{t_1}dt_2 [ \tilde{G}_v(t_1), \tilde{G}_v(t_2)] + \dots,
\end{equation}
where $[ \cdot, \cdot ]$ denotes the matrix commutator. 

How robust a control sequence is to a variation is then analyzed perturbatively using one of the above expansions. Furthermore, robust control sequences are designed specifically to optimize the above terms.

\subsection{General Form of Perturbation Terms} \label{section:general_aht_form}
In this manuscript, we will be concerned with integrals of the form
\begin{equation}
	U(t) \int_0^t dt_1 \dots \int_0^{t_{m-1}}dt_m f(t_1, \dots, t_m) U^{-1}(t_1)A_1(t_1)U(t_1) \dots U^{-1}(t_m)A_m(t_m)U(t_m),\label{equation:general_aht}
\end{equation}
for $U(t) = \mathcal{T}\exp\left(\int_0^t dt_1 G(t_1)\right)$, and where $f$ is some scalar valued function. As a shorthand, we denote the above integral as $\Dyson_U^f(A_1, \dots, A_m)(t)$, and when $f=1$ (i.e. $f$ is a constant), we will write $\Dyson_U(A_1, \dots, A_m)(t)$. $\Dyson$ should be read as \emph{Dyson term}.

Terms arising from either the Dyson series or Magnus series may be constructed out of integrals of the above form. In application, the function $f$ will often be a correlation function of a time-dependent stochastic noise source.

\subsection{The Dyson Series and Directional Derivatives}
In this manuscript, we will use the Dyson series expansion, as its terms have a direct interpretation as directional derivatives. As an example, we consider the directional derivative of $U(t)$ as a result of variation in $G(t)$ in the direction $\Gv(t)$, given by
\begin{equation}
	\frac{d}{d\epsilon}\Big|_{\epsilon = 0}\mathcal{T}\exp\left( \int_0^t dt_1 [G(t_1) + \epsilon \Gv(t_1)]\right) .
\end{equation}
If we expand $\Utog(t)$ via the Dyson series, the result is a power series for $\Utot(t)$ in $\epsilon$:
\begin{equation}
	\Utot(t) = U(t) + \epsilon U(t)\int_0^t dt_1 \tGv(t_1) + \epsilon^2 U(t)\int_0^t dt_1 \int_0^{t_1} dt_2 \tGv(t_1)\tGv(t_2) + \dots,
\end{equation}
and from this we may directly read off the directional derivative as the matrix corresponding to the $\epsilon$ term:
\begin{equation}
\begin{aligned}
	\frac{d}{d\epsilon}\Big|_{\epsilon = 0}\mathcal{T}\exp\left( \int_0^t dt_1 [G(t_1) + \epsilon \Gv(t_1)]\right) &= U(t)\int_0^t dt_1 \tGv(t_1)\\ 
	&= U(t) \int_0^t dt_1 U^{-1}(t_1) \Gv(t_1) U(t_1). \label{equation:var_first_deriv}
\end{aligned}
\end{equation}
Similarly, the second derivative is
\begin{equation}
\begin{aligned}
	\frac{d^2}{d\epsilon^2}\Big|_{\epsilon = 0}\mathcal{T}\exp&\left( \int_0^t dt_1 [G(t_1) + \epsilon \Gv(t_1)]\right) = 2U(t)\int_0^t dt_1\int_0^{t_1}dt_2 \tGv(t_1)\tGv(t_2)\\ 
	&= 2U(t) \int_0^t dt_1\int_0^{t_1}dt_2 U^{-1}(t_1) \Gv(t_1) U(t_1)U^{-1}(t_2) \Gv(t_2) U(t_2).
\end{aligned}
\end{equation}

The same analysis applies with respect to multiple variations:
\begin{equation}
\begin{aligned}
	\mathcal{T}\exp&\left( \int_0^t dt_1 [G(t_1) + \epsilon_1\Gvo(t_1) + \epsilon_2 \Gvt(t_1)]\right)\\ 
	&= U(t) + U(t)\int_0^t dt_1(\epsilon_1\tGvo(t_1) + \epsilon_2 \tGvt(t_1))\\ 
	&\quad + U(t)\int_0^tdt_1 \int_0^{t_1}dt_2 (\epsilon_1 \tGvo(t_1) + \epsilon_2 \tGvt(t_1))(\epsilon_1 \tGvo(t_2) + \epsilon_2 \tGvt(t_2)) + \dots,
\end{aligned}
\end{equation}
from which we may conclude that
\begin{equation}
\begin{aligned}
	\frac{d}{d\epsilon_1}\Big|_{\epsilon_1 = 0}&\frac{d}{d\epsilon_2}\Big|_{\epsilon_2 = 0}\mathcal{T}\exp\left( \int_0^t \mathop{dt_1} [G(t_1) + \epsilon_1\Gvo(t_1) + \epsilon_2 \Gvt(t_1)]\right)\\ 
	&= U(t)\int_0^t\mathop{dt_1}\int_0^{t_1}\mathop{dt_2}\tGvo(t_1)\tGvt(t_2) + U(t)\int_0^t\mathop{dt_1}\int_0^{t_1}\mathop{dt_2}\tGvt(t_1)\tGvo(t_2).
\end{aligned}
\end{equation}

Within the context of quantum control, directional derivatives with respect to variations in the generator are typically viewed in two ways: (1) when $\Gv(t)$ arises from a variation in an underlying control sequence, the directional derivatives are used to make informed decisions about how to modify the control sequence to make it better, and (2) when $\Gv(t)$ represents uncertainty in $G(t)$, or even known but unwanted terms in the generator, these terms represent how \emph{robust} the propagator under $G(t)$ is to the variation $\Gv(t)$.

\section{\label{sec:controlSetup}Setup of the Control Problem}
\begin{center}\begin{figure}
   \includegraphics[width=\textwidth]{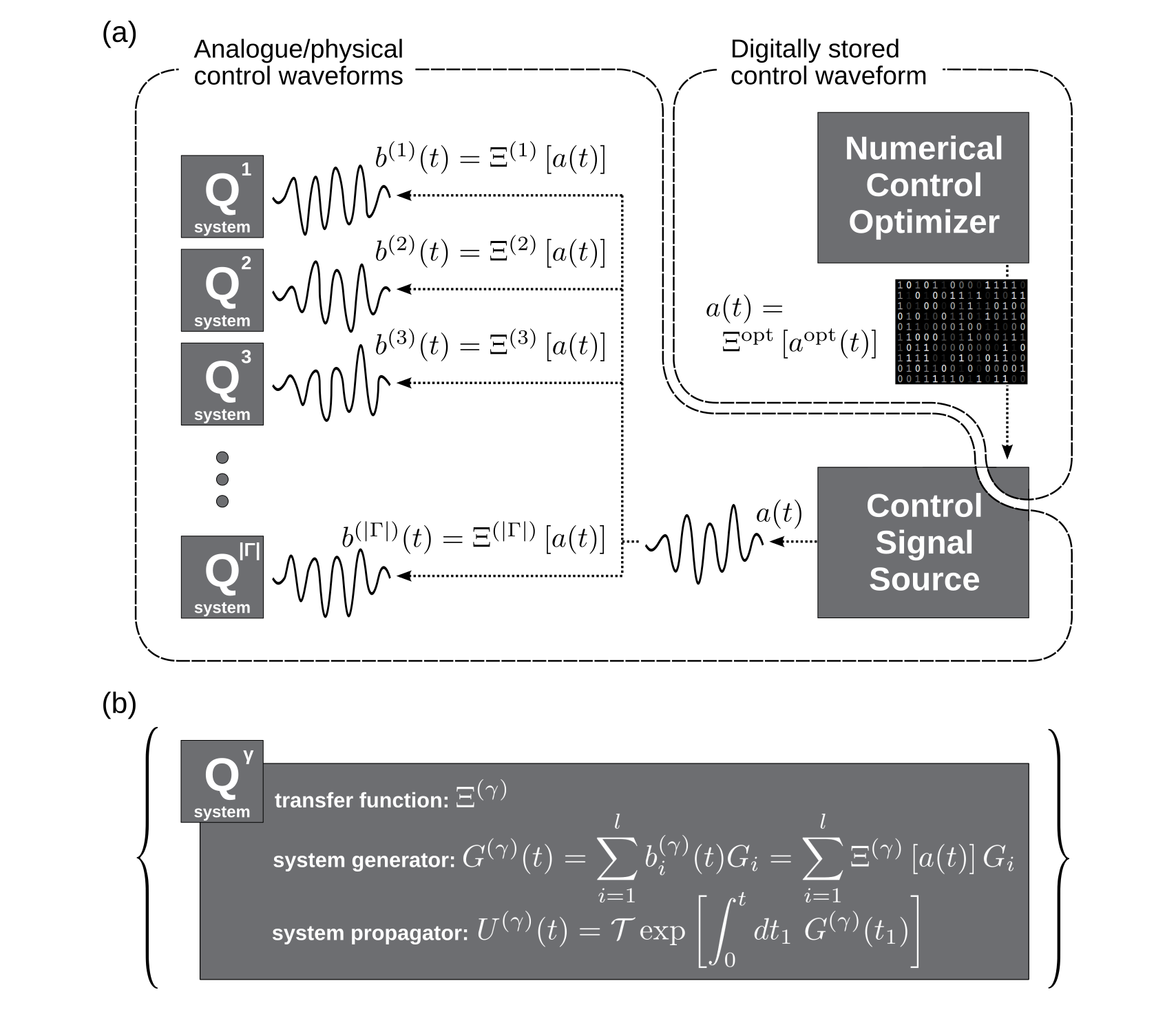}
   \caption{
      \label{fig:controlSetup}(a) Illustration of the control setting considered in this manuscript. We say that we have an ensemble $\Gamma$ of quantum systems, the unique characteristics of each quantum system $\gamma \in \Gamma$ are captured by the transfer function $\Xi^{(\gamma)}$ associated with it. We carry out our numerical control finding searches on the optimization control sequence $a^\text{opt}(t)$ that is transformed into an experimentally implementable sequence $a(t)$ through the application of the optimization transfer function $\Xi^\text{opt}$. $\Xi^\text{opt}$ is used for imposing the experimentally necessary constraints on $a(t)$, while $a^\text{opt}(t)$ need not adhere to such restrictions. When performing experiments, the sequence $a(t)$ is fed into a control signal source, typically an arbitrary waveform generator, in digital form that results from numerical control optimization. The control signal source outputs $a(t)$ as an analogue waveform. $a(t)$ is transformed by the set of transfer functions $\lbrace \Xi^{(\gamma)} \rbrace$ to a set of control amplitudes $\lbrace b^{(\gamma)}(t) = \Xi^{(\gamma)} \left[ a(t) \right] \rbrace$ which dictate the evolution of each quantum system. (b) Each quantum system $\gamma \in \Gamma$ is identified by its unique transfer function $\Xi^{(\gamma)}$, whereas the evolution of it is determined by the system propagator $U^{(\gamma)}(t)$ generated by the system generator $G^{(\gamma)}(t)$.
   }
\end{figure}\end{center}

In this section, we give a full, albeit abstract, description of the control problems the block matrix Van~Loan differential equation framework is capable of addressing. The description that will be outlined maps almost one to one to our implementation. In fact, a lot of our treatment and notation has been chosen specifically to simplify the implementation process while retaining full generality. In Figure~\ref{fig:controlSetup}, we illustrate the general control setting addressed. We say that we have a finite set of quantum systems labelled by a single compound label $\gamma \in \Gamma$. In principle, this is true for any control setting, although in many practical cases one approximates macroscopic ensembles of quantum systems as being parametrized by some set of continuous variables -- Rabi field strengths, resonance offsets, etc -- in such cases, we think of $\Gamma$ as a representative sample of the real ensemble. The ensemble $\Gamma$ could in some cases denote the same quantum system under different conditions for distinct experimental realizations, i.e., it might stand for an ensemble in time rather than a spatial ensemble of physical systems.
The ensemble $\Gamma$ of quantum systems is controlled by a \emph{control sequence} $a(t)$ -- a real vector valued function specified over an interval $[0,T]$ and delivered by some control signal source. The source should be thought of as a physical device which outputs $a(t)$, $a: [0,T] \rightarrow \real^k$; usually, we think of it as an arbitrary waveform generator. In the context of this manuscript, we regard $a(t)$ as the waveform generated by our numerical pulse search routines. Each quantum system labelled by $\gamma$ has an associated \emph{transfer function} $\Xi^{(\gamma)}$, $\Xi^{(\gamma)}: \phi^k \rightarrow \phi^l$, where we use $\phi^i$ to denote the space of real vector valued functions $\phi^i : [0,T] \rightarrow \real^i$. $\Xi^{(\gamma)}$ is an analytic deterministic map which transforms the control sequence $a(t)$ to system specific \emph{control amplitudes} $b^{(\gamma)}(t) = \Xi^{(\gamma)} \left[ a(t) \right]$, $b^{(\gamma)} : [0,T] \rightarrow \real^l$. The components of $b^{(\gamma)}(t)$ are the real valued functions that appear in the matrix differential equation determining the evolution of system $\gamma \in \Gamma$. In Appendix~\ref{app:transferfn}, we demonstrate how to construct $\Xi^{(\gamma)}$ for piecewise constant control sequences and control amplitudes in the case of linear transfer functions.

All quantum control problems boil down to engineering quantum state trajectories with certain desired properties. Mathematically this corresponds to generating a \emph{system propagator} $U^{(\gamma)}(t)$, $U^{(\gamma)}: [0,T] \rightarrow \M_n$, which satisfies some set of conditions. We emphasize that the properties wanted from $\lbrace U^{(\gamma)}(t) \rbrace$ need not be merely its value at time $T$, they could also be various integral expressions of $U^{(\gamma)}(t)$ over $0 \le t \le T$, which describe the trajectory of $U^{(\gamma)}(t)$. Here, a quantum system should be understood simply as a finite level system or one that can be treated as such; the time dependent state of the quantum system is determined by $U^{(\gamma)}(t)$. The time dependent value of $U^{(\gamma)}(t)$ itself is determined by a first order linear matrix differential equation which we refer to as the \emph{system differential equation}:
\begin{equation}
    \label{eq:systemEquation}
    \dot{U}^{(\gamma)}(t) = G^{(\gamma)}(t) U^{(\gamma)}(t) ,
\end{equation}
where $G_\gamma : [0,T] \rightarrow \M_n$ is the \emph{system generator}, while $U^{(\gamma)}(0) = \I_n$. The value of $G^{(\gamma)}(t)$ at each instant is determined by the control amplitudes 
\begin{equation}
    \label{eq:systemGenerator}
    G^{(\gamma)}(t) = \sum_{i=1}^l b^{(\gamma)}_i(t) G_i,
\end{equation}
where $G_i \in \M_n$ is a constant matrix for all $i$. This implies that the problem is a bilinear control theory problem \cite{elliott_2009}. The system differential equation should be understood as the Schr{\"o}dinger equation or some generalization of it, e.g., the Liouville-von Neumann equation for vectorized density matrices.

We note we assume all $G^{(\gamma)}(t)$ to have identical generators for all $\gamma \in \Gamma$. Even though this may not be the case for all quantum control problems, one can always use our problem description by employing a direct sum of different sets of $\lbrace G_i \rbrace$. Such an approach is computationally not the most efficient, but it does substantially simplify implementing the algorithm while retaining total generality.

As we said in Section~\ref{sec:aht}, Equation~\eqref{eq:systemEquation} has a formal solution
\begin{equation}
  \label{eq:systemPropagator}
  U^{(\gamma)}(t) = \Texp \left( \int_0^t dt_1 G^{(\gamma)}(t_1) \right) .
\end{equation}
We are typically interested in finding an $a(t)$, $0 \le t \le T$, that yields the wanted final unitary operations $\lbrace U^{(\gamma)}(T) \rbrace$ as well as some desired values for nested integral expressions of the following form:
\begin{equation}
  \begin{aligned}
   \label{eq:genNestedIntegrals}
  \int_0^T dt_1 \dots \int_0^{t_{n-1}} dt_n& f(t_1,\dots,t_n) \\
  &\times \left[U^{(\gamma)}(t_1)\right]^{-1} A(t_1) U^{(\gamma)}(t_1) \dots \left[U^{(\gamma)}(t_n)\right]^{-1} A(t_n) U^{(\gamma)}(t_n) .
  \end{aligned}
\end{equation}

In this manuscript, we designate $\M_k(\M_n)$ as a set of $k \times k$ block matrices composed of $n \times n$ complex matrices, hence, an element $A_{i,j}$ of $A \in \M_k(\M_n)$ is an element of $\M_n$. The main and the most significant result of this manuscript is demonstrating that control problems of this kind can still be written as bilinear control theory problems \cite{elliott_2009} that involve the same control amplitudes $\lbrace b^{(\gamma)}(t) \rbrace$ that appear in the system differential equation. In order to find $a(t)$ that yields the desired $\lbrace U^{(\gamma)}(T) \rbrace$ and the desired values for any set of integral expressions for $\lbrace U^{(\gamma)}(t) \rbrace$, $U^{(\gamma)}: [0,T] \rightarrow \M_n$, we can always construct a block matrix differential equation -- called the \emph{Van Loan differential equation} -- which comprises the system generators $\lbrace G_i \rbrace$ and the objects that appear in the integral expressions for $\lbrace U^{(\gamma)}(t) \rbrace$. The Van Loan differential equation is expressed as
\begin{equation}
    \label{eq:vanLoanEquation}
    \dot{V}^{(\gamma)}(t) = L^{(\gamma)}(t) V^{(\gamma)}(t) ,
\end{equation}
where $V^{(\gamma)}(t)$ is the \emph{Van Loan propagator} and $L^{(\gamma)}(t)$ is the \emph{Van Loan generator}
\begin{equation}
    \label{eq:vanLoanGenerator}
    L^{(\gamma)}(t) =L_0 + \sum_{i=1}^l b^{(\gamma)}_i(t) L_i ,
\end{equation}
for $L_i \in \M_m(\M_n)$. It will be shown that the integral expressions of interest appear as various blocks of $V^{(\gamma)}(T)$. The benefit of such block matrix methods is two-fold: it enables an accurate and efficient way for evaluating the integral expressions for piecewise constant $\lbrace b^{(\gamma)}(t) \rbrace$ and is readily deployable within control finding routines that take advantage of the linear differential equation structure of the problem.

Having constructed the Van Loan differential equation that enables the evaluation of all terms of interest, we can always define a target function $\Phi^{(\gamma)}$ for each system in the ensemble. $\Phi^{(\gamma)}$ being a function of the final Van Loan propagator $V^{(\gamma)}(T)$ for system $\gamma$, i.e., $\Phi^{(\gamma)}: \M_m(\M_n) \rightarrow [0,1]$, where $\Phi^{(\gamma)} = 1$ corresponds to having the desired properties from the system propagator $U^{(\gamma)}(T)$ and from any number of nested integral terms of interest. Finally, we combine $\lbrace \Phi^{(\gamma)} \rbrace$ into a target function $\Phi$ for the whole ensemble $\Gamma$: $\Phi = \sum_{\gamma \in \Gamma} p^{(\gamma)} \Phi^{(\gamma)}$, where $\lbrace p^{(\gamma)} \rbrace$ are the relative weights assigned to each member of $\Gamma$. We have assumed that $0 \le p^{(\gamma)} \le 1$, for all $\gamma \in \Gamma$, and that $\sum_{\gamma \in \Gamma} p^{(\gamma)} = 1$. Of course, the linear form of $\Phi$ is not necessary but it does simplify the implementation. It is clear that $\Phi$ is a functional of $a(t)$ and its derivatives with respect to the control sequence are given as
\begin{equation}
  \label{eq:GammaPhiDerivs}
  \frac{\partial}{\partial a(t)} \Phi = \sum_{\gamma \in \Gamma} p^{(\gamma)} \frac{\partial}{\partial a(t)} \Phi^{(\gamma)} \left[ V^{(\gamma)}(T) \left( \Xi^{(\gamma)} \left[ a(t) \right] \right) \right] .
\end{equation}

Throughout this manuscript, we will only deal with piecewise constant control amplitudes $a(t)$, for which, we split the interval $[0,T]$ into $N$ subintervals with respective durations $\Delta T_j$ such that $\Delta T_j \ge 0$, for all $j \in \lbrace 1,2,\dots,N \rbrace$, and $\sum_{j=1}^N \Delta T_j = T$. We say that the control sequence $a(t)$ takes a constant value over each of the subintervals, i.e., 
\begin{align}
   \label{eq:alphaElement}
   a_i\left(\sum_{s=1}^{j-1} \Delta T_s \le t < \sum_{s=1}^{j-1} \Delta T_s + \Delta T_j\right) = \alpha_{i,j} ,
\end{align}
where $\alpha \in \M_{k,N}(\R)$ is a real valued $k \times N$ matrix that contains all piecewise control elements $\lbrace \alpha_{i,j} \rbrace$. Given a non-identity transfer function $\Xi^{(\gamma)}$, which is likely the case for any experimental setting, we have to use the chain rule to evaluate $\frac{\partial}{\partial a(t)} V^{(\gamma)}(T) \left( \Xi^{(\gamma)} \left[ a(t) \right] \right)$. For piecewise constant control settings, we define matrices $\lbrace \beta^{(\gamma)} \rbrace$ that specify the piecewise constant amplitudes of $\lbrace b^{(\gamma)}(t) \rbrace$ just like $\alpha$ for $a(t)$ above. We split the interval $[0,T]$ into $M$ subintervals with respective durations $\delta T_j$ such that $\delta T_j \ge 0$, for all $j \in \lbrace 1,2,\dots,M \rbrace$, and $\sum_{j=1}^M \delta T_j = T$. We note that $M$ does not necessarily have to match $N$. We can now specify $\beta^{(\gamma)} \in \M_{l,M}(\R)$ the components of which correspond to the piecewise constant values of $b^{(\gamma)}(t)$:
\begin{align}
   \label{eq:betaElement}
   b^{(\gamma)}_i\left(\sum_{s=1}^{j-1} \delta T_s \le t < \sum_{s=1}^{j-1} \delta T_s + \delta T_j\right) = \beta^{(\gamma)}_{i,j} .
\end{align}
Because we treat the control sequence and the control amplitudes as piecewise constant functions, we will also, from now on, regard the transfer functions $\lbrace \Xi^{(\gamma)} \rbrace$ as matrix valued functions, i.e., $\Xi^{(\gamma)}: \M_{k,N}(\R) \rightarrow \M_{l,M}(\R)$, such that $\beta^{(\gamma)} =  \Xi^{(\gamma)}(\alpha)$. We write $\Phi(\alpha) = \sum_{\gamma \in \Gamma} p^{(\gamma)} \Phi^{(\gamma)} \left( V^{(\gamma)}(T) \left[ \Xi^{(\gamma)} (\alpha) \right] \right)$. Solving the Van Loan differential equations enables us to find the values of $\frac{\partial}{\partial \beta^{(\gamma)}_{i,j}} V^{(\gamma)}(T)$, as is shown in Appendix~\ref{sec:GRAPE}. Hence, we can evaluate $\frac{\partial}{\partial \alpha_{i,j}} V^{(\gamma)}(T) = \sum_{s=1}^l\sum_{t=1}^M \frac{\partial}{\partial \alpha_{i,j}} \left(\Xi^{(\gamma)}(\alpha)\right)_{s,t} \frac{\partial}{\partial \beta^{(\gamma)}_{s,t}} V^{(\gamma)}(T)$. Since we assume $\Xi^{(\gamma)}$ to be an analytic matrix valued function, we can always evaluate the elements of its Jacobian $\lbrace \frac{\partial}{\partial \alpha_{i,j}} \left(\Xi^{(\gamma)}(\alpha)\right)_{s,t} \rbrace$ in order to implement the maximization of $\Phi$. For all examples considered in this manuscript $\lbrace \Xi^{(\gamma)} \rbrace$ are taken to be linear, meaning that their Jacobians are trivial.

Finally, in most practical cases the experimentalist needs $a(t)$ to adhere to some constraints, e.g., pulse waveform bandwidth and amplitude constraints or periods for which $a(t) = 0$. As constrained optimization is more technically challenging than unconstrained optimization, it is common to try to enforce the constraints on $a(t)$ in a way that keeps the overall optimization problem unconstrained. Our preferred method for doing this is an \emph{optimization transfer function} $\Xi^\text{opt}$, $\Xi^\text{opt}: \M_{k,N^\text{opt}}(\R) \rightarrow \M_{k,N}(\R)$, where $N^\text{opt}$ is the number of time steps for the piecewise constant \emph{optimization control sequence} $a^\text{opt}(t)$. The idea is that the mapping $\Xi^\text{opt}$ is constructed in a way that ensures all control sequences in its output space adhere to either all or some of the constraints. Having constructed $\Xi^\text{opt}$, the numerical pulse searches are then carried out over its input $a^\text{opt}(t)$, which need not adhere to all constraints on $a(t)$. We represent the piecewise-constant $a^\text{opt}(t)$ by a matrix $\alpha^\text{opt} \in \M_{k,N^\text{opt}}(\R)$, just as we did for $a(t)$ and $\lbrace b^{(\gamma)}(t) \rbrace$ above. For finding a suitable control sequence using gradient based algorithms, one then needs to evaluate
\begin{equation}
  \label{eq:PhichainRuleOpt}
  \frac{\partial}{\partial \alpha^\text{opt}_{i,j}} \Phi = \sum_{\gamma \in \Gamma} p^{(\gamma)} \frac{\partial}{\partial \alpha^\text{opt}_{i,j}} \Phi^{(\gamma)} \left[ V^{(\gamma)}(T) \left[ \Xi^{(\gamma)} \left( \Xi^\text{opt} \left[ \alpha^\text{opt} \right] \right) \right] \right] ,
\end{equation}
which means evaluating the elements of the Jacobian for $\lbrace \Xi^{(\gamma)} \rbrace$ as well as $\Xi^\text{opt}$. After finding an $\alpha^\text{opt}$ that yields a high enough $\Phi$ value, the control sequence that is to be implemented experimentally is calculated simply as $\alpha = \Xi^\text{opt} \left( \alpha^\text{opt} \right)$. In Appendix~\ref{app:transferfn}, we demonstrate explicitly how to construct $\Xi^\text{opt}$ that introduces zero pulse amplitudes to the beginning and the end of the control sequence and how to construct $\Xi^\text{opt}$ that limits the bandwidth of the waveform $a(t)$ in the frequency (Fourier) domain. We note that our use of optimization transfer function is similar to the method in \cite{motzoi_optimal_2011}; i.e. it implements the mapping from the parameters of a control sequence to the control sequence itself.

\section{\label{sec:diff_comp_methods}Computational Methods}
In this section, we outline the framework for computing Dyson terms of general form, $\Dyson^f_U(A_1, \dots, A_m)(t)$, defined in Section~\ref{section:general_aht_form}. The general idea is that these terms may be written as solutions of first order matrix differential equations of the same form as the base differential equation. The generators for the new differential equations are block matrices with blocks consisting of pieces from the original differential equation.

To illustrate the approach, we consider the simplest case, a first order integral:
\begin{equation}
	\Dyson_U(B)(t) = U(t)\int_0^t dt_1U^{-1}(t_1)B(t_1)U(t_1),
\end{equation}
for $U(t) = \mathcal{T}\exp\left(\int_0^t dt_1 G(t_1)\right)$. We may further simplify this by assuming that $G(t) = A$ and $B(t) = B$, i.e. they are time-independent. In this case, the expression reduces to
\begin{equation}
	\Dyson_U(B)(t) = e^{At}\int_0^t dt_1e^{-At_1}Be^{At_1}.
\end{equation}
A priori, computing the above for a particular time $t$ requires an integral approximation method. However, it was originally observed by Van Loan \cite{vanloan_1978} that this expression can be computed using a single matrix exponential: 
\begin{equation}
	\exp\left[\left(\begin{array}{cc}
	A & B \\
	0 & A
	\end{array}\right)t\right] = \left(\begin{array}{cc}
	e^{At} & e^{At}\int_0^t dt_1 e^{- At_1}Be^{ At_1} \\
	0 & e^{At}
	\end{array}\right). \label{equation:first_order_constant}
\end{equation}
Van Loan showed \cite{vanloan_1978} more generally how nested integrals up to order $4$ involving matrix exponentials can be computed by exponentiating a single upper triangular block matrix, and \cite{carbonell_2008} extended this to arbitrary order.

This has found application in physics where such expressions often arise \cite{najfeld_1995,goodwin_2015}, and in particular it has been used to compute directional derivatives for pulse finding \cite{goodwin_2016}. Here, we extend this idea to the case of time-dependent matrices, to compute integrals involving \emph{time-ordered} exponentials. The simplest case of this extension is the time-dependent version of Equation~\eqref{equation:first_order_constant}. For two matrix-valued functions $A(t)$, and $B(t)$, it holds that
\begin{equation}
	\mathcal{T}\exp\left[\int_0^t dt_1 \left(\begin{array}{cc}
	A(t_1) & B(t_1) \\
	0 & A(t_1)
	\end{array}\right)\right] = \left(\begin{array}{cc}
	U(t) & U(t)\int_0^t dt_1 U^{-1}(t_1)B(t_1)U(t_1) \\
	0 & U(t)
	\end{array}\right). \label{equation:first_order_time_dependent}
\end{equation}
The above formula may be verified by differentiating both sides of the equation, and verifying they are both solutions to the same initial value problem. Hence, we may compute $D_U(B)(t)$ via propagation of the differential equation 
\begin{equation}\dot{V}(t) =
	\left(\begin{array}{cc}
	A(t) & B(t) \\
	0 & A(t)
	\end{array}\right)V(t)\textnormal{, with } V(0) = \left(\begin{array}{cc}
	\I_n & 0 \\
	0 & \I_n
	\end{array}\right).
\end{equation}
We note that this particular formula has been observed in \cite{machnes_2018} in the context of pulse finding for derivative evaluation. 

In this section, we present a generalization of the previous work to arbitrary order in the time-dependent case, including scalar functions $f$. The general idea is the same as for the first order example; a Dyson term may be rephrased as part of the solution to a first order matrix differential equation. In the context of control, this rephrases controlling Dyson terms as a bilinear control theory problem \cite{elliott_2009}. In all cases, the generator of the differential equation is an upper triangular block matrix, and hence we also develop general tools for analyzing the structure of the time-ordered exponential of arbitrary upper triangular block matrices.

This section is organized as follows.
\begin{itemize}
	\item In Section~\ref{section:aht_theorem}, we generalize the theorems in \cite{vanloan_1978,carbonell_2008}, giving the general structure of time-ordered exponentials of upper triangular block matrices. As described therein, Appendix~\ref{app:mathematica} describes code for symbolically simplifying this structure 
	\item In Section~\ref{section:f_1}, we give a differential equation computing $\Dyson_U(A_1, \dots, A_m)(t)$, i.e. the case when no scalar function appears in the integral.
	\item In Section~\ref{sec:generalizations}, we provide a similar construction for terms $\Dyson_U^f(A_1, \dots, A_m)(t)$ when $f$ is either a linear combination of exponentials, or is a polynomial.
\end{itemize}
We note that in this manuscript we are concerned specifically with terms arising in effective Hamiltonian treatments, but Section~\ref{section:aht_theorem} describes a much more general class of integrals involving time-ordered exponentials that this approach may be applied to. Hence, this method may find application in control design beyond optimization of Dyson terms.

\subsection{Integrals Involving Time-Ordered Matrix Exponentials} \label{section:aht_theorem}

Here, we present a full time-ordered generalization of the theorems of Van Loan \cite{vanloan_1978} and Carbonell et al. \cite{carbonell_2008}. First, we introduce some notation. Let $B_{i,j} : [0,T] \rightarrow \M_n$ for $1 \leq i \leq j \leq m$. For $1 \leq i \leq m$, denote
\begin{equation}
	U_i(t) = \mathcal{T}\exp\left(\int_0^t dt_1 B_{i,i}(t_1)\right).
\end{equation} 
For $s \geq 2$ and indices $i_1, \dots i_s$, denote
\begin{align}
	&\Inte_{(i_1, \dots, i_s)}(t)\\ 
	\nonumber &=U_{i_1}(t)\int_0^t dt_1 \dots \int_0^{t_{s-2}} dt_{s-1} U^{-1}_{i_1}(t_1) B_{i_1,i_2}(t_1)U_{i_2}(t_1) \dots U^{-1}_{i_{s-1}}(t_{s-1}) B_{i_{s-1},i_s}(t_{s-1}) U_{i_s}(t_{s-1}),
\end{align}
and for a single index $i$, denote
\begin{align}
	\Inte_{(i)}(t) = U_i(t).
\end{align}
Note that for $s \geq 2$ and indices $i_1, \dots, i_s$, these definitions satisfy the recursive expression
\begin{equation}
	\Inte_{(i_1, \dots, i_s)}(t) = U_{i_1}(t) \int_0^t dt_1 U^{-1}_{i_1}(t_1) B_{i_1, i_2}(t_1) \Inte_{(i_2, \dots, i_s)}(t_1). \label{equation:int_recursive}
\end{equation}

\begin{theorem} \label{theorem:HP_theorem}
Let $B_{i,j} : [0,T] \rightarrow \M_n$ for $1 \leq i \leq j \leq m$. Define $C_{i,j}:[0,T] \rightarrow \M_n$ implicitly by the equation
\begin{equation}
\begin{aligned}
	&\left(\begin{array}{cccc}
    			C_{1,1}(t) & C_{1,2}(t) & \dots & C_{1,m}(t) \\
			0 & C_{2,2}(t) & \dots & C_{2,m}(t) \\
			\vdots & \ddots & \ddots & \vdots \\
			0 & 0 & \dots & C_{m,m}(t)
			\end{array}\right)\\
	&\qquad\qquad\qquad\qquad\qquad=
	\mathcal{T}\exp \left[ \int_0^t dt_1 \left(\begin{array}{cccc}
    			B_{1,1}(t_1) & B_{1,2}(t_1) & \dots & B_{1,m}(t_1) \\
			0 & B_{2,2}(t_1) & \dots & B_{2,m}(t_1) \\
			\vdots & \ddots & \ddots & \vdots \\
			0 & 0 & \dots & B_{m,m}(t_1)
			\end{array}\right)  \right],
\end{aligned}
\end{equation}
and assume that the $B_{i,j}(t)$ are such that the solution to the IVP associated to the above time-ordered exponential exists and is unique\footnote{We omit an explicit statement of conditions under which existence and uniqueness holds. As we are concerned only with applications of these expressions in physical settings, finding the most general and exact technical statements for which this assumption holds is of no real interest; most reasonable physical assumptions, such as piecewise continuity, suffice (see, e.g. Theorem 3.1 in Section I.3 in \cite{hale_2009}).}.

For all $1 \leq s \leq m$, $1 \leq j \leq m-s$, and $t \in [0,T]$, it holds that
\begin{equation}
	C_{s,s}(t) = U_s(t), \label{equation:unitary_explicit}
\end{equation}
and 
\begin{equation}
	C_{s,s+j}(t) = \Inte_{(s, s+j)}(t) +  \sum_{r=1}^{j-1} \sum_{s < i_1 < \dots < i_r < s+j} \Inte_{(s, i_1, \dots, i_r, s+j)}(t), \label{equation:block-upper-explicit}
\end{equation}
where the inner sum is over all indices $i_1, \dots, i_r$ satisfying the relations, and $U_i$ and $\Inte$ are as defined before the theorem. Alternatively, these matrices can be given recursively as
\begin{equation}
	C_{s,s+j}(t) = \sum_{i=1}^{j}U_{s}(t) \int_0^t dt_1U^{-1}_{s}(t_1) B_{s, s+i}(t_1) C_{s+i, s+j}(t_1). \label{equation:block-upper-recursive}
\end{equation}
\end{theorem}

The proof is given in Appendix~\ref{app:proof}. In Appendix~\ref{app:mathematica} we describe code that symbolically simplifies the structure arising from this theorem. That is, in general the above expressions are quite complicated, but in the constructions we will see in the following sections, many of the blocks $B_{i,j}(t)$ will be $0$ or proportional to the identity, in which case the expressions of the above theorem can simplify dramatically.

\subsection{The $f=1$ Case} \label{section:f_1}

First, we show how to compute expressions of the form
\begin{equation}
\begin{aligned}
	\Dyson_U(A_1, &\dots, A_m)(t)\\ &= U(t) \int_0^t dt_1 \dots \int_0^{t_{m-1}}dt_m U^{-1}(t_1) A_1(t_1) U(t_1) \dots U^{-1}(t_m) A_m(t_m) U(t_m),
\end{aligned}
\end{equation}
where $U(t) = \mathcal{T}\exp\left(\int_0^t dt_1 G(t_1)\right)$, i.e., perturbation theory terms without a time-dependent scalar function. In this case, we may observe that for
\begin{equation}
L(t) = 	\left(\begin{array}{cccccc}
					G(t) & A_1(t) & 0 & \dots & 0 & 0 \\
					0 & G(t) & A_2(t) & \dots &0 & 0 \\
					0 & 0 & G(t) & \dots & 0 & 0 \\
					\vdots & \vdots &\vdots & \ddots & \vdots & \vdots \\
					0 & 0 & 0 & \dots & G(t) & A_m(t) \\
					0 & 0 & 0 & \dots & 0 & G(t)
					\end{array} \right)
\end{equation}
it holds that
\begin{align}
	&\mathcal{T}\exp\left[\int_0^t dt_1 L(t_1)\right] = \\ 
					\nonumber &\left(\begin{array}{cccccc}
					U(t) & \Dyson_U(A_1)(t) & \Dyson_U(A_1,A_2)(t) & \dots & \Dyson_U(A_1,\dots, A_{m-1})(t) & \Dyson_U(A_1,\dots, A_m)(t) \\
					0 & U(t) & \Dyson_U(A_2)(t) & \dots &\Dyson_U(A_2,\dots, A_{m-1})(t) & \Dyson_U(A_2,\dots, A_m)(t) \\
					0 & 0 & U(t) & \dots & \Dyson_U(A_3,\dots, A_{m-1})(t) & \Dyson_U(A_3,\dots, A_m)(t) \\
					\vdots & \vdots &\vdots & \ddots & \vdots & \vdots \\
					0 & 0 & 0 & \dots & U(t) & \Dyson_U(A_m)(t) \\
					0 & 0 & 0 & \dots & 0 & U(t)
					\end{array} \right).
\end{align}
That is, the generator for this system $L(t)$ is in $\M_{m+1}(\M_n)$, where all blocks are $0$ except: 
\begin{itemize}
	\item All diagonal blocks are $G(t)$, and
	\item The first off-diagonal is given by $(A_1(t), \dots, A_m(t))$.
\end{itemize}
The time ordered exponential $\mathcal{T}\exp\left(\int_0^t dt_1 L(t_1)\right)$ has upper triangular structure with:
\begin{itemize}
	\item All diagonal blocks are $U(t)$, and 
	\item For $i < j$, the $(i,j)$ block is given by $\Dyson_U(A_i, \dots, A_j)(t)$.
\end{itemize}
Hence, propagating the differential equation associated with the generator $L(t)$ computes the desired term $\Dyson_U(A_1, \dots, A_m)(t)$, as well as many other terms that will likely be of interest.

To see this, one may simply apply Theorem~\ref{theorem:HP_theorem} to the generator $L(t)$. Alternatively, one may purposefully \emph{construct} this differential equation using the procedure described in the next section.

\subsection{Including Scalar Functions} \label{sec:generalizations}

Next, we consider integrals of the form
\begin{equation}
	\Dyson^f_U(A_1, \dots, A_m)(t) =U(t)\int_0^T dt_1 \dots \int_0^{t_{m-1}} dt_m f(t_1, \dots, t_m) \tilde{A}_1(t_1) \dots\tilde{A}_m(t_m), \label{equation:more_general_integral} 
\end{equation}
for $\tilde{A}_i(t) = U^{-1}(t)A_i(t)U(t)$, where $f$ is a scalar valued function, which may represent, for example, a correlation function for stochastic noise.

For a given $f$, it is not immediately clear how to write the integral in Equation~\eqref{equation:more_general_integral} as a part of the solution to a linear matrix differential equation, in the way we have done in the $f=1$ case. Certainly, it will not be possible for most functions. However, we will show here that it is possible for a very large class of functions; in particular $f$ satisfying the following properties:
\begin{itemize}
	\item $f$ is a linear sum in product form: $f(t_1, \dots, t_m) = \sum_i c_i f^{(i)}_1(t_1) \dots f^{(i)}_m(t_m)$, with
	\item Each function $f^{(i)}_r(t)$ is drawn from a finite dimensional vector space of functions closed under differentiation.
\end{itemize}
Note that polynomials and linear combinations of products of exponentials fall into this class, and we will cover these particular cases in this section\footnote{In general, the second point restricts the functions $f^{(i)}_r(t)$ to be linear combinations of functions of the form $t^s e^{dt}$, for $s$ a natural number and $d$ an arbitrary complex constant, i.e., linear combinations of products of polynomials with exponentials.}. These special cases have the benefit that they can approximate arbitrary continuous functions. In experiment, we will take them to approximate correlation functions, or, as the correlation functions themselves arise from fits of experimental data, we could simply fit a function in these classes to the data directly.

Here, we outline a procedure for constructing a Van Loan differential equation to compute $\Dyson^f_U(A_1, \dots, A_m)(t)$ for $f$ drawn from the above class. Note that, for functions of product form, one approach is to simply absorb $f_i(t)$ into the definition of $A_i(t)$ and apply the method from the original $f=1$ case. This is valid, however it will generally introduce explicit time-dependence into $A_i(t)$. In quantum control problems, $A_i(t)$ will usually only depend on time as a function of the control amplitudes, and it is computationally preferable that the generators in the newly constructed Van Loan differential equations also depend on time only through the control amplitudes.

For $f$ satisfying the above conditions, we construct a Van Loan differential equation to compute $\Dyson_U^f(A_1, \dots, A_m)(t)$ using the following algorithm.
\begin{enumerate}
	\item Define a variable $x_0(t) = \Dyson_U(f_1(t)A_1, \dots, f_m(t)A_m)(t)$. This is the term we wish to compute.
	\item Differentiate $x_0(t)$ with respect to time. The result will be a linear combination of expressions of the same form as the original integral. 
	\item Add any newly appearing expressions into the list of variables.
	\item Differentiate the new variables from the previous step.
	\item Repeat steps $3$ and $4$ until no new expressions appear.
\end{enumerate}
The assumption that the function pieces $f_r^{(i)}$ are drawn from a finite dimensional vector space of functions closed under differentiation ensures that this procedure terminates after a finite number of steps. Once the procedure terminates, we write down the resulting coupled differential equation for the defined variables. The generator for this differential equation will be an upper triangular block matrix, i.e., the generator is of the form amenable to analysis via Theorem~\ref{theorem:HP_theorem}.

We do this procedure when the $f^{(i)}_r$ are exponentials, and when they are polynomials.
\subsubsection{Products of Exponentials}

First, consider the case 
\begin{equation}
	f(t_1, \dots, t_m) = \exp(d_1 t_1 + \dots + d_m t_m) = \exp(d_1 t_1) \dots \exp(d_m t_m),
\end{equation}
where $d_1, \dots, d_m \in \complex$. That is, $f$ is a product of exponentials in each time variable. Hence, the goal is to write 
\begin{equation}
\begin{aligned}
	\Dyson_U(e^{d_1 t}A_1,& \dots, e^{d_m t}A_m)(t)= \\
	&U(t) \int_0^t dt_1 \dots \int_0^{t_{m-1}}dt_m \exp\left(\sum_{i=1}^m d_i t_i \right) \tilde{A}_1(t_1) \dots \tilde{A}_m(t_m),
\end{aligned}
\end{equation}
where $\tilde{A}_i(t) = U^{-1}(t)A_i(t)U(t)$, as part of the solution to a linear matrix differential equation. 

To do this, we follow the algorithm constructing Van Loan differential equations given at the beginning of Section \ref{sec:generalizations}. First, we denote the function:
\begin{equation}
	x_0(t) = \Dyson_U(e^{d_1 t}A_1, \dots, e^{d_m t}A_m)(t).
\end{equation}
Differentiating, we find
\begin{align}
\nonumber	\dot{x}_0(t) &= G(t) x_0(t) + A_1(t)\left(e^{d_1t}U(t)\int_0^t dt_2 \dots \int_0^{t_{m-1}}dt_m e^{d_2 t_2 + \dots + d_m t_m} \tilde{A}_2(t_2) \dots \tilde{A}_m(t_m)\right) \\
	&=G(t) x_0(t) + A_1(t) \left(e^{d_1 t}\Dyson_U(e^{d_2 t} A_2, \dots, e^{d_m t}A_m)\right) .
\end{align}
The new expression appearing here is the second term in the brackets. Hence, we define this as a new variable:
\begin{equation}
	x_1(t) = e^{d_1 t}\Dyson_U(e^{d_2 t} A_2, \dots, e^{d_m t}A_m)(t). 
\end{equation} 
Next, differentiating $x_1(t)$, we find:
\begin{equation}
	\dot{x}_1(t) = (d_1 \I + G(t))x_1(t) + A_2(t)e^{(d_1+d_2)t}\Dyson_U(e^{d_3 t }A_3, \dots, e^{d_m t}A_m)(t),
\end{equation}
and again we define a new variable for the newly appearing term:
\begin{equation}
	x_2(t) = e^{(d_1+d_2)t}\Dyson_U(e^{d_3 t }A_3, \dots, e^{d_m t}A_m)(t).
\end{equation}

Continuing this procedure until no new variables appear results in the following family of functions:
\begin{equation}
\begin{aligned}
	x_0(t) &= \Dyson_U(e^{d_1 t}A_1, \dots, e^{d_mt}A_m)(t)\\
	x_1(t) &= e^{d_1 t}\Dyson_U(e^{d_2 t}A_2, \dots, e^{d_m t}A_m)(t) \\
	x_2(t) &= e^{(d_1 + d_2)t}\Dyson_U(e^{d_3 t} A_3, \dots, e^{d_m t}A_m)(t) \\
	&\vdots \\
	x_m(t) &= e^{(d_1 + \dots + d_m)t}U(t),
\end{aligned}
\end{equation}
which evolve according to the coupled differential equations:
\begin{equation}
\begin{aligned}
	\dot{x}_0(t) &= G(t) x_0(t) + A_1(t) x_1(t) \\
	\dot{x}_1(t) &= (G(t) + d_1 \I_n) x_1(t) + A_2(t)x_2(t) \\
	\dot{x}_2(t) &= (G(t) +(d_1 + d_2)\I_n)x_2(t) + A_3(t)x_3(t)\\
	& \vdots\\
	\dot{x}_m(t) &= (G(t) + (d_1 + \dots + d_m) \I_n) x_m(t)
\end{aligned}
\end{equation}
with initial conditions $x_0(0) = \dots = x_{m-1}(0) = 0$, and $x_m(0) = \I_n$. Note that the generator for this system has upper triangular block form. In particular, the generator lies in $\M_{m+1}(\M_n)$ and has all blocks equal to $0$ except:
\begin{itemize}
	\item The diagonal is given by $(G(t), e^{d_1 t}G(t), e^{(d_1 + d_2)t} G(t), \dots, e^{(d_1 + \dots + d_m)t}G(t))$, and
	\item The first off diagonal is $(A_1(t), \dots, A_m(t))$.
\end{itemize}
For example, when $m=2$, we have:
\begin{equation}
	\left(\begin{array}{c}
    			\dot{x}_0(t) \\
			\dot{x}_1(t) \\
			\dot{x}_2(t) 
			\end{array}\right)
	=
	\left(\begin{array}{ccc}
    			G(t) 	& A_1(t) 	& 0\\
			0 	& G(t) + d_1 \I_n & A_2(t) \\
			0 & 0  & G(t) + (d_1 + d_2) \I_n 
			\end{array}\right)
	\left(\begin{array}{c}
    			x_0(t)\\
			x_1(t) \\
			x_2(t) 
			\end{array}\right).
\end{equation}
Hence, if we take the time ordered exponential of the above generator, the desired integral $\Dyson(e^{d_1 t}A_1, e^{d_2 t} A_2)$ will be in the top right block. Denoting the generator as $L_2(t)$, we may also explicitly determine the blocks of the time ordered exponential using Theorem~\ref{theorem:HP_theorem}:
\begin{equation}
\label{eq:expVLDE}
\mathcal{T}\exp\left(\int_0^t dt_1 L_2(t_1)\right) = \left(\begin{array}{ccc}
    			U(t) 	& \Dyson_U(e^{d_1t}A_1) 	& \Dyson_U(e^{d_1 t}A_1, e^{d_2 t} A_2) \\
			0 	& e^{d_1 t}U(t) & e^{d_1 t}\Dyson_U(e^{d_2t}A_2) \\
			0 & 0  & e^{(d_1+d_2) t}U(t) 
			\end{array}\right).
\end{equation}

\subsubsection{Polynomials for Second Order Integrals} \label{section:polynomial_system}

Next, we consider polynomials, and in particular exhibit the procedure for second order integrals involving polynomials. That is, second order integrals of the form
\begin{equation}
	\Dyson^p_U(A_1,A_2)(t) = U(t)\int_0^t dt_1 \int_0^{t_1} dt_2 p(t_1,t_2) \tilde{A}_1(t_1)\tilde{A}_2(t_2), \label{equation:second_polynomial}
\end{equation} 
where $p(t_1,t_2)$ is a polynomial in $t_1$ and $t_2$ with either real or complex coefficients, and again $\tilde{A}_i(t) = U^{-1}(t)A_i(t)U(t)$ with $U(t) = \mathcal{T}\exp\left(\int_0^t dt_1 G(t_1)\right)$. Let $s_1$ and $s_2$ be the respective highest powers of $t_1$ and $t_2$ occurring in $p$, so that it may be decomposed as: 
\begin{equation}
	p(t_1,t_2) = \sum_{i=0}^{s_1}\sum_{j=0}^{s_2} c_{ij}t_1^i t_2^j.
\end{equation}
With respect to this decomposition, the integral becomes the linear combination
\begin{equation}
	\Dyson^p_U(A_1,A_2)(t) = \sum_{i=0}^{s_1}\sum_{j=0}^{s_2} c_{ij} \Dyson_U(t^i A_1, t^j A_2).
\end{equation}
	
Here, we show how the terms $\Dyson_U(t^i A_1, t^j A_2)$ for all $0 \leq i \leq s_1$ and $0 \leq j \leq s_2$ can be computed using a \emph{single} Van Loan differential equation. Hence, all terms $\Dyson^p_U(A_1,A_2)(t)$ with $p(t_1,t_2)$ a polynomial of degree at most $s_1$ in $t_1$ and $s_2$ in $t_2$ may computed using this single generator. We construct this generator by applying the procedure to the highest order term $\Dyson_U(t^{s_1} A_1, t^{s_2} A_2)$, and find by chance that the solution contains all terms of lower order. In particular, the solution to the Van Loan differential equation for computing $\Dyson_U(t^{s_1} A_1, t^{s_2} A_2)$ contains a basis for the vector space of expressions
\begin{equation}
	\textnormal{span}\{ \Dyson_U(t^i A_1,t^j A_2) : 0 \leq i \leq s_1, 0 \leq j \leq s_2\}.
\end{equation}
Note that we have not proven this, but we conjecture it to be true, and have computationally verified this conjecture for all pairs $\{(s_1, s_2) : 0 \leq s_1 \leq 15, 0 \leq s_2 \leq 15\}$.

Applying the procedure for generating the Van Loan differential equation to the term $\Dyson_U(t^{s_1} A_1, t^{s_2} A_2)$, we arrive at the following set of functions:
\begin{equation}
\begin{aligned}
	x_0(t) &= U(t),\\
	x_j(t) &= t^j x_0(t) = t x_{j-1}(t), \text{ for } j \in \{1, \dots, s_1 + s_2\},\\
	y_0(t) &= \Dyson_U(t^{s_2} A_2)(t) = U(t)\int_0^t dt_2 t_2^{s_2} \tilde{A}_2(t_2) = U(t) \int_0^t dt_2 U^{-1}(t_2)A_2(t_2)x_{s_2}(t_2), \\
	y_j(t) & = t^jy_0(t) = t y_{j-1}(t), \text{ for } j \in \{1, \dots,  s_1\}, \\
	z_j(t) & = \Dyson_U(t^jA_1, t^{s_2} A_2)= U(t) \int_0^t dt_1 U^{-1}(t_1)A_1(t_1)y_j(t),\text{ for } j \in \{0, \dots, s_1\}.
\end{aligned}
\end{equation}
This set of variables evolves in time according to the following first order coupled differential equation:
\begin{equation}
\begin{aligned}
	\dot{x}_0(t) &= G(t) x_0(t), \\
	\dot{x}_j(t) &= j x_{j-1}(t) + G(t) x_j(t) \text{ for } j \in \{1, \dots, s_1+s_2\}, \\
	\dot{y}_0(t) &= G(t) y_0(t) + A_2(t) x_{s_2}(t), \\
	\dot{y}_j(t) &= j y_{j-1}(t) + G(t) y_j(t) + A_2(t)x_{j+s_2}(t), \text{ for } j \in \{1, \dots, s_1\}, \\
	\dot{z}_j(t) &= A_1(t) y_j(t) + G(t) z_j(t), \text{ for } j \in \{0, \dots, s_1\},
\end{aligned}
\end{equation}
with initial conditions of all variables being $0$ at $t=0$ other than $x_0(0) = \I_n$. Again, note that the derivative of each function only depends on the functions coming before it in the ordering 
\begin{equation}
	(x_0, x_1, \dots, x_{s_1+s_2}, y_0, y_1, \dots, y_{s_1}, z_0, \dots, z_{s_1}),
\end{equation}
and hence the generator for this system, which we denote $L_{s_1,s_2}(t)$ is an upper triangular block matrix in $\M_{3s_1+s_2+3}(\M_n)$. An explicit description of how to construct $L_{s_1,s_2}(t)$ is:
\begin{itemize}
	\item Every diagonal block is $G(t)$,
	\item For the first off diagonal, the first $s_1+1$ blocks are $0$, and the remaining blocks are
	\begin{equation}
		(s_1, s_1-1, \dots, 0, s_1 + s_2, s_1+s_2-1, \dots, 1),
	\end{equation}
	and
	\item The $(s_1+1)^{th}$ off diagonal is given by $s_1 +1$ repetitions of $A_1(t)$, then $s_1+1$ repetitions of $A_2(t)$, followed by zeros.
\end{itemize}

By construction, the upper right block of $\mathcal{T}\exp\left(\int_0^t dt_1 L_{s_1,s_2}(t_1)\right)$ is $\Dyson_U(t^{s_1}A_1, t^{s_2}A_2)$. Furthermore, we conjecture the following.
\begin{conjecture} \label{conjecture:polynomial}
It holds that the top right $(s_1 + 1) \times (s_2 + 1)$ blocks of
\begin{equation}
	\mathcal{T}\exp\left(\int_0^t dt_1 L_{s_1,s_2}(t_1)\right)
\end{equation}
is a basis for the vector space of expressions
\begin{equation}
	\textnormal{span}\{ \Dyson_U(t^i A_1,t^j A_2) : 0 \leq i \leq s_1, 0 \leq j \leq s_2\}.
\end{equation}
\end{conjecture}

To get a sense for this claim, examine the special case $s_1= s_2 = 1$. By applying Theorem~\ref{theorem:HP_theorem}, we find the top $2 \times 2$ blocks of $\mathcal{T}\exp\left(\int_0^t dt_1L_{1,1}(t_1)\right)$ are given by:
\begin{equation}
	\left(\begin{array}{cc}
    			\Dyson_U(A_1,tA_2) + \Dyson_U(tA_1, A_2)& \Dyson_U(tA_1,tA_2) \\
			 \Dyson_U(A_1,A_2)& \Dyson_U(A_1,tA_2)
			\end{array}\right),
\end{equation}
and it can be checked that these blocks form a basis for the desired set. 

We have computationally verified this conjecture for all pairs $s_1,s_2 \in \{0, \dots, 15\}$ the details of which can be found in Appendix~\ref{app:mathematica}.

\section{\label{sec:examples}Examples}
With this section, we give five examples of increasing complexity for numerical engineering of effective Hamiltonians using the Van~Loan differential equation framework. First, we set up two rather standard decoupling problems with known analytical solutions and arrive at control sequences which resemble ones that have been known for some time. Our aim is not to reiterate these solutions, rather it is to demonstrate that the length of the control sequences found using block matrix numerical tools does not significantly exceed that of the sequences that have been derived analytically based on physical insights. This demonstration provides an encouraging starting point for employing the same tools to tackle far harder control problems, for which the search of analytical solutions is intractable. With the third example we provide an illustration for a problem that demands simultaneous minimization of some Dyson terms, while preserving or maximizing other Dyson terms. Such control problems are very common in many sensing and spectroscopy applications. For the first three examples we apply no transfer functions, ensemble effects nor pulse waveform bandwidth constraints etc. Only maximum amplitude constraints are used which generally yield pulses of rather jagged form, however, such constraints are typically the only constraints considered when deriving analytical solutions. With the fourth and the fifth example, we give two experimentally realistic control search examples for which we demand that the pulse ends go smoothly to zero and that the spectral width of the pulse waveform be limited. The fifth example also employs a non-trivial set of experimentally determined transfer functions $\lbrace \Xi^{(\gamma)} \rbrace$.

First, we introduce the matrix norm and the matrix fidelity function that are used throughout this section. We take $\|A\|$ to stand for the Frobenius norm \cite{Watrous_2018} for a matrix $A \in \M_n$, defined as $\|A\| = \sqrt{\Tr \left( A^\dagger A \right)}$. We also define a fidelity function $\F(U,V)$ for a pair of matrices $U,V \in \M_n$: $\F(U,V) = \sqrt{
  \frac{\Tr\left(U^\dagger V\right) \Tr\left(V^\dagger U\right)}{\Tr\left(U^\dagger U\right) \Tr\left(V^\dagger V\right)}
  }$. Furthermore, we introduce a shorthand that makes the definitions of our target functions in this section more concise. For any nested operator integral $\Dyson_{U}(A_1, \dots, A_n)$, we denote its maximum Frobenius norm maximized over all permissible control sequences $a(t)$, $t \in [0,T]$, as $\max_{a(t)} \|\Dyson_{U}(A_1, \dots, A_n)\|$. For the numerical pulse searches, we always use a target function $\Phi$ that is a linear combination of different matrix norms for various Dyson terms and the fidelity of $U(T)$ with the target unitary $U_\text{target}$. A typical target function takes the following form:
\begin{equation}
   \label{eq:sampleTargetFunction}
 \Phi = \sum_{\gamma \in \Gamma} p^{(\gamma)} \left[ p_0 \left(\F[U^{(\gamma)}(T),U_\text{target}]\right)^2 + \sum_{i>0} p_i \left(1 - \frac{\| \Dyson_{U^{(\gamma)}}(A_i) \|^2}{\left[\max_{a(t)} \|\Dyson_{U^{(\gamma)}}(A_i)\| \right]^2} \right) \right],
\end{equation}
where $0 \le p_i \le 1$, for all $i$, such that $\sum_i p_i = 1$. $\lbrace p_i \rbrace$ denote various weights of constituent optimization targets. It is easy to see that $0 \le \Phi(\alpha) \le 1$, for all $\alpha$, and equal to $1$ if and only if $U^{(\gamma)}(T)=U_\text{target}$ as well as $\Dyson_{U^{(\gamma)}}(A_i)=0$, for all $i$. The linear form of the target function in Equation~\eqref{eq:sampleTargetFunction} is, of course, not strictly necessary, but it does greatly simplify some calculations.

All control searches were undertaken using the modified GRAPE algorithm for evaluating partial derivatives with respect to piecewise constant control amplitudes, the details of which are given in Appendix~\ref{sec:GRAPE}. Our general procedure for finding a working control sequence is to first pick a pulse length $T$ and thereafter a number of time steps $N$. We kept all subintervals of $[0,T]$ of equal length $\Delta T = T/N$ and always picked an $N$ for which $\Delta T < \tau_\text{Rabi} / 30$, where $\tau_\text{Rabi}$ is the length of a Rabi cycle. For each example we picked $\lbrace p_i \rbrace$, appearing in Equation~\eqref{eq:sampleTargetFunction}, that yielded relatively equal convergence rates for each constituent of the total target function during the control optimization procedure. Since this was control problem specific we generally determined the particular weights $\lbrace p_i \rbrace$ by running the optimization for a large number ($\sim 100$) of random initial seeds and monitoring the convergence of each constituent target for a given $\lbrace p_i \rbrace$. We adjusted these weights until we observed roughly equal convergence from all constituent targets. For the examples in this section, this procedure yielded relatively equal weights for the Dyson terms in Equation~\eqref{eq:sampleTargetFunction} and a $p_0$ value slightly lower than $\lbrace p_i \rbrace, i>0$. We state the specific target functions used for our optimizations in the upcoming subsections.

Given some $T$ and $N$ and having determined the suitable $\lbrace p_i \rbrace$, we generated $\sim 40$ seed waveforms $\alpha^{(0)}$, the pulse amplitudes $\lbrace \alpha^{(0)}_{i,j} \rbrace$ of which were drawn from independent uniform distributions. We used \mathematica{}'s \texttt{FindMaximum} function for multivariate conjugate-gradients optimization on these seeds with the maximum number of target function evaluations set to 1000. If none of the $\sim 40$ searches yielded $\Phi > 0.9999$ we increased $T$ and $N$ and repeated the same procedure. For the seeds that reached $\Phi > 0.9999$ in under 1000 $\Phi$ evaluations, we let the optimization run until FindMaximum was terminated by machine precision.

\subsection{Dipolar Decoupling}
The simplest numerical effective Hamiltonian example that we consider is the problem of dipolar decoupling. For such a problem, we imagine a pair of spins coupled via the dipolar Hamiltonian $D = 3 \sigma_z \otimes \sigma_z - \sum_{i \in \lbrace x,y,z \rbrace} \sigma_i \otimes \sigma_i$. Here, we use no optimization transfer function and we assume the ensemble size $|\Gamma|$ to be equal to one with the transfer function $\Xi$ acting as an identity, i.e., $b(t) = a(t)$. We take the spin control to be global, such that the Rabi fields for either spin are identical. Also, for notational convenience, we assign $a_x(t) = a_1(t)$ and $a_y(t) = a_2(t)$ such that the two\=/spin system generator becomes
\begin{equation}
   \label{eq:twoSpinGenerator}
   G_2(t) = -i\frac{a_x (t)}{2} \left( \sigma_x \otimes \I_2 + \I_2 \otimes \sigma_x \right) - i\frac{a_y (t)}{2} \left( \sigma_y \otimes \I_2 + \I_2 \otimes \sigma_y \right) ,
\end{equation}
which generates the following system propagator $U_2(t) = \Texp \left[\int_0^t dt_1 G_2(t_1) \right]$. The control task is to engineer a sequence $a(t)$ that enables the spins to evolve effectively uncoupled. First order perturbative solution to the problem dictates setting the first derivative of $U_2(T)$ in the direction of $D$ to zero, i.e., we want $\Dyson_{U_2}(D) = 0$. The dipolar decoupling problem is a simple yet non-trivial problem. It is easy to show that the desired $U_2(t)$, $t \in [0,T]$, that yields $\Dyson_{U_2}(D)=0$ cannot be generated with either $a_x (t) = 0$ for all $t \in [0,T]$ or $a_y (t) = 0$ for all $t \in [0,T]$.

Here, we are not concerned about the final unitary $U_2(T)$ generated on either of the spins, as our first and foremost aim is to demonstrate an ability to engineer control sequences that yield desired values for various Dyson terms. With the subsequent examples we tackle the problem of simultaneous engineering of various Dyson terms and final unitaries. Accordingly, our target function for the optimization is
\begin{equation}
   \label{eq:dipolarTarget}
   \Phi = 1 - \frac{\|\Dyson_{U_2}(D)\|^2}{\left[\max_{a(t)}\|\Dyson_{U_2}(D)\|\right]^2} ,
\end{equation}
where the denominator is a normalization factor ensuring that $0 \le \Phi \le 1$. We now set up a block matrix differential equation for $V(t) \in \M_2(\M_4)$, that will be used for evaluating $\Phi$. It easy to show either by differentiation or by employing Theorem~1 that
\begin{align}
   \label{eq:dipolarVLDE}
V(t) &= \Texp \left[\int_0^t dt_1 \left(\begin{array}{cc}
    			G_2(t_1)	& D \\
			0 	& G_2(t_1)
\end{array}\right) \right] = \left(\begin{array}{cc}
    			U_2(t)	& \Dyson_{U_2}(D) \\
			0 	& U_2(t)
\end{array}\right) .
\end{align}
Consequently, our target could also be given as
\begin{align}
  \label{eq:dipolarVTarget}
  \Phi = 1 - \frac{\Tr\left[V_{1,2}^\dagger(T) V_{1,2}(T)\right]}{24 T^2} .
\end{align}
Following the procedure outlined in Appendix~\ref{sec:GRAPE}, we can evaluate $V(T)$ as a function of $\alpha$ along with the partial derivatives $\lbrace \frac{\partial}{\partial \alpha_{i,j}} V(T) \rbrace$ with respect to the piecewise constant control amplitudes $\lbrace \alpha_{i,j} \rbrace$.

\begin{center}\begin{figure}
   \includegraphics[width=\textwidth]{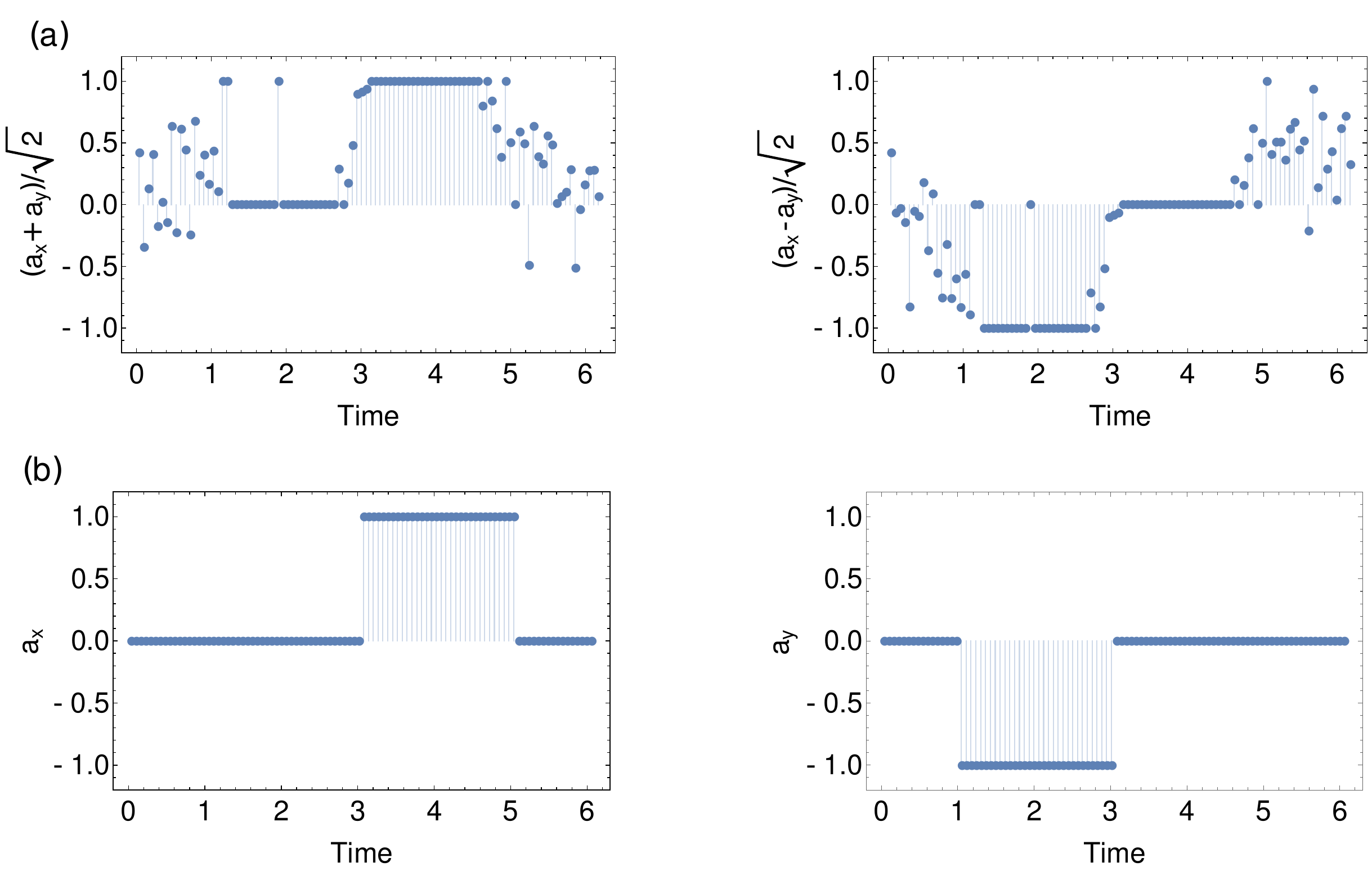}
   \caption{
      \label{fig:dipolarPulse} The numerically found dipolar coupling sequence is shown in (a), and the analytic dipolar decoupling sequence introduced by Mehring \cite{mehring_1973} is shown in (b). For the numerically found sequence, we plot $\left[a_x(t)+a_y(t)\right]/\sqrt{2}$ and $\left[a_x(t)-a_y(t)\right]/\sqrt{2}$, rather than $a_x(t)$ and $a_y(t)$ to highlight the similarities with the analytic sequence. This corresponds to a basis change $\sigma_x \rightarrow \frac{1}{\sqrt{2}} \left( \sigma_x + \sigma_y \right)$ and $\sigma_y \rightarrow \frac{1}{\sqrt{2}} \left( \sigma_x - \sigma_y \right)$, which does not affect the value of $\|\Dyson_{U_2}(D)\|$}
\end{figure}\end{center}
Given $V(T)$ and $\lbrace \frac{\partial}{\partial \alpha_{i,j}} V(T) \rbrace$, we can write
\begin{equation}
\frac{\partial}{\partial \alpha_{i,j}} \Phi = -\frac{2}{24 T^2} \text{Re}\left(\Tr\left[V_{1,2}^\dagger(T) \left(\frac{\partial}{\partial \alpha_{i,j}} V(T) \right)_{1,2}\right] \right)
\end{equation}
for all $i$ and $j$. We employ these partial derivatives in the optimization protocol described above. For the elements of $\alpha$ we use amplitude constraints: $-\frac{1}{\sqrt{2}} \le \alpha_{i,j} \le \frac{1}{\sqrt{2}}$, for all $i$ and $j$, to ensure that $|a (t)| \le 1$, for all $t \in [0,T]$. We find a pulse with a total length of $T=6.2$ consisting of $N=100$ subintervals shown in Figure~\ref{fig:dipolarPulse}(a) which yields $\frac{\|\Dyson_{U_2}(D)\|}{\max_{a(t)}\|\Dyson_{U_2}(D)\|} = 3.1 \times 10^{-7}$. We point out the rough similarity to the dipolar sequence proposed by Mehring \cite{mehring_1973} consisting of two square $116^\circ 14'$ pulses with orthogonal phases depicted in Figure~\ref{fig:dipolarPulse}(b). The similarity is noticeable after a particular basis rotation, described in the figure caption, is performed on our numerically found sequence which is only $1.02$ times longer than the known analytical solution.

\subsection{Universal Decoupling with Control Variations}
With this example we turn to single spin control. Again, we use no optimization transfer function and we assume $|\Gamma| = 1$ with the transfer function $\Xi$ acting as an identity. The system generator is 
\begin{equation}
   \label{eq:singleSpinGenerator}
   G_1(t) = -i\frac{a_x (t)}{2} \sigma_x - i\frac{a_y (t)}{2} \sigma_y 
\end{equation}
and generates a system propagator $U_1(t) = \Texp \left[\int_0^t dt_1 G_1(t_1) \right]$. We consider a universal decoupling sequence which would decouple all non-identity operators $\sigma_x, \sigma_y, \sigma_z$ acting on a single spin, which translates to setting the respective lowest order directional variations of $U_1(T)$ to zero, i.e., $\Dyson_{U_1}(\sigma_x)=\Dyson_{U_1}(\sigma_y)=\Dyson_{U_1}(\sigma_z)=0$. In addition, we also demand that the sequence is robust against variations in the control control amplitudes, such that $\Dyson_{U_1}\left[a_x(t) \sigma_x\right] = 0$ and $\Dyson_{U_1}\left[a_y(t) \sigma_y\right] = 0$. A sequence know to have such properties is called an XY8 sequence \cite{gullion_1990}, which implements an identity operation. To demonstrate the ability of our numerical control searches in incorporating averaging of time dependent operators, we set up a search that would simultaneously set $\Dyson_{U_1}(\sigma_x)=\Dyson_{U_1}(\sigma_y)=\Dyson_{U_1}(\sigma_z)=\Dyson_{U_1}\left[a_x(t) \sigma_x\right] = \Dyson_{U_1}\left[a_y(t) \sigma_y\right]=0$ while implementing an identity operation. We search for a pulse with the following target:
\begin{equation}
\begin{aligned}
  \Phi &= \frac{2}{5} \left( 1 - \frac{1}{3} \sum_{i \in \lbrace x,y,z \rbrace} \frac{\|\Dyson_{U_1}(\sigma_i)\|^2}{\left[\max_{a(t)}\|\Dyson_{U_1}(\sigma_i)\|\right]^2} \right) \\
  &+ \frac{2}{5} \left( 1 -\frac{1}{2} \sum_{i \in \lbrace x,y \rbrace} \frac{\|\Dyson_{U_1}\left[a_i(t) \sigma_i\right]\|^2}{\left[\max_{a(t)}\|\Dyson_{U_1}\left[a_i(t) \sigma_i\right]\|\right]^2} \right) + \frac{1}{5} \left( \F \left[ \I_2, U_1(T) \right] \right)^2 .
\end{aligned}
\end{equation}

A suitable Van Loan generator $L(t) \in \M_6(\M_2)$ for such a target function is
\begin{equation}
L(t) = \left(\begin{array}{cccccc}
    			G_1(t) 	& \sigma_x 	& 0 		& 0	 	& 0	& 0 \\
			0 	& G_1(t) 	& \sigma_y	& 0 		& 0	& 0 \\
			0	& 0		& G_1(t)	& \sigma_z	& 0	& 0\\
			0	& 0		& 0		& G_1(t)	& a_x(t) \sigma_x	& 0 \\
			0	& 0		& 0		& 0		& G_1(t)	& a_y(t) \sigma_y \\
			0	& 0		& 0		& 0		& 0	& G_1(t)
			\end{array}\right) ,
\end{equation}
that generates the following Van Loan propagator:
\begin{equation}
\begin{aligned}
\label{eq:FirstOrderIdentityControlVLDE}
V(t) &= \Texp \left(\int_0^t dt_1 L(t_1) \right) \\
&=\left(\begin{array}{cccccc}
    			U_1(t)	& \Dyson_{U_1}(\sigma_x)	& \cdot & \cdot	& \cdot & \cdot	\\
			0	& U_1(t)		& \Dyson_{U_1}(\sigma_y)	& \cdot	& \cdot	& \cdot	\\
			0	& 0			& U_1(t)		& \Dyson_{U_1}(\sigma_z)	& \cdot	& \cdot \\
			0	& 0			& 0			& U_1(t)	& \Dyson_{U_1}\left[ a_x(t)\sigma_x \right]	& \cdot \\
			0	& 0			& 0			& 0		& U_1(t)	& \Dyson_{U_1}\left[ a_y(t)\sigma_y \right] \\
			0	& 0			& 0			& 0		& 0		& U_1(t)
			\end{array}\right) .
\end{aligned}
\end{equation}
Note that, we have not specified the $V(t)$ elements that are not relevant for our control problem. Given the $V(t)$ above, $\Phi$ can be determined as
\begin{equation}
\begin{aligned}
  \label{eq:FirstOrderIdentityControlVTarget}
  \Phi &= \frac{2}{5} \left(1 - \frac{1}{3} \sum_{i=1}^3 \frac{\Tr\left[V_{i,i+1}^\dagger(T) V_{i,i+1}(T)\right]}{2 T^2} \right) 
  +\frac{2}{5} \left(1 - \frac{1}{2} \sum_{i=4}^5 \frac{\Tr\left[V_{i,i+1}^\dagger(T) V_{i,i+1}(T)\right]}{T^2} \right) \\
   &+ \frac{1}{5} \frac{\Tr \left[V_{1,1}^\dagger(T)\right] \Tr \left[V_{1,1}(T)\right]}{4} .
\end{aligned}
\end{equation}

\begin{center}\begin{figure}
   \includegraphics[width=\textwidth]{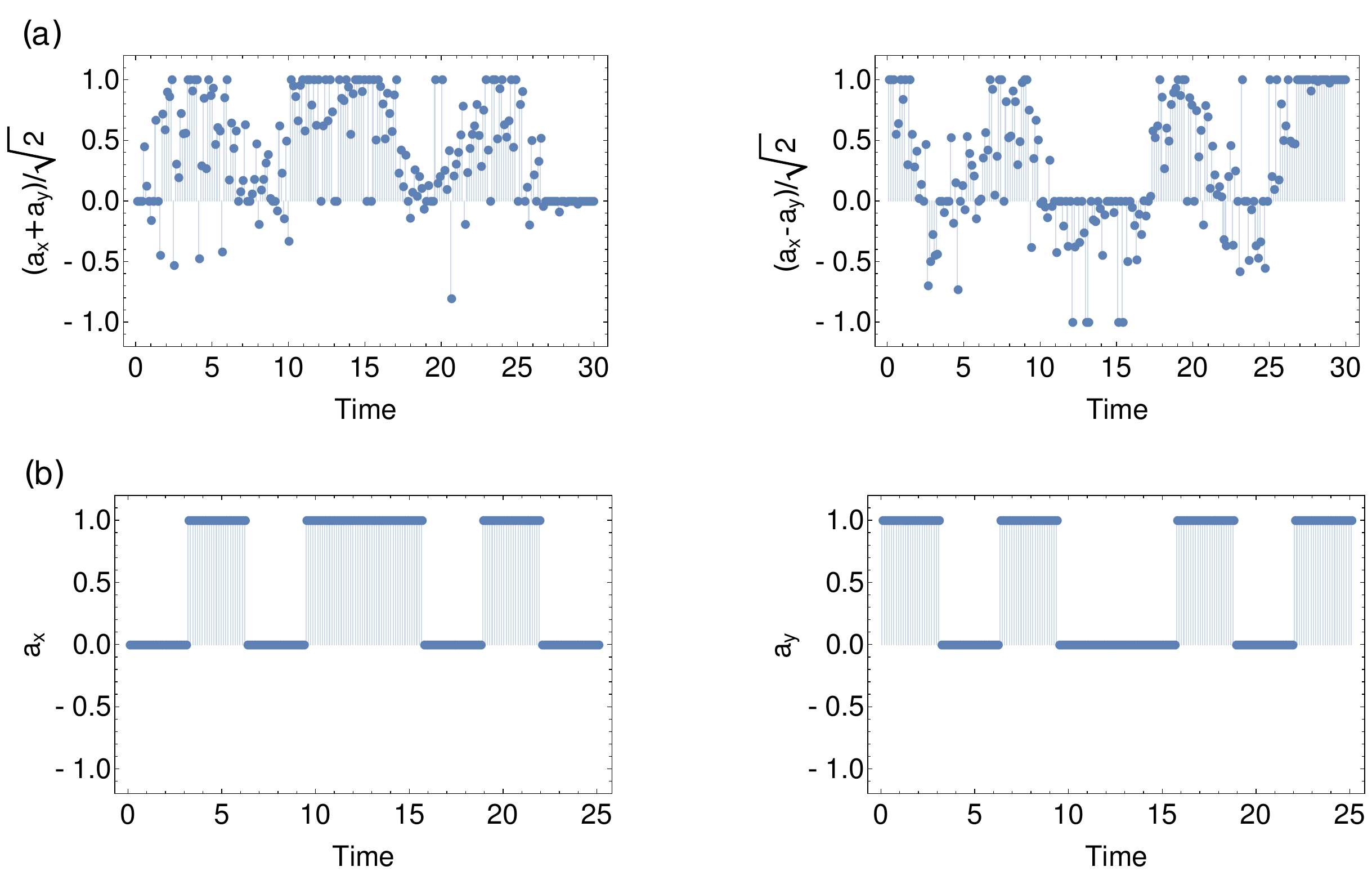}
   \caption{
      \label{fig:qubitDecouplingControlPulse} (a) Numerically found universal decoupling pulse robust to control variations implementing an $\I_2$ gate; (b) The analytic XY8 sequence \cite{gullion_1990} satisfying the same properties. As in Figure~\ref{fig:dipolarPulse}, for the numerically found sequence, we plot $\left[a_x(t)+a_y(t)\right]/\sqrt{2}$ and $\left[a_x(t)-a_y(t)\right]/\sqrt{2}$ to highlight the similarities to the analytic solution. Again, this can be thought of as a basis change which leave the values of Dyson term norms and $U_1(T)$ invariant because $U_1(T) = \I_2$.
   }
\end{figure}\end{center}
For the elements of $\alpha$ we again use amplitude constraints: $-\frac{1}{\sqrt{2}} \le \alpha_{i,j} \le \frac{1}{\sqrt{2}}$, for all $i$, $j$; and using our gradient optimization scheme, we find a pulse with a total length of $T=30$ consisting of $N=200$ subintervals with its pulse metrics given as: $1-\F \left[ \I_2, U_1(T) \right] = 2.8 \times 10^{-14}$, $\frac{\|\Dyson_{U_1}(\sigma_x)\|}{\max_{a(t)}\|\Dyson_{U_1}(\sigma_x)\|} = 2.2 \times 10^{-6}$, $\frac{\|\Dyson_{U_1}(\sigma_y)\|}{\max_{a(t)}\|\Dyson_{U_1}(\sigma_y)\|} = 2.4 \times 10^{-6}$, $\frac{\|\Dyson_{U_1}(\sigma_z)\|}{\max_{a(t)}\|\Dyson_{U_1}(\sigma_z)\|} = 1.6 \times 10^{-7}$, $\frac{\|\Dyson_{U_1}\left[a_x(t) \sigma_x\right]\|}{\max_{a(t)}\|\Dyson_{U_1}\left[a_x(t) \sigma_x\right]\|} = 6.2 \times 10^{-6}$, $\frac{\|\Dyson_{U_1}\left[a_y(t) \sigma_y\right]\|}{\max_{a(t)}\|\Dyson_{U_1}\left[a_y(t) \sigma_y\right]\|} = 6.2 \times 10^{-6}$. We present the pulse waveform in Figure~\ref{fig:qubitDecouplingControlPulse}(a). We note that, while the pulse found is $1.19$ times the length of the XY8 sequence, it does display definite similarities to the latter shown in Figure~\ref{fig:qubitDecouplingControlPulse}(b) after a particular basis rotation, described in the figure caption, is performed.

\subsection{\label{subsec:exchange}Exchange Interaction Recoupling}
With the following example, we wish to highlight that the block matrix method does not only enable the removal of unwanted Hamiltonian terms; it is equally easy to set up optimization targets which retain or reshape parts of the Hamiltonian, while possibly removing others. A problem which arises in many situations involving ensembles of spins, is removing pairwise dipolar interactions between all members of the ensemble as well as inhomogeneities in their energy level splittings, while retaining exchange interaction with some other system or systems the spins are interacting with \cite{Bienfait_2016, Wood_2016, Ansel_2018}.

Such a situation would be described by the following Hamiltonian:
\begin{equation}
   \label{eq:spinExchangeHam}
   H_\text{exchange} = \sum_i \Delta \omega_i \sigma_z^{(i)} + \sum_{\left\langle i,j \right\rangle} \xi_{i,j} D^{(i,j)} + \sum_i g_i \left(\sigma_+^{(i)} \otimes q^{(i)} + \sigma_-^{(i)} \otimes \left[q^{(i)}\right]^\dagger \right) ,
\end{equation}
where $\sigma_\pm = \left(\sigma_x \pm i\sigma_y\right)/2$ and $\left\langle i,j \right\rangle$ denotes a sum over all spin pairs. The first sum in Equation~\eqref{eq:spinExchangeHam} specifies these spin dependent energy level splitting variations $\lbrace \Delta \omega_i \rbrace$, the second sum gives all dipolar interaction strengths $\lbrace\xi_{i,j}\rbrace$ that correspond to the dipolar Hamiltonian $D^{(i,j)}$ for a pair of spins. The third sum contains the aforementioned exchange interactions $\sigma_+^{(i)} \otimes q^{(i)} + \sigma_-^{(i)} \otimes \left[q^{(i)}\right]^\dagger$, that can often be substantially weaker than the other two terms, yet this is frequently the term in the Hamiltonian that leads to desirable spin dynamics.

Once again, we use no optimization transfer function and we assume $|\Gamma| = 1$ with the transfer function $\Xi$ acting as an identity. Here, we consider two system propagators $U_1(t)$ and $U_2(t)$ that are generated by $G_1(t)$ and $G_2(t)$ defined by Equation~\eqref{eq:singleSpinGenerator} and \eqref{eq:twoSpinGenerator}, respectively. Our block matrix tools enable us to search for control sequences that would effectively remove the spin-spin dipolar couplings and variations in level splittings, while retaining the form of the exchange interaction and performing an identity operation. To achieve this, we need to set $\Dyson_{U_1}(\sigma_z) = \Dyson_{U_2}(D) = 0$, $U_1(T) = \I_2$ and $U_1^{-1}(T) \Dyson_{U_1}(\sigma_+) = c \sigma_+$, where $c \in \R$ is a constant. In order to set up a target function $\Phi$ that can reach its maximum value, we do not set up the optimization with any specific value for $c$. Instead, we merely enforce that the integral $I = U_1^{-1}(T) \Dyson_{U_1}(\sigma_+)$ is proportional to $\sigma_+$. This is achieved by demanding that $I$ is orthogonal to $\sigma_z$ and $\sigma_-$, i.e., $\Tr\left(\sigma_z^\dagger I\right) = \Tr\left(\sigma_-^\dagger I\right)=0$. Correspondingly, the optimization target for this problem is 
\begin{equation}
\begin{aligned}
   \label{eq:spinCavityTarget}
  \Phi &= \frac{2}{5} \left( 1  - \frac{\|\Dyson_{U_1}(\sigma_z)\|^2}{2\left[\max_{a(t)}\| \Dyson_{U_1}(\sigma_z)\|\right]^2} - \frac{\|\Dyson_{U_2}(D)\|^2}{2\left[\max_{a(t)}\|\Dyson_{U_2}(D)\|\right]^2} \right) \\
  &+ \frac{2}{5} \left( 1 - \frac{\left| \Tr \left[\sigma_z^\dagger \Dyson_{U_1}(\sigma_+) \right]\right|^2}{2\left[\max_{a(t)}\|\Dyson_{U_1}(\sigma_+)\|\right]^2} - \frac{\left| \Tr \left[\sigma_-^\dagger \Dyson_{U_1}(\sigma_+) \right] \right|^2}{2\left[\max_{a(t)}\|\Dyson_{U_1}(\sigma_+)\|\right]^2} \right) \\
  &+ \frac{1}{5} \left( \F \left[ \I_2, U_1(T) \right] \right)^2 .
\end{aligned}
\end{equation}

Staying completely consistent with the formalism of Section~\ref{sec:controlSetup} in the case of current example would require expressing $\Dyson_{U_1}\left(\sigma_i\right) = \Tr_2 \left[ \Dyson_{U_2}\left(\sigma_i \otimes \I_2\right) \right] /2$, for $\sigma_i \in \lbrace \sigma_+, \sigma_z \rbrace$, where $\Tr_2 \left[ \cdot \right]$ denotes partial trace over the (identity) operator acting on the second Hilbert space. Such identification would enable evaluating all terms in Equation~\eqref{eq:spinCavityTarget} by propagating a single $L(t) \in \M_4(\M_4)$, the diagonal elements of which are all $G_2(t)$ whereas the first off diagonal elements are $D$, $\sigma_+ \otimes \I_2$ and $\sigma_+ \otimes \I_2$. Nevertheless, in order to reduce the dimension of the Van~Loan generator $L(t)$, and hence the computational cost of evaluating $V(T)$, instead we define a block matrix generator $L(t) \in \M_{14}$, which decomposes into a direct sum of $\M_3(\M_2)$ and $\M_2(\M_4)$ and helps us evaluate $\Phi$:
\begin{equation}
\begin{aligned}
L(t) = \left(\begin{array}{ccccc}
			G_1(t)		& \sigma_z	& 0	& 0	& 0 \\
			0	& G_1(t)	& \sigma_+	& 0	& 0 \\
			0	& 0	& G_1(t)	& 0	& 0 \\    			
    			0	& 0	& 0	& G_2(t)	& D	\\
			0	& 0	& 0	& 0	&	G_2(t)
			\end{array}\right) ,
\end{aligned}
\end{equation}
which generates
\begin{equation}
\begin{aligned}
\label{eq:exchangeVLDE}
V(t) &= \Texp \left(\int_0^t dt_1 L(t_1) \right) \\
	 &= \left(\begin{array}{ccccc}
    			U_1(t)	& \Dyson_{U_1}(\sigma_z)	& \Dyson_{U_1}(\sigma_z, \sigma_+)		& 0 &	0\\
			0	& U_1(t)		& \Dyson_{U_1}(\sigma_+)	& 0 	& 0	\\
			0	& 0			& U_1(t)		& 0	& 	0\\
			0	& 0			& 0			& U_2(t)		& \Dyson_{U_2}(D) \\
			0	& 0			& 0			& 0		& U_2(t)
			\end{array}\right) .
\end{aligned}
\end{equation}
Simplifications and reductions of $L(t)$ matrix dimension of this kind are frequent and can provide significant speed ups for the numerical pulse engineering routine. We will employ a similar simplification for the example in Section~\ref{sec:broadBandDipolar}. In the following, when writing our target function $\Phi$ in terms of $V(T)$, we will slightly abuse our notation for specifying the components of a block matrix. We take $V_{i,j}(T)$ to mean the $i$th row and $j$th column of $V(T)$ as it is specified above. However, note that not all blocks of $V(T)$ are of the same size, e.g, $V_{4,5}(T) = \Dyson_{U_2}(D) \in \M_4$ while  $V_{2,3}(T) = \Dyson_{U_1}(\sigma_+) \in \M_2$. It can now be seen that
\begin{align}
   \label{eq:spinCavityTargetV}
   \nonumber \Phi &= \frac{2}{5} \left( 1  - \frac{1}{2}\frac{\Tr\left[V_{1,2}^\dagger(T) V_{1,2}(T)\right]}{2 T^2} - \frac{1}{2}\frac{\Tr\left[V_{4,5}^\dagger(T) V_{4,5}(T)\right]}{24 T^2} \right) + \frac{1}{5} \frac{\Tr \left[V_{1,1}^\dagger(T)\right] \Tr \left[V_{1,1}(T)\right]}{4}\\
   &+ \frac{2}{5} \left( 1 - \frac{1}{2}\frac{\Tr \left[\sigma_z^\dagger V_{2,3}(T) \right] \Tr \left[\sigma_z V_{2,3}^\dagger(T) \right]}{T^2} - \frac{1}{2}\frac{\Tr \left[\sigma_-^\dagger  V_{2,3}(T) \right] \Tr \left[\sigma_- V_{2,3}^\dagger(T) \right]}{T^2} \right) .
\end{align}

\begin{center}\begin{figure}
   \includegraphics[width=\textwidth]{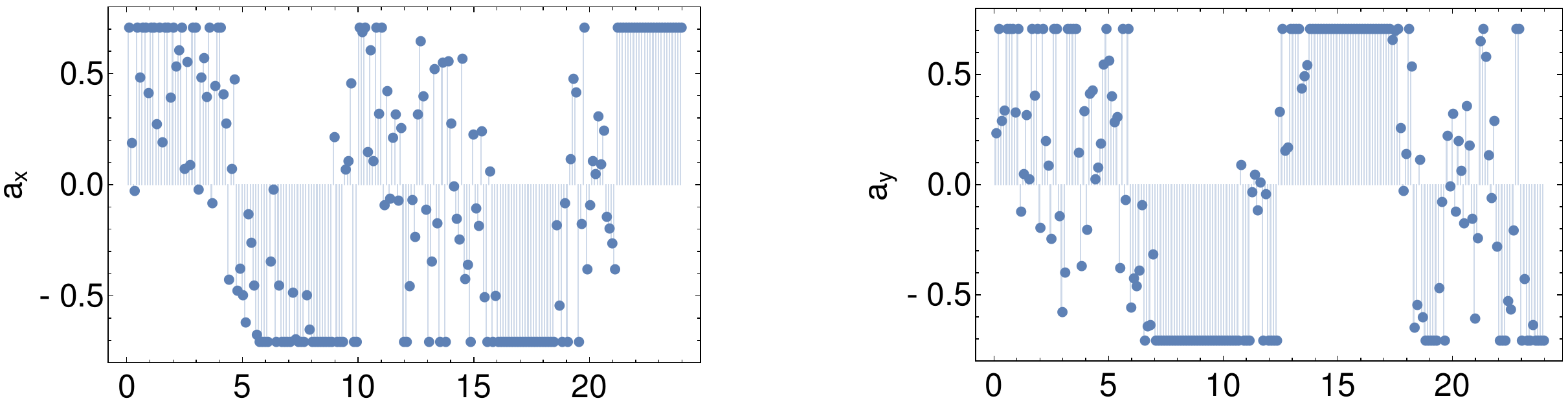}
   \caption{
      \label{fig:coolingPulse} Exchange recoupling pulse sequence: $a_x(t)$ on the left and $a_y(t)$.
   }
\end{figure}\end{center}

For the elements of $\alpha$ we again use amplitude constraints: $-\frac{1}{\sqrt{2}} \le \alpha_{i,j} \le \frac{1}{\sqrt{2}}$, for all $i$, $j$; and using our optimization scheme, we arrive at a control sequence with a total length of $T=24$ and $N=200$ subintervals with the following characteristics: $1-\F \left[ \I_2, U_1(T) \right] <10^{-16}$, $\frac{\|\Dyson_{U_1}(\sigma_+)\|}{\max_{a(t)}\|\Dyson_{U_1}(\sigma_+)\|} = 0.48$, $\frac{\|\Dyson_{U_2}(D)\|}{\max_{a(t)}\|\Dyson_{U_2}(D)\|} = 5.4 \times 10^{-6}$, $\frac{\|\Dyson_{U_1}(\sigma_z)\|}{\max_{a(t)}\|\Dyson_{U_1}(\sigma_z)\|} = 3.1 \times 10^{-7}$, $\frac{\left| \Tr \left[\sigma_-^\dagger \Dyson_{U_1}(\sigma_+) \right] \right|}{\max_{a(t)}\|\Dyson_{U_1}(\sigma_+)\|} = 1.5 \times 10^{-7}$, $\frac{\left| \Tr \left[\sigma_z^\dagger \Dyson_{U_1}(\sigma_+) \right] \right|}{\max_{a(t)}\|\Dyson_{U_1}(\sigma_+)\|} = 1.7 \times 10^{-8}$. The sequence is presented in Figure~\ref{fig:coolingPulse}. The pulse does virtually remove the dipolar and $\sigma_z$ Hamiltonians while rescaling the exchange coupling $\Dyson_{U_1}(\sigma_+)$ by a factor of $0.48$ with extremely small ($<2 \times 10^{-7}$) unwanted orthogonal components.

\subsection{\label{sec:1overfExample}1/f Noise Decoupling}
In this section we demonstrate our ability to engineer control sequences that are designed to decouple stochastic noise characterised by its spectral density function. For that, we employ the tools for that were developed in the previous section for evaluating generalized nested integrals of Equation~\eqref{equation:more_general_integral} kind. We pick 1/f noise due to its ubiquity in solid state devices, including superconducting qubits. Noise spectroscopy experiments on flux qubits \cite{bylander_2011} have clearly revealed a 1/f-like spectral density function for the level splitting variations. We proceed by evaluating the first non-zero term in the perturbative cumulant expansion for the Liouville-von Neumann equation. We then demonstrate how such a toggling frame term can be approximated and consequently minimized using Van Loan differential equations.

When treating the evolution of quantum systems under stochastic operators one has to consider ensemble behaviour averaged over many realizations of the noise process, consequently it is necessary to work with some form of Liouville-von Neumann equation \cite{kubo_1963, Gamliel_1995, green_2012, cappellaro_2006, green_2012, pazsilva_2014}, which determines the evolution of density matrices, as the system differential equation rather than the Schr{\"o}dinger equation. Accordingly, we start with a single qubit Liouville-von Neumann generator which includes a stochastic noise term $G_\text{n}(t) = \varepsilon (t) G_z$ and dictates the evolution of the system
\begin{align}
   \label{eq:1overfHamiltonian}
   G(t) + G_\text{n}(t) = a_x(t) G_x + a_y(t) G_y + \varepsilon (t) G_z,
\end{align}
where $G_i = -i\left(\frac{\sigma_i}{2} \otimes \I-\I\otimes\frac{\sigma_i^\T}{2}\right)$, for $i \in \lbrace x,y,z \rbrace$, and $\varepsilon(t)$ is a stationary, zero mean, Gaussian stochastic function capturing the fluctuations in the qubit level spacing. This implies that $\left\langle \varepsilon(t) \right\rangle = 0$, where the angle brackets denote an ensemble average over noise realizations. Here, we take the power spectral density of $\varepsilon(t)$ to be given by $P(\nu) = \frac{2}{\pi \nu} \left[ \arctan \left(\frac{\nu}{\Lambda_1} \right) - \arctan \left(\frac{\nu}{\Lambda_2} \right) \right]$, where $\Lambda_1$ and $\Lambda_2$ are the smooth low and high frequency cutoffs for $P(\nu)$, respectively. It is easy to show that $\lim_{\Lambda_1 \rightarrow 0, \Lambda_2 \rightarrow \infty} P(\nu) = \frac{1}{\left| \nu \right|}$. For this example we use $\Lambda_1 = 2 \pi~\text{Hz}$ and $\Lambda_2 = 2 \pi \cdot 10^{10}~\text{Hz}$. According to the Wiener-Khinchin theorem:
\begin{align}
\left\langle \delta \varepsilon(t_1) \varepsilon(t_2) \right\rangle = \int_{-\infty}^{\infty} d\nu~P(\nu)e^{i\nu \left|t_1-t_2\right|} = -2\left[\text{Ei}\left(-\Lambda_1 \left|t_1-t_2\right|\right)-\text{Ei}\left(-\Lambda_2 \left|t_1-t_2\right|\right)\right] ,
\end{align}
where $\text{Ei}(z) = -\int_{-z}^\infty dt~ e^{-t}/t$ stands for the exponential integral function.

Using stochastic Liouville theory \cite{kubo_1963, cappellaro_2006}, we can treat the noise perturbatively and show that the toggling frame propagator $U_\text{tog}(T)$ is given as
\begin{equation}
\begin{aligned}
   \label{eq:1overfUtog}
   U_\text{tog}(T) &= \left\langle \Texp \left[\int_0^T dt_1 ~U^{-1}(t_1) G_\text{n}(t_1) U(t_1) \right] \right\rangle \\
   &= \I_4 + \int_0^T dt_1 \int_0^{t_1} dt_2~ \left\langle \varepsilon(t_1) \varepsilon(t_2) \right\rangle U^{-1}(t_1) G_z U(t_1) U^{-1}(t_2) G_z U(t_2) + ... ,
\end{aligned}
\end{equation}
where $U(t) = \mathcal{T}\exp\left( \int_0^t dt_1~G(t_1)\right)$ and we have made use of the fact that
\begin{equation}
\int_0^t dt_1~ \left\langle \varepsilon(t_1)\right\rangle U^{-1}(t_1) G_z U(t_1) = 0 .
\end{equation}
We remark that $U_\text{tog}(T)$ is not a unitary matrix, in fact, it is precisely the operator that encapsulates the non-unitary decoherence effects induced by $G_\text{n}(t)$. In order to reduce such decoherence at its lowest perturbative order, we need to minimize the nested double integral term in Equation~\eqref{eq:1overfUtog}. Because the generator $G(t)$ is Liouville-von Neumann generator, it also holds that $U(t) = \mathcal{T}\exp\left( \int_0^t dt_1~G(t_1)\right) = U_1(t) \otimes \overline{U_1(t)}$, where the bar denotes entry-wise matrix conjugation, and $U_1(t) = \mathcal{T}\exp\left(-i\int_0^t dt_1 \left[a_x(t_1) \frac{\sigma_x}{2} + a_y(t_1) \frac{\sigma_y}{2} \right] \right)$.

To employ the method developed in the previous section, we first note that a linear combination of exponential functions provides a good approximation for $\left\langle \varepsilon(t_1) \varepsilon(t_2) \right\rangle$ over a region of integration $0 \leq t_2 \leq t_1 \leq T$, with $T =400~\text{ns}$. Using a least squares fit to the correlation function over that region, we find an approximation that combines seven exponential functions
\begin{equation}
   \label{eq:1overfApprox}
   \left\langle \epsilon(t_1) \varepsilon(t_2) \right\rangle = -2\left[\text{Ei}\left(-\Lambda_1 \left|t_1-t_2\right|\right)-\text{Ei}\left(-\Lambda_2 \left|t_1-t_2\right|\right)\right] \approx \sum_{i=1}^7 c_i~e^{d_i (t_1-t_2)} ,
\end{equation}
where $c_1 = 7.49448$, $d_1 =-1.11796 \cdot 10^8~\text{Hz}$, $c_2 = 0.947027$, $d_2 =-3.37122 \cdot 10^7~\text{Hz}$, $c_3 =-0.490555$, $d_3 = -4.69721 \cdot 10^6~\text{Hz}$, $c_4 =-0.163987$, $d_4 = -3.77087 \cdot 10^6~\text{Hz}$, $c_5 =29.83$, $d_5 = -577865~\text{Hz}$, $c_6 =-0.102058$, $d_6 =122339~\text{Hz}$, $c_7 =0.00035238$ and $d_7 =2.05605 \cdot 10^7~\text{Hz}$. Given the approximation, we can now write
\begin{equation}
\begin{aligned}
   \label{eq:1overfIntegralApprox}
   I_{1/f}(t)&=\int_0^t dt_1 \int_0^{t_1} dt_2~\left\langle \varepsilon(t_1) \varepsilon(t_2) \right\rangle U^{-1}(t_1) G_z U(t_1) U^{-1}(t_2) G_z U(t_2) \\
   &\approx \sum_{i=1}^7 c_i \int_0^t dt_1 \int_0^{t_1} dt_2~ e^{d_i( t_1-t_2)} U^{-1}(t_1) G_z U(t_1) U^{-1}(t_2) G_z U(t_2) \\
   &= \sum_{i=1}^7 c_i U^{-1}(T)\Dyson_{U} \left( e^{d_i t} G_z, e^{-d_i t} G_z \right) .
\end{aligned}
\end{equation}

It is important to realize that in the case of actual experimental scenarios either the noise correlation function or its power spectral density would be characterized before embarking on control engineering. In such cases, $\left\langle \varepsilon(t_1) \varepsilon(t_2) \right\rangle$ is extremely unlikely to fit to some simple and specific analytic function. Therefore, our procedure, for fitting a set of functions to a set of data -- in this case an analytic function -- in order to approximate the noise correlation, closely matches a real control engineering protocol.

We search for a pulse implementing a Y gate, i.e., we wish to set $U(T) = e^{-i \pi \sigma_y/2} \otimes \overline{e^{-i \pi \sigma_y/2}}$. Consequently, we use the following target function:
\begin{equation}
   \label{eq:1overfTarget}
   \Phi = \frac{4}{5} \left( 1 - \frac{\|U(T) I_{1/f}(t)\|^2}{\left[\max_{a(t)}\|U(T) I_{1/f}(t)\|\right]^2} \right)
 + \frac{1}{5} \F \left[ e^{-i \pi \sigma_y/2} \otimes \overline{e^{-i \pi \sigma_y/2}}, U(T) \right].
\end{equation}
Combining Equation~\eqref{eq:1overfIntegralApprox} with Equation~\eqref{eq:expVLDE} in the previous section, we set up a Van Loan differential equation for $V(t) \in \M_{21}(\M_4)$. $V(t)$ will be generated by
\begin{align}
L(t) = \left(\begin{array}{ccccccc}
    			G(t)	& G_z		& 0			& \hdots		& 0	& 0	& 0	\\
			0		& G(t)	+d_1 \I_4	& G_z		& \hdots		& 0	& 0	& 0 	\\
			0		& 0			& G(t)		& \hdots		& 0	& 0	& 0 \\
			\vdots	& \vdots		& \vdots		& \ddots		& \vdots		& \vdots		& \vdots \\
			0		& 0			& 0			& \hdots		& G(t)	& G_z	& 0 \\
			0		& 0			& 0			& \hdots		& 0	& G(t)	+d_7 \I_4	& G_z \\
			0		& 0			& 0			& \hdots		& 0	& 0	& G(t)
			\end{array}\right) ,
\end{align}
such that
\begin{align}
&V(t) = \Texp \left(\int_0^t dt_1 L(t_1) \right) = \\
\nonumber &\left(\begin{array}{ccccccc}
    			U(t)		& \Dyson_{U} \left( e^{d_1 t} G_z \right)		& \Dyson_{U} \left( e^{d_1 t} G_z, e^{-d_1 t} G_z \right)			& \hdots		& 0	& 0	\\
			0		& e^{d_1 t} U(t)	& e^{d_1 t} \Dyson_{U} \left( e^{-d_1 t} G_z \right)		& \hdots		& 0	& 0 	\\
			0		& 0			& U(t)		& \hdots		& 0	& 0 \\
			\vdots	& \vdots		& \vdots		& \ddots		& \vdots		& \vdots \\
			0		& 0			& 0			& \hdots		& \Dyson_{U} \left( e^{d_7 t} G_z \right)	& \Dyson_{U} \left( e^{d_7 t} G_z, e^{-d_7 t} G_z \right) \\
			0		& 0			& 0			& \hdots		& e^{d_7 t} U(t)	& e^{d_7 t} \Dyson_{U} \left( e^{-d_7 t} G_z \right) \\
			0		& 0			& 0			& \hdots		& 0	& U(t)
			\end{array}\right) .
\end{align}
We can now approximate the target as a function of $V(T)$:
\begin{equation}
\begin{aligned}
   \label{eq:1overfTargetVL}
   \Phi &= \frac{4}{5} \left(1 - \frac{\Tr\left(\left[\sum_{i=1}^7 c_i V_{3(i-1)+1,3i}(T) \right]^\dagger \sum_{i=1}^7 c_i V_{3(i-1)+1,3i}(T)\right)}{2 \left( \int_0^T dt_1 \int_0^{t_1} dt_2~\left\langle \varepsilon(t_1) \varepsilon(t_2) \right\rangle \right)^2} \right) \\
   &+ \frac{1}{5} \frac{\Tr \left[(e^{-i \pi \sigma_y/2})^\dagger \otimes (e^{-i \pi \sigma_y/2})^\T V_{1,1}(T)\right]}{4},
\end{aligned}
\end{equation}
where $\int_0^T dt_1 \int_0^{t_1} dt_2~\left\langle \varepsilon(t_1) \varepsilon(t_2) \right\rangle$ is evaluated numerically for any particular $T$. The partial derivatives of $\Phi$ with respect to the control amplitudes $\lbrace \beta_{i,j} \rbrace$ are given as
\begin{equation}
\begin{aligned}
   \label{eq:1overfTargetDerivatives}
   \frac{\partial}{\partial \beta_{i,j}}\Phi &=-\frac{4}{5}\frac{\text{Re}\left[\Tr\left(\left[\sum_{i=1}^7 c_i V_{3(i-1)+1,3i}(T) \right]^\dagger \sum_{i=1}^7 c_i \left(\frac{\partial}{\partial \beta_{i,j}}V(T) \right)_{3(i-1)+1,3i} \right)\right]}{\left( \int_0^T dt_1 \int_0^{t_1} dt_2~\left\langle \varepsilon(t_1) \varepsilon(t_2) \right\rangle \right)^2} \\
   &+ \frac{1}{10} \text{Re}\left( \Tr \left[(e^{-i \pi \sigma_y/2})^\dagger \otimes (e^{-i \pi \sigma_y/2})^\T \left(\frac{\partial}{\partial \beta_{i,j}}V(T) \right)_{1,1}\right] \right).
\end{aligned}
\end{equation}

Here, we attempt to closely mimic a control search procedure that would be undertaken when searching for an experimentally implementable sequence. Hence, we impose three distinct constraint on the pulse waveform: maximum amplitude constraint, bandwidth limitations for the pulse waveform frequency components and zero amplitude periods at the beginning and at the end of the sequence. The last two constraints are implemented by introducing an optimization transfer function $\Xi^\text{opt}$, as was described in Section~\ref{sec:controlSetup}, and the explicit construction of $\Xi^\text{opt}$ is given in Appendix~\ref{app:transferfn}. We do not consider any ensemble effects, i.e., $|\Gamma| = 1$, and we take the only experimental transfer function to act as an identity, such that $\beta = \beta^{(1)} = \Xi^{(1)} \left(\alpha\right) = \alpha$. Accordingly, the numerical control searches are conducted for $\alpha^\text{opt}$, with $\alpha = \Xi^\text{opt}\left(\alpha^\text{opt}\right)$.

For the searches, we limit the Rabi frequency $|a^\text{opt}(t)|/(2\pi)$ to be less than or equal to $200~\text{MHz}$ by enforcing that
\begin{equation}
-\frac{1}{\sqrt{2}} 2 \pi \cdot 200 \cdot 10^6~\text{Hz}\le a^\text{opt}_i (t) \le \frac{1}{\sqrt{2}} 2 \pi \cdot 200 \cdot 10^6~\text{Hz}
\end{equation}
for $i=\lbrace x,y \rbrace$. We take the pulse length $T$ to be $10$ Rabi cycles or $50~\text{ns}$, which is divided into $N=300$ intervals of equal length $\Delta T = 1.67 \cdot 10^{-10}~\text{s}$, whereas the zero amplitude periods at the beginning and at the end of the pulse $a(t)$ have a length of $8.33~\text{ns}$, corresponding to $N_0 = 50$. Therefore, the numerical control search is conducted on $33.33~\text{ns}$\=/long $a^\text{opt}(t)$ that is divided into $N^\text{opt} = N-2N_0 = 200$ equal steps. Using $\Xi^\text{opt}$, we constrain all spectral components of the pulse $a(t)$ to lie within a $\Delta \nu = 400~\text{MHz}$ bandwidth around the carrier frequency. The optimization transfer function $\Xi^\text{opt}(N,N_0,\Delta T,\Delta \nu)$, that is employed to enforce the constraints, is defined by Equation~\eqref{eq:optTransferFn} in Appendix~\ref{app:transferfn}.

Because our control optimization was carried out on the optimization control sequence $a^\text{opt}(t)$ we needed to evaluate $\lbrace \frac{\partial}{\partial \alpha^\text{opt}_{i,j}}\Phi \rbrace$ for $i=\lbrace 1,2 \rbrace$ and $j=\lbrace 1,\dots,N^\text{opt} \rbrace$. With Equation~\eqref{eq:1overfTargetDerivatives} we evaluate $\lbrace \frac{\partial}{\partial \beta_{i,j}}\Phi \rbrace$, and identify that $\frac{\partial}{\partial \alpha_{i,j}}\Phi = \frac{\partial}{\partial \beta_{i,j}}\Phi$ for all $i$ and $j$. Finally, we link $\lbrace \frac{\partial}{\partial \alpha^\text{opt}_{i,j}}\Phi \rbrace$ with $\lbrace \frac{\partial}{\partial \beta_{i,j}}\Phi \rbrace$ through Equation~\eqref{eq:optimizationTransferFnJacobian} in Appendix~\ref{app:transferfn}:
\begin{equation}
\begin{aligned}
\frac{\partial}{\partial \alpha^\text{opt}_{1,j}}&\Phi = \\
&\sum_{t=1}^N \left( \text{Re} \left[\Xi^\text{opt}(N,N_0,\Delta T,\Delta \nu) \right]_{t,j} \frac{\partial}{\partial \beta_{1,t}}\Phi -\text{Im} \left[ \Xi^\text{opt}(N,N_0,\Delta T,\Delta \nu) \right]_{t,j} \frac{\partial}{\partial \beta_{2,t}}\Phi \right)
\end{aligned}
\end{equation}
and
\begin{equation}
\begin{aligned}
\frac{\partial}{\partial \alpha^\text{opt}_{2,j}}&\Phi = \\
&\sum_{t=1}^N \left(\text{Im} \left[ \Xi^\text{opt}(N,N_0,\Delta T,\Delta \nu) \right]_{t,j} \frac{\partial}{\partial \beta_{1,t}}\Phi + \text{Re} \left[ \Xi^\text{opt}(N,N_0,\Delta T,\Delta \nu) \right]_{t,j} \frac{\partial}{\partial \beta_{2,t}}\Phi \right) .
\end{aligned}
\end{equation}

\begin{center}\begin{figure}
   \includegraphics[width=\textwidth]{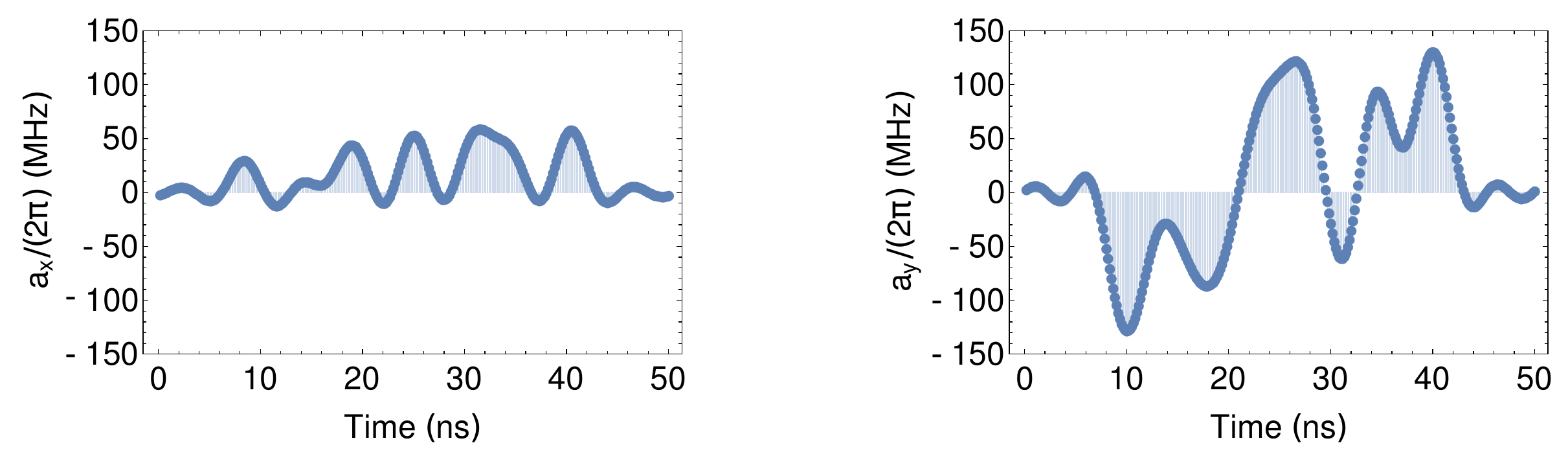}
   \caption{
      \label{fig:1overFpulse} Control sequence robust to 1/f noise that implements a Y gate: $a_x(t)$ on the left and $a_y(t)$ on the right.
   }
\end{figure}\end{center}

Given the target function and the partial derivatives above, we search for a control sequence as it was described at the beginning of the section. The resulting waveform $a(t)$ is shown in Figure~\ref{fig:1overFpulse} and the pulse characteristics are  $1-\F \left( \sigma_y, U(T) \right) = 1.25 \times 10^{-7}$ and $\left. \|I_{1/f}(T)\| \middle/ \sqrt{2\left( \int_0^T dt_1 \int_0^{t_1} dt_2~\left\langle \varepsilon(t_1) \varepsilon(t_2) \right\rangle \right)} \right. = 0.0127$. In the case of a stochastic operator $G_\text{n}(t)$, one cannot expect to be able to set the integral $I_{1/f}(t)$ in Equation~\eqref{eq:1overfIntegralApprox} equal to zero, since the high frequency components of the noise always retain their decoherence inducing effect. Nevertheless, for a reasonably low amplitude noise, our sequence in Figure~\ref{fig:1overFpulse} would extend the qubit coherence time by a factor of $1/0.0127 \approx 80$.

\subsection{\label{sec:broadBandDipolar}Broadband Dipolar Pulse}
\begin{center}\begin{figure}
   \includegraphics[width=\textwidth]{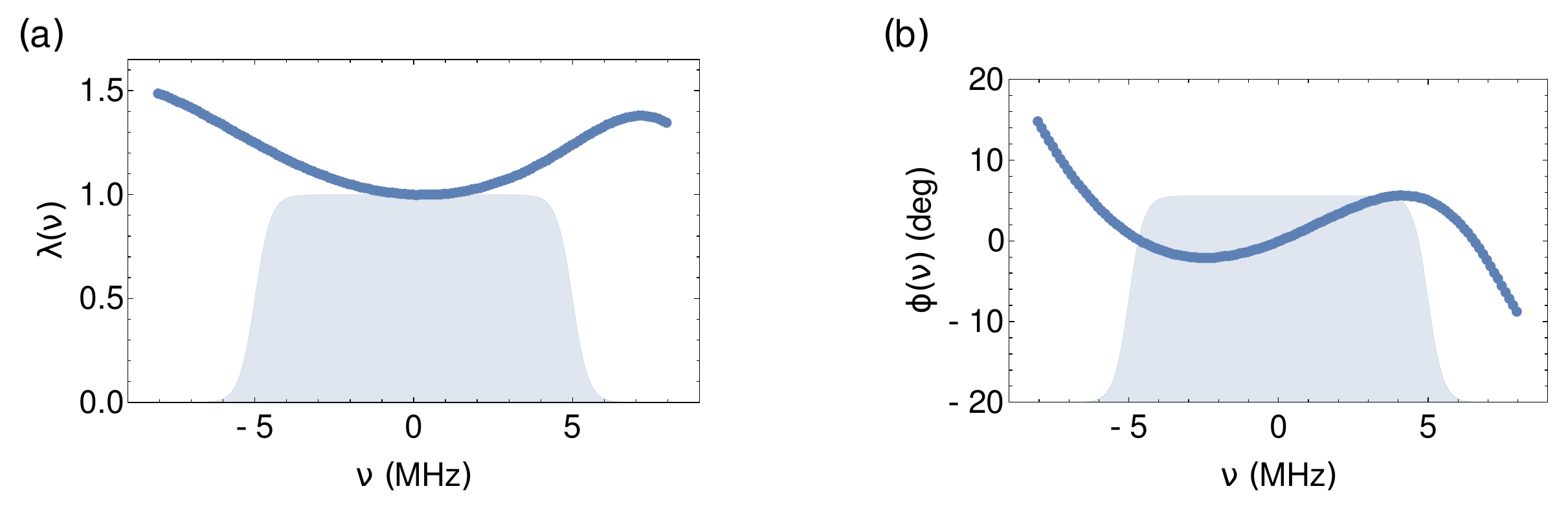}
   \caption{
      \label{fig:TransferFns} (a) Experimentally determined amplitude transfer function $\lambda(\nu)$ as a function of frequency $\nu$, with $\nu = 0$ corresponding to the carrier frequency. The shaded area illustrates the low-pass filter defined by Equation~(\ref{eq:BandPass}) that was incorporated into the optimization transfer function. (b) Experimentally determined phase transfer function $\phi(\nu)$ as a function of frequency $\nu$, with $\nu = 0$ corresponding to the carrier frequency.
   }
\end{figure}\end{center}

The control sequence presented here was engineered for nanoscale nuclear magnetic resonance experiments \cite{rose_2018} and was used to increase the proton spin phase coherence time by a factor of 500 under rather difficult control conditions. The resonant control fields $b^{(\gamma)}(t)$ for the experiment had a vast range of $\gamma$ dependent maximum values $[0.9~\text{MHz}, 1.7~\text{MHz}]$. Furthermore, the the strong dipolar interactions between the proton spins as well as chemical shifts limited the spin coherence time to $11~\mu\text{s}$. Furthermore, it had been determined that the amplitude and phase transfer functions for the electronics -- $\lambda(\nu)$ and $\phi(\nu)$ , respectively -- had non-trivial character as it can be seen in Figure~\ref{fig:TransferFns}.

We will define our control problem exactly according to the ensemble control setup abstraction that was laid out in Section~\ref{sec:controlSetup}. We say that we have an ensemble of proton spins labelled by $\gamma \in \Gamma$; here, we consider this ensemble to be a representative ensemble of spins in the sample volume of interest. Each $\gamma$ has an associated unique transfer function $\Xi^{(\gamma)}$ that determines the control amplitudes $b^{(\gamma)}(t)$ as a function of the control sequence $a(t)$. We constrain the maximum amplitude $|a(t)|$ of the control sequence to be equal to one. Our transfer functions $\lbrace \Xi^{(\gamma)} \rbrace$ reflect the effects of Rabi field distribution and the amplitude and phase transfer functions shown in Figure~\ref{fig:TransferFns}.

In addition to the transfer functions $\lbrace \Xi^{(\gamma)} \rbrace$ defined by Equation~\eqref{eq:dipolarTransferFunctions}, we also use an optimization transfer function $\Xi^\text{opt}$ in order to limit the range of frequency components in $a(t)$ to within a bandwidth of $\Delta \nu$ as well as to enforce that the pulse starts and ends with zero amplitude. The control sequence $a(t)$ has piecewise constant pulse amplitudes for $N$ equal periods of duration $\Delta T$, so that the total pulse length is $T=N \Delta T$. The low-pass filter that is incorporated into the optimization transfer function is illustrated by the shaded area in Figure~\ref{fig:TransferFns}(a). $\Xi^\text{opt}(N,N_0,\Delta T,\Delta \nu)$ is defined by Equation~\eqref{eq:optTransferFn} in Appendix~\ref{app:transferfn}. The control sequence is then determined by $\alpha^\text{opt} \in \M_{2,N^\text{opt}}(\R)$, where $N^\text{opt} = N-2N_0$, and $N_0$ determines the number of zero amplitude intervals of duration $\Delta T$ at the beginning and at the end of the sequence.

It is natural to label the transfer functions $\lbrace \Xi^{(\gamma)} \rbrace$ according to the maximum Rabi field strengths they yield on the nuclear spins, i.e., according to the $|b^{(\gamma)}(t)|/(2\pi)$ value corresponding to $a(t)=1$ for all $0 \leq t \leq T$. The transfer functions $\lbrace \Xi^{(\gamma)} \rbrace$ could then be written explicitly as
\begin{equation}
   \label{eq:dipolarTransferFunctions}
   \Xi^{(\gamma)}(N,\Delta T) = 2\pi \gamma~ \Xi(N,\Delta T, \lambda,\phi) ,
\end{equation}
where the function $\Xi(N,\Delta t, \lambda,\phi)$ is given by Equation~\eqref{eq:linearTransferFn} in Appendix~\ref{app:transferfn}; $\lambda(\nu)$ and $\phi(\nu)$ being evaluated by interpolating the experimental transfer function data in Figure~\ref{fig:TransferFns}.

The system generators $\lbrace G^{(\gamma)}(t) \rbrace$ are given by
\begin{equation}
   \label{eq:dipolarMRFMSystemGenerator}
   G^{(\gamma)}(t) = -i\frac{b^{(\gamma)}_1 (t)}{2} \sigma_x - i\frac{b^{(\gamma)}_2 (t)}{2} \sigma_y ,
\end{equation}
while the system propagators evaluate to $U^{(\gamma)}(t) = \Texp \left[\int_0^t dt_1 G^{(\gamma)}(t_1) \right]$ for all $\gamma \in \Gamma$. The control amplitudes $\lbrace b^{(\gamma)}(t) \rbrace$ that are specified by a matrix $\beta^{(\gamma)} \in \M_{2,N}(\R)$ for each $\gamma \in \Gamma$ equate to
\begin{equation}
\label{eq:MRFMbeta1Values}
\beta^{(\gamma)}_{1,j} = \text{Re} \left[\Xi^{(\gamma)}(N,\Delta t)~\Xi^\text{opt}(N,N_0,\Delta T,\Delta \nu)(\alpha^\text{opt}_1 - i\alpha^\text{opt}_2) \right]_j
\end{equation}
and
\begin{equation}
\label{eq:MRFMbeta2Values}
\beta^{(\gamma)}_{2,j} = -\text{Im} \left[\Xi^{(\gamma)}(N,\Delta t)~\Xi^\text{opt}(N,N_0,\Delta T,\Delta \nu)(\alpha^\text{opt}_1 - i\alpha^\text{opt}_2) \right]_j ,
\end{equation}
where $\alpha^\text{opt} \in \M_{2,N^\text{opt}}(\R)$ is the matrix specifying the optimization waveform $a^\text{opt}(t)$. In order to enforce $|a(t)| \leq 1$ for all $0 \leq t \leq T$, we constrain $-1/\sqrt{2} \leq \alpha_i^\text{opt} \leq 1/\sqrt{2}$ for $i \in \lbrace 1,2 \rbrace$. The control sequence $a(t)$ that is implemented experimentally is specified by another matrix $\alpha \in \M_{2,N}(\R)$ the elements of which are calculated as
\begin{equation}
\alpha_{1,j} = \text{Re} \left[\Xi^\text{opt}(N,N_0,\Delta T,\Delta \nu)(\alpha^\text{opt}_1 - i\alpha^\text{opt}_2) \right]_j
\end{equation}
and
\begin{equation}
\alpha_{2,j} = -\text{Im} \left[\Xi^\text{opt}(N,N_0,\Delta T,\Delta \nu)(\alpha^\text{opt}_1 - i\alpha^\text{opt}_2) \right]_j ,
\end{equation}
after a suitable $\alpha^\text{opt}$ is found.

Our objective is to find a control sequence that would yield $U^{(\gamma)}(T) = U_\text{target} = \exp \left(-i \frac{\pi}{2} \frac{\sigma_x}{2} \right)$ as well as
\begin{equation}
\Dyson_{U^{(\gamma)} \otimes U^{(\gamma)}}(D) = \Dyson_{U^{(\gamma)}}(\sigma_z) = 0
\end{equation}
for all $\gamma \in \Gamma$. We assign equal weights $p^{(\gamma)}=\frac{1}{|\Gamma|}$ to each member of the ensemble and define the following combined target function
\begin{equation}
\begin{aligned}
   \label{eq:dipolarMRFMCostFunction}
   \Phi &= 1 - \frac{1}{|\Gamma|} \sum_{\gamma \in \Gamma} \frac{5}{9} \sqrt{1-\left(\F\left[ U_\text{target}, U^{(\gamma)}(T) \right]\right)^2} \\
   &-\frac{1}{|\Gamma|} \sum_{\gamma \in \Gamma} \left( \frac{3}{9} \frac{\|\Dyson_{U^{(\gamma)} \otimes U^{(\gamma)}}(D)\|}{\max_{a(t)}\|\Dyson_{U^{(\gamma)} \otimes U^{(\gamma)}}(D)\|} + \frac{1}{9} \frac{\|\Dyson_{U^{(\gamma)}}(\sigma_z)\|}{\max_{a(t)}\|\Dyson_{U^{(\gamma)}}(\sigma_z)\|}\right) .
\end{aligned}
\end{equation}

We construct a set of Van Loan generators $L^{(\gamma)} \in \M_{12}$ that decompose into a direct sum of $\M_{2}(\M_{2})$ and $\M_{2}(\M_{4})$:
\begin{align}
L^{(\gamma)}&(t) = \\
 \nonumber &\left(\begin{array}{cccc}
			G^{(\gamma)}(t)	& \sigma_z	& 0	& 0 \\
			0	& G^{(\gamma)}(t)	& 0	& 0 \\    			
    			0	& 0	& G^{(\gamma)}(t) \otimes \I + \I \otimes G^{(\gamma)}(t)	& D	\\
			0	& 0	& 0	&	G^{(\gamma)}(t) \otimes \I + \I \otimes G^{(\gamma)}(t)
			\end{array}\right) ,
\end{align}
the corresponding Van Loan propagators of which are given as
\begin{equation}
\begin{aligned}
V^{(\gamma)}(t) &= \Texp \left[\int_0^t dt_1 L(t_1) \right] \\
&= \left(\begin{array}{cccc}
    			U^{(\gamma)}(t)	& \Dyson_{U^{(\gamma)}}(\sigma_z)	& 0 &	0\\
			0	& U^{(\gamma)}(t)		& 0 	& 0	\\
			0	& 0			& U^{(\gamma)}(t) \otimes U^{(\gamma)}(t)	& \Dyson_{U^{(\gamma)} \otimes U^{(\gamma)}}(D) \\
			0	& 0			& 0		& U^{(\gamma)}(t) \otimes U^{(\gamma)}(t)
			\end{array}\right) .
\end{aligned}
\end{equation}
Here, we will slightly abuse our sub-matrix index notation, just as it was done in Section~\ref{subsec:exchange}, and write the target function in Equation~\eqref{eq:dipolarMRFMCostFunction} as a function of $\lbrace V^{(\gamma)}(T) \rbrace$:
\begin{align}
   \label{eq:dipolarMRFMCostFunction2}
   \Phi &= 1 - \frac{1}{|\Gamma|} \sum_{\gamma \in \Gamma} \frac{5}{18} \sqrt{4-\Tr \left[V^{(\gamma)}_{1,1}(T)~e^{i \frac{\pi}{2} \frac{\sigma_x}{2}}\right] \Tr\left[e^{-i \frac{\pi}{2} \frac{\sigma_x}{2}} \left(V^{(\gamma)}_{1,1}(T) \right)^\dagger\right]} \\
   \nonumber &-\frac{1}{|\Gamma|} \sum_{\gamma \in \Gamma} \left( \frac{3}{18\sqrt{6}T} \sqrt{\Tr \left[V^{(\gamma)}_{3,4}(T) \left(V^{(\gamma)}_{3,4}(T)\right)^\dagger \right]} + \frac{1}{18\sqrt{2}T} \sqrt{\Tr \left[V^{(\gamma)}_{1,2}(T) \left(V^{(\gamma) \dagger}_{1,2}(T) \right)^\dagger\right]} \right) .
\end{align}

We now evaluate the partial derivatives of Equation~\eqref{eq:dipolarMRFMCostFunction2} with respect to the elements of $\beta^{(\gamma)}$ for each $\gamma \in \Gamma$:
\begin{equation}
\begin{aligned}
   \label{eq:dipolarMRFMDerivatives}
   \frac{\partial}{\partial \beta^{(\gamma)}_{i,j}} \Phi &= \frac{5}{18|\Gamma|} \frac{\text{Re}\left( \Tr \left[ \left(\frac{\partial}{\partial \beta^{(\gamma)}_{i,j}} V^{(\gamma)}(T) \right)_{1,1}e^{i \frac{\pi}{2} \frac{\sigma_x}{2}}\right] \Tr\left[e^{-i \frac{\pi}{2} \frac{\sigma_x}{2}} \left(V^{(\gamma)}_{1,1}(T) \right)^\dagger\right] \right)}{\sqrt{4-\Tr \left[V^{(\gamma)}_{1,1}(T)~e^{i \frac{\pi}{2} \frac{\sigma_x}{2}}\right] \Tr\left[e^{-i \frac{\pi}{2} \frac{\sigma_x}{2}} \left(V^{(\gamma)}_{1,1}(T) \right)^\dagger\right]}} \\
   &-\frac{3}{18\sqrt{6}T |\Gamma|} \frac{\text{Re}\left(\Tr \left[ \left(\frac{\partial}{\partial \beta^{(\gamma)}_{i,j}} V^{(\gamma)}(T) \right)_{3,4} \left(V^{(\gamma)}_{3,4}(T)\right)^\dagger \right]\right)}{\sqrt{\Tr \left[V^{(\gamma)}_{3,4}(T) \left(V^{(\gamma)}_{3,4}(T)\right)^\dagger \right]}} \\
   &-\frac{1}{9\sqrt{2}T |\Gamma|} \frac{\text{Re}\left(\Tr \left[ \left(\frac{\partial}{\partial \beta^{(\gamma)}_{i,j}} V^{(\gamma)}(T) \right)_{1,2} \left(V^{(\gamma)}_{1,2}(T)\right)^\dagger \right]\right)}{\sqrt{\Tr \left[V^{(\gamma)}_{1,2}(T) \left(V^{(\gamma)}_{1,2}(T)\right)^\dagger \right]}} .
\end{aligned}
\end{equation}
In order to carry out the gradient ascent searches to find $\alpha^\text{opt}$ that yields $\Phi \approx 1$, we evaluate partial derivatives of $\Phi$ with respect to the elements of $\alpha^\text{opt}$:
\begin{equation}
\begin{aligned}
\label{eq:MRFMpartial1Values}
\frac{\partial}{\partial \alpha^\text{opt}_{1,j}}\Phi &= \sum_{\gamma \in \Gamma} \sum_{t=1}^N \text{Re} \left[\Xi^{(\gamma)}(N,\Delta T)~\Xi^\text{opt}(N,N_0,\Delta T,\Delta \nu) \right]_{t,j} \frac{\partial}{\partial \beta^{(\gamma)}_{1,t}}\Phi \\
&-\sum_{\gamma \in \Gamma} \sum_{t=1}^N \text{Im} \left[ \Xi^{(\gamma)}(N,\Delta t)~\Xi^\text{opt}(N,N_0,\Delta T,\Delta \nu) \right]_{t,j} \frac{\partial}{\partial \beta^{(\gamma)}_{2,t}}\Phi
\end{aligned}
\end{equation}
and
\begin{equation}
\begin{aligned}
\label{eq:MRFMpartial2Values}
\frac{\partial}{\partial \alpha^\text{opt}_{2,j}}\Phi &= \sum_{\gamma \in \Gamma} \sum_{t=1}^N \text{Im} \left[ \Xi^{(\gamma)}(N,\Delta t)~\Xi^\text{opt}(N,N_0,\Delta T,\Delta \nu) \right]_{t,j} \frac{\partial}{\partial \beta^{(\gamma)}_{1,t}}\Phi \\
&+\sum_{\gamma \in \Gamma} \sum_{t=1}^N \text{Re} \left[ \Xi^{(\gamma)}(N,\Delta t)~\Xi^\text{opt}(N,N_0,\Delta T,\Delta \nu) \right]_{t,j} \frac{\partial}{\partial \beta^{(\gamma)}_{2,t}}\Phi .
\end{aligned}
\end{equation}

Our control searches are conducted in the way it was described as it was described at the beginning of the section. The three individual quantities that appear in Equation~\eqref{eq:dipolarMRFMCostFunction2} are the unitary metric $\Psi_U^{(\gamma)}$, dipolar metric $\Psi_D^{(\gamma)}$, and $\sigma_z$ metric  $\Psi_{\sigma_z}^{(\gamma)}$, which are defined as
\begin{align}
   \label{eq:unitaryMetric}
   \Psi_U^{(\gamma)} &= \sqrt{1-\left(\F\left[ \exp \left(-i \frac{\pi}{2} \frac{\sigma_x}{2} \right), U^{(\gamma)}(T) \right]\right)^2}, \\
  \label{eq:dipolarMetric}
    \Psi_D^{(\gamma)} &= \frac{\|\Dyson_{U^{(\gamma)} \otimes U^{(\gamma)}}(D)\|}{\max_{a(t)}\|\Dyson_{U^{(\gamma)} \otimes U^{(\gamma)}}(D)\|}, \\
    \label{eq:chemicalMetric}
    \Psi_{\sigma_z}^{(\gamma)} &= \frac{\|\Dyson_{U^{(\gamma)}}(\sigma_z)\|}{\max_{a(t)}\|\Dyson_{U^{(\gamma)}}(\sigma_z)\|} .
\end{align}

\begin{center}\begin{figure}
   \includegraphics[width=\textwidth]{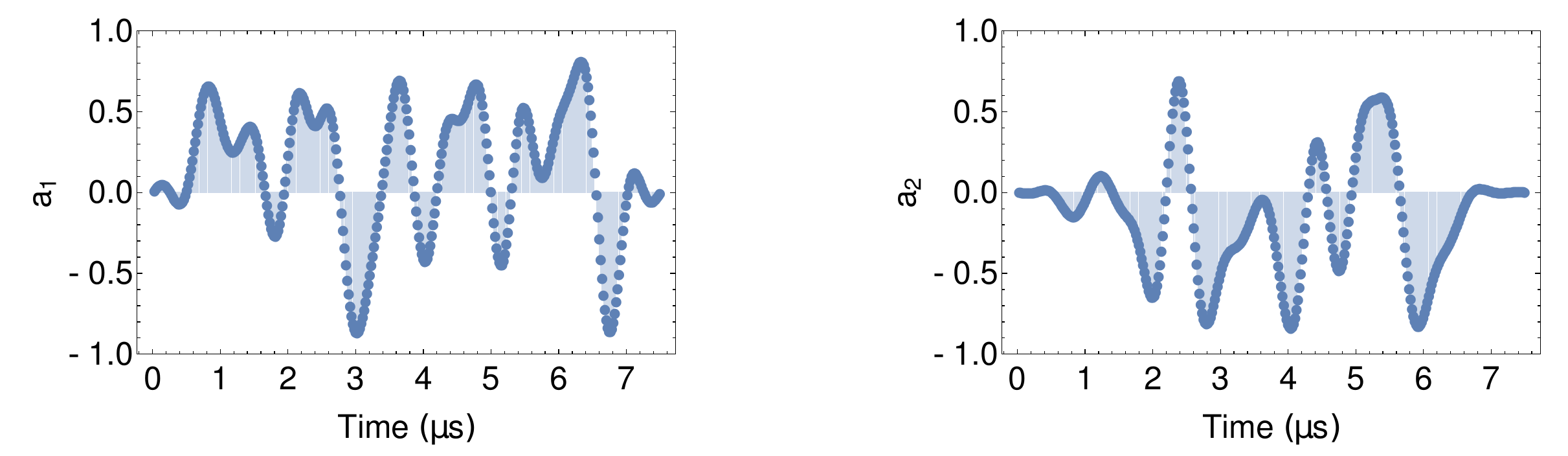}
   \caption{
      \label{fig:7usPulse} Broadband decoupling pulse: $a_1(t)$ on the left, $a_2(t)$ on the right
      }
\end{figure}\end{center}
\begin{center}\begin{figure}
   \includegraphics[width=\textwidth]{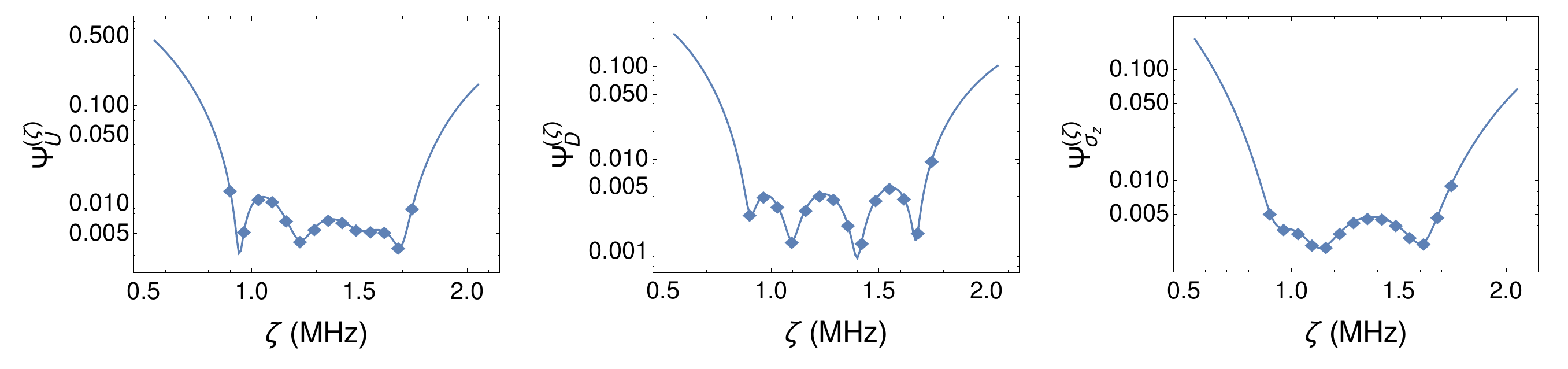}
   \caption{
      \label{fig:7usMetrics} Unitary metric $\Psi_U^{(\zeta)}$ defined by Equation~\eqref{eq:unitaryMetric}, dipolar metric $\Psi_D^{(\zeta)}$ defined by Equation~\eqref{eq:dipolarMetric} and $\sigma_z$ metric $\Psi_{\sigma_z}^{(\zeta)}$ defined by Equation~\eqref{eq:chemicalMetric} as functions of frequency $\zeta$. The blue diamonds in the figures denote the 14 elements of the ensemble $\Gamma$ used during control optimization, the ensemble elements $\gamma \in \Gamma$ were each identified by a particular maximum Rabi strength which corresponds to the particular $\zeta$ value in the figures. It can be seen that all three metrics take their lowest values within the target range of $0.9~ \text{MHz}$ to $1.7~ \text{MHz}$. Furthermore, it can also be seen that all three metrics take roughly equal values over that range, this is because of our choice for the weights in the target function $\Phi$ given by Equation~\eqref{eq:dipolarMRFMCostFunction}. The process for choosing the weights was described at the beginning of this section.
   }
\end{figure}\end{center}

Our representative set for the Rabi strengths is $\gamma \in \left\lbrace 0.9~\text{MHz}\right.$, $0.965~\text{MHz}$, $1.03~\text{MHz}$, $1.095~\text{MHz}$, $1.16~\text{MHz}$, $1.225~\text{MHz}$, $1.29~\text{MHz}$, $1.355~\text{MHz}$, $1.42~\text{MHz}$, $1.485~\text{MHz}$, $1.55~\text{MHz}$, $1.615~\text{MHz}$, $1.68~\text{MHz}$, $\left.1.745~\text{MHz} \right\rbrace$. The parameters for the control search were $N=360$, $N_0=30$, $\Delta \nu=10~\text{MHz}$ and $\Delta T = 0.0208~\mu\text{s}$. The control sequence and its figures of merit are shown in Figures~\ref{fig:7usPulse} and \ref{fig:7usMetrics}. The sequence length of $\sim 7.5~\mu\text{s}$ corresponds to $6.75$ Rabi cycles for the spins experiencing $\omega_1/(2\pi) = 0.9~\text{MHz}$.

\section{Conclusions} \label{sec:conclusions}
We have developed a general method for computing perturbation integrals arising in numerous effective Hamiltonian control schemes, which are generally of the form:
\begin{align}
    \label{eq:conclusionsIntegrals}
	U(T)\int_0^T dt_1 \dots \int_0^{t_{n-1}} dt_n~ f(t_1, \dots, t_n) U^{-1}(t_1) A_1(t_1) U(t_1) \dots U^{-1}(t_n) A_n(t_n) U(t_n) ,
\end{align}
where the system propagator $U(t)$, $0 \leq t \leq T$, is generated by piecewise constant control sequences $a(t)$. Our method is based on the observation that these integrals can be written as solutions to first order matrix differential equations, which we call the Van Loan equations, that have the same general form as the Schr\"{o}dinger equation. Consequently, the same optimization algorithms that have been developed for quantum control may be applied to optimizing these expressions, and importantly, this formulation enables the immediate application of methods that also compute and make use of gradient information (e.g. GRAPE \cite{khaneja_2005} or GOAT \cite{machnes_2018}).

Gradient evaluation for Dyson terms is crucial for ensuring fast convergence to high accuracy solutions, as gradient information increases the efficiency of the search. Here, we have only used first order derivatives, but as is described in \cite{de_fouquieres_2011}, incorporating second order derivative information into standard quantum control problems can ensure fast convergence rates near the optimum, and lead to better performance than first order methods \cite{goodwin_2016}. To reiterate, as we rephrase the effective Hamiltonian control problem into a bilinear control theory problem, all of the higher order algorithms and methods outlined in \cite{goodwin_2016} may be applied. Consequently, we expect the methods presented here to be useful tools for efficiently exploring the space of control sequences, to find solutions satisfying large number of design criteria involving Dyson terms.

As a computational task, this method has the benefits of both implementation efficiency, as well as computationally efficiency. With regards to implementation, given any software package for solving general bilinear control theory problems, the only new ingredient necessary to optimize integrals of the type in Equation~\eqref{eq:conclusionsIntegrals} is the construction of the relevant Van Loan differential equations.\footnote{Indeed, the reason our implementation of GRAPE described in Appendix \ref{sec:GRAPE} uses the commutator method for gradients, as opposed to the exponential method, as in \cite{goodwin_2016}, is exactly due to the utilization of an already-existing code base for bilinear control problems.} In terms of efficiency, Van Loan \cite{vanloan_1978} found --- in the context of constant generators --- that the number of operations in this approach is favourable as compared to other methods (requiring either fewer, or a comparable number of operations). In particular, for a given accuracy, the method is much faster than using numerical integral rules. This has been corroborated by numerical experiments in \cite{jimenez_simple_2002}, and has been observed in \cite{goodwin_2015}, in which the method significantly outperformed numerical integration methods for computing spin relaxation theory expressions, enabling simulation of larger systems.

In this manuscript, we have also demonstrated the use of this method in a variety of scenarios, including the successful application in nanoscale magnetic resonance experiments \cite{rose_2018}. We note that the examples presented in this manuscript should not be understood as comprehensive; in order to keep our treatment illustrative and concise, we did not present control searches involving Dyson terms higher than first order. However, such extensions follow naturally from the given examples, and in our experience optimizations with higher order terms have also converged to desired solutions. Moreover, we wish to highlight that these block matrix methods are applicable to terms of even more general form than those appearing in perturbative control schemes, as are captured in Equation \eqref{eq:conclusionsIntegrals}. Any nested integral expressible as an output to Theorem \ref{theorem:HP_theorem} could be incorporated into control optimization.\footnote{As a concrete example, a unitary trajectory $U : [0,T] \rightarrow \M_n$ forms a continuous-time unitary $k$-design if and only if the integral $\int_0^T dt U(t)^{\otimes^k} \otimes (U(t)^\dagger)^{\otimes^k}$ takes a specific value. This integral may be computed, and hence optimized, using the methods presented here.}

The methods developed here could, in principle, be used in any setting requiring robust coherent control of quantum systems, whether it be quantum computation, sensing, or spectroscopy. The main requirement for successful implementation is an accurate model of the system generators, and a precise knowledge of the control amplitudes seen by the quantum system, i.e. sufficiently good characterization of the experimental transfer function. Quantum control problems that could be addressed with these tools in the future include, but are not limited to: minimizing the effect of cross-talk in the case of simultaneous multiple qubit control, reducing the effect of counter-rotating terms in the cases where the Rabi strength approaches the qubit level spacing (i.e. Bloch-Siegert type effects), and preventing leakage to higher levels. Another potential application of this framework is the inclusion of non-Hamiltonian Lindblad terms for cross-polarization problems.

\subsection*{Acknowledgements}
This work was undertaken thanks in part to funding from the the Canada First Research Excellence Fund (CFREF). Further support was provided by the Natural Sciences and Engineering Research Council of Canada (NSERC), the Canada Excellence Research Chairs (CERC) program (215284), the Canadian Institute for Advanced Research (CIFAR), the province of Ontario and Industry Canada. H. H. and D.P. would like to thank Ian Hincks, Raffi Budakian and William Rose for their insights and fruitful discussions. F. Z. would like to acknowledge Guilu Long for his insights and inspiring discussions.

\appendix

\section{\label{sec:GRAPE}GRAPE Algorithm}
In this appendix, we describe the Gradient Ascent Pulse Engineering (GRAPE) algorithm introduced by Khaneja \emph{et al} \cite{khaneja_2005} for numerically solving bilinear control theory problems for piecewise constant control amplitudes. A control problem over a time interval $[0,T]$ is bilinear if the control amplitudes $b: [0,T] \rightarrow \R^l$ determine the system evolution through a first order matrix differential equation:
\begin{align}
   \label{eq:bilinearProblem}
   \dot{V}(t) = \left[ L_0 + \sum_{i=1}^l b_i(t) L_i \right] V(t),
\end{align}
where $L_i \in \M_n$ for all $i \in \lbrace 0,1,\dots,l \rbrace$ and $V(0)=\I_n$. It is clear that the Schr{\"o}dinger equation is a special instance of bilinear control with $\lbrace L_i \rbrace$ being anti-Hermitian matrices, as are Van~Loan equations.

Throughout this manuscript, we only deal with piecewise constant control amplitudes $b(t)$, for which, we split the interval $[0,T]$ into $M$ subintervals with respective durations $\delta T_j$ such that $\delta T_j \ge 0$, for all $j \in \lbrace 1,2,\dots,M \rbrace$, and $\sum_{j=1}^M \delta T_j = T$. As we said in Section~\ref{sec:controlSetup} we store the piecewise constant values of $b(t)$ in a real valued matrix $\beta \in \M_{l,M}(\R)$, the elements of which $\lbrace \beta_{i,j} \rbrace$ are determined by Equation~\eqref{eq:betaElement}. As such,
\begin{align}
   \label{eq:bilinearProblemPiecewiseConst}
   V(T) = \prod_{j=1}^M \exp \left[ \left( L_0 + \sum_{i=1}^l \beta_{i,j} L_i \right) \delta T_j \right] ,
\end{align}
where the product symbol denotes a sequential matrix multiplication.

A bilinear control theory problem for piecewise constant control amplitudes is stated as: find control amplitudes $b: [0,T] \rightarrow \R^l$, or equivalently the corresponding matrix $\beta$, that adhere to certain problem specific constraints and yields a $V(T)$ with some desired properties. Given the desired properties for $V(T)$, we can always write down a target function $\Phi: \M_n \rightarrow [0,1]$ that is an analytic function and takes the value $1$ if and only if its argument has the properties that we want from $V(T)$. Having defined such a target function, the problem of finding a $\beta$ that yields $\Phi\left[V(T)\right] = \Phi(\beta) = 1$ becomes a multivariable optimization problem.

Because $\Phi$ is an analytic function of $V(T)$, knowing $V(T)$ and its partial derivatives $\lbrace \frac{\partial}{\partial \beta_{i,j}}V(T) \rbrace$ for $i \in \lbrace 1, \dots,l \rbrace$ and $j \in \lbrace 1, \dots, M \rbrace$, is enough for determining both $\Phi\left[V(T)\right]$ and $\lbrace \frac{\partial}{\partial \beta_{i,j}}\Phi\left[V(T)\right] \rbrace$. The key insight by Khaneja \emph{et al} \cite{khaneja_2005} was to point out that the computational cost of the simultaneous evaluation of $V(T)$ and $\lbrace \frac{\partial}{\partial \beta_{i,j}}V(T) \rbrace$ is not much more than evaluating $V(T)$ alone. As such, it is advantageous to use optimization algorithms that make use of the gradient information, e.g. the conjugate gradient algorithm.

Here, we proceed by evaluating a partial derivative
\begin{equation}
\begin{aligned}
   \label{eq:partialOfU}
   \frac{\partial}{\partial \beta_{r,s}} & V(T) = \\
   &\left( \prod_{j=s+1}^M \exp \left[ \left( L_0 + \sum_{i=1}^l \beta_{i,j} L_i \right) \delta T_j \right] \right) \Upsilon_{rs} \prod_{j=1}^{s-1} \exp \left[ \left( L_0 + \sum_{i=1}^l \beta_{i,j} L_i \right) \delta T_j \right] ,
\end{aligned}
\end{equation}
with
\begin{equation}
\begin{aligned}
   \label{eq:partialOfExp}
   \Upsilon_{rs} &= \frac{\partial}{\partial \beta_{r,s}} \exp \left[ \left( L_0 + \sum_{i=1}^l \beta_{i,s} L_i \right) \delta T_s \right] \\
   &= \left. \frac{d}{d \epsilon} \right|_{\epsilon = 0} \exp \left[ \left( L_0 + \sum_{i=1}^l \beta_{i,s} L_i \right) \delta T_s + \epsilon L_r \delta T_s \right] = \exp \left[ \left( L_0 + \sum_{i=1}^l \beta_{i,s} L_i \right) \delta T_s \right] \\
   &\cdot \int_0^{\delta T_s} dt~ \exp \left[ -\left( L_0 + \sum_{i=1}^l \beta_{i,s} L_i \right) t \right] L_r \exp \left[ \left( L_0 + \sum_{i=1}^l \beta_{i,s} L_i \right) t \right] .
\end{aligned}
\end{equation}
The integral term in Equation~\eqref{eq:partialOfExp} can be evaluated either through block matrix techniques introduced in Section~\ref{sec:diff_comp_methods} with Equation~\eqref{equation:first_order_constant} or by noticing that
\begin{equation}
\begin{aligned}
   \label{eq:derivativeApproximation}
   \int_0^{\delta T_s} dt~ &\exp \left[ -\left( L_0 + \sum_{i=1}^l \beta_{i,s} L_i \right) t \right] L_r \exp \left[ \left( L_0 + \sum_{i=1}^l \beta_{i,s} L_i \right) t \right] \\
   &= L_r \delta T_s + \frac{1}{2} \left[ L_r \delta T_s, \left( L_0 + \sum_{i=1}^l \beta_{i,s} L_i \right) \delta T_s \right] \\
   &+ \frac{1}{6} \left[ \left[ L_r \delta T_s, \left( L_0 + \sum_{i=1}^l \beta_{i,s} L_i \right) \delta T_s \right], \left( L_0 + \sum_{i=1}^l \beta_{i,s} L_i \right) \delta T_s \right] + \dots ,
\end{aligned}
\end{equation}
and approximating the integral with a finite sum of commutators. During the optimization of the examples presented in this manuscript, we always approximate the integral with a finite sum of 15 commutators since the speed of the optimization was not our main concern in this work. We note that a very comprehensive analysis of the performance of various algorithms for bilinear control optimization is given in \cite{Goodwin_2017}, which also investigates the use of block matrix methods for evaluating the integral expression in Equation~\eqref{eq:derivativeApproximation}.

Our scheme for control searches is much the same as the one by Khaneja \emph{et al} \cite{khaneja_2005}. We always define target functions that are analytic functions of $V(T)$, i.e., $\Phi(\beta)=\Phi[V(T)]$, $\Phi: \M_n \rightarrow [0,1]$. The partial derivatives of $\Phi(\beta)$ with respect to $\lbrace \beta_{i,j} \rbrace$ are then evaluated in terms of $\lbrace \frac{\partial}{\partial \beta_{i,j}}V(T) \rbrace$ using the chain rule. We use an off-the-shelf gradient ascent optimizer to maximize $\Phi(\beta)$ starting from a set of initial control sequences $\beta^{(0)} \in \M_{l,M}(\R)$ until the algorithm yields a $\Phi(\beta)$ value sufficiently close to one.

\section{\label{app:transferfn}Transfer Functions}
In this appendix, we describe the matrix methods used for performing the control searches for two examples in Section~\ref{sec:examples}. In these cases, the control sequences $a(t)$, $a: [0,T] \rightarrow \R^k$, are piecewise constant over intervals of equal length $\Delta T$ such that $T = N \Delta T$. Furthermore, in these cases, the control vectors $a(t)$ are of dimension two, i.e., $k=2$. Consequently, we map the real valued control vectors onto a complex scalar function $a'(t): [0,T] \rightarrow \complex$, with $a'(t) = a_1(t) - i a_2(t) = a_x(t) - i a_y(t)$. Given the complex vector representation and the fact that all transfer functions $\lbrace \Xi^{(\gamma)} \rbrace$ in this work are linear functions that treat $a_x(t)$ and $a_y(t)$ symmetrically, we represent each $\Xi^{(\gamma)}$ as an $N \times N$ matrix with complex entries such that
\begin{align}
   \label{eq:controlElements1}
   \beta^{(\gamma)}_{1,j} = \text{Re} \left[\Xi^{(\gamma)} (\alpha_1 - i\alpha_2) \right]_j,
\end{align}
and
\begin{align}
   \label{eq:controlElements2}
   \beta^{(\gamma)}_{2,j} = -\text{Im} \left[\Xi^{(\gamma)} (\alpha_1 - i\alpha_2) \right]_j,
\end{align}
where $\lbrace \beta^{(\gamma)}_{i,j} \rbrace$ are the piecewise constant control amplitudes as defined in Section~\ref{sec:controlSetup} and $\alpha_i$ denotes the $i$th row of the matrix $\alpha \in \M_{2,N}(\R)$, that specifies the control sequence $a(t)$.

Moreover, for all examples in this manuscript $\lbrace \Xi^{(\gamma)} \rbrace$ are diagonal in the Fourier domain and fully specified by two real valued scalar functions -- the amplitude transfer function $\lambda^{(\gamma)}\left(\nu\right)$ and the phase transfer function $\phi^{(\gamma)}\left(\nu\right)$. For the example in Section~\ref{sec:broadBandDipolar} the $\lbrace \Xi^{(\gamma)} \rbrace$ were constructed from $\lambda^{(\gamma)}\left(\nu\right)$ and $\phi^{(\gamma)}\left(\nu\right)$ given in Figure~\ref{fig:TransferFns}(a) and \ref{fig:TransferFns}(b), respectively. In order to construct each $\Xi^{(\gamma)}$, we first calculate the discrete Fourier transform matrix, which is a unitary transformation $W^\text{Fourier}(N) \in \M_{N}$, with its elements given as
\begin{align}
   \label{eq:FourierElement}
   \left[W^\text{Fourier}(N)\right]_{s,t} = \frac{1}{\sqrt{N}} \exp \left[ \frac{2 \pi i (s-1)(t-1)}{N} \right] ,
\end{align}
for $1 \leq s,t \leq N$. We then construct a diagonal matrix $\Lambda^{(\gamma)} \in \M_{N}$, the diagonal elements of which are given as $\Lambda^{(\gamma)}_{j,j} = \lambda^{(\gamma)}(\nu_j)~ \exp \left[ i \phi^{(\gamma)}(\nu_j) \right]$, with
\begin{equation}
\nu_j = \frac{1}{N \Delta T} \left(2 \left[(j-1)~\text{mod}~ \frac{N}{2} \right] - (j-1)\right) ,
\end{equation}
for $1 \leq j \leq N$. Here, $x~ \text{mod}~y$ denotes the remainder from diving $x$ by $y$. Finally,
we evaluate
\begin{align}
   \label{eq:linearTransferFn}
   \Xi^{(\gamma)}(N,\Delta T, \lambda^{(\gamma)}, \phi^{(\gamma)}) = \left[W^\text{Fourier} (N) \right]^{-1} \Lambda^{(\gamma)} (N,\Delta T, \lambda^{(\gamma)}, \phi^{(\gamma)}) W^\text{Fourier} (N)
\end{align}
for each $\gamma \in \Gamma$. The control amplitudes $\lbrace \beta^{(\gamma)} \rbrace$ are then evaluated via Equation \eqref{eq:controlElements1} and \eqref{eq:controlElements2}. The elements of the Jacobian $\lbrace \frac{\partial}{\partial \alpha_{i,t}} \beta^{(\gamma)}_{j,s} \rbrace$, that are also necessary for the control searches, evaluate to
\begin{equation}
\begin{aligned}
   \label{eq:linearTransferFnJacobian}
   \frac{\partial}{\partial \alpha_{1,t}} \beta^{(\gamma)}_{1,s} &= \text{Re} \left( \Xi^{(\gamma)} \right)_{s,t} \\
   \frac{\partial}{\partial \alpha_{2,t}} \beta^{(\gamma)}_{1,s} &= \text{Im} \left( \Xi^{(\gamma)} \right)_{s,t} \\   
   \frac{\partial}{\partial \alpha_{1,t}} \beta^{(\gamma)}_{2,s} &= -\text{Im} \left( \Xi^{(\gamma)} \right)_{s,t} \\
   \frac{\partial}{\partial \alpha_{2,t}} \beta^{(\gamma)}_{2,s} &= \text{Re} \left( \Xi^{(\gamma)} \right)_{s,t},
\end{aligned}
\end{equation}
for all $1 \leq s,t \leq N$.

\subsection{Optimization Transfer Functions}
In Section~\ref{sec:controlSetup}, we argue that to implement certain constraints on the control sequence $a(t)$, one can use an optimization transfer function $\Xi^\text{opt}$ such that $\alpha = \Xi^\text{opt}\left[\alpha^\text{opt}\right]$, where $\alpha$ and $\alpha^\text{opt}$ specify the piecewise constant control sequences $a(t)$ and $a^\text{opt}(t)$, respectively. $\Xi^\text{opt}$ is constructed in such a way that its output functions always adhere to the necessary constraints. In this subsection, we demonstrate explicitly how to construct $\Xi^\text{opt}$ that introduces periods of zero pulse amplitudes at the beginning and at the end of the control sequence $a(t)$ and limits the bandwidth of $a(t)$ in the Fourier domain. This optimization transfer function is used for two control searches in Section~\ref{sec:examples}.

First, we construct an optimization transfer function that ensures that the control sequence $a(t)$ has equal periods of zero amplitude at the beginning and at the end of the sequence. To introduce such periods, we define an optimization transfer function $\Xi^\text{0}$. Just like above, we map both $\alpha \in \M_{2,N}(\R)$ and $\alpha^\text{opt} \in \M_{2,N^\text{opt}}(\R)$ onto complex vectors $\alpha' = \alpha_1 - i\alpha_2$ and $\alpha^{\text{opt}'} = \alpha^\text{opt}_1 - i\alpha^\text{opt}_2$, respectively. $N^\text{opt} = N-2N_0$ is the number of piecewise constant elements of $a^\text{opt}(t)$, where $N_0$ is the number of zero amplitude elements that are introduced at the beginning and at the end of $a(t)$. The action of $\Xi^\text{0} \in \M_{N,N^\text{opt}}$ is then implicitly defined as
\begin{align}
   \alpha' = \Xi^\text{0} \alpha^{\text{opt}'} = \left(\begin{array}{c}
    			0^{N_0}	\\
    			\alpha^{\text{opt}'}	\\
    			0^{N_0}
\end{array}\right) ,
\end{align}
for any $\alpha^{\text{opt}'}$; $0^{N_0}$ denotes a zero vector of length $N_0$. It is easy to see that a $\Xi^\text{0}$, which has the above property, can be constructed from three blocks:
\begin{align}
   \label{eq:ZeroPulseExtension}
\Xi^\text{0}(N,N_0) = \left(\begin{array}{c}
    			0	\\
    			\I_{N-2N_0}	\\
    			0
\end{array}\right) ,
\end{align}
where $0 \in M_{N_0,N-2N_0}$ is a rectangular matrix with all its entries being zero. The elements of $\alpha$ are then given by
\begin{equation}
\alpha_{1,j} = \text{Re} \left[\Xi^\text{0}(N,N_0)(\alpha^\text{opt}_1 - i\alpha^\text{opt}_2) \right]_j
\end{equation}
and
\begin{equation}
\alpha_{2,j} = -\text{Im} \left[\Xi^\text{0}(N,N_0)(\alpha^\text{opt}_1 - i\alpha^\text{opt}_2) \right]_j .
\end{equation}
Of course, the construction of $\Xi^\text{0}(N,N_0)$ generalizes easily for introducing an arbitrary number of zero amplitude periods of an arbitrary duration into $a(t)$.

Now, to limit the bandwidth of $a(t)$ frequency components, we construct an optimization transfer function $\Xi^\text{bp}$ that acts as a low-pass filter. Here, a low-pass filter should be understood simply as some amplitude transfer function $\lambda^\text{bp}: \R \rightarrow \R$ in Equation \eqref{eq:linearTransferFn}, that takes non-zero values only over some limited range $\Delta \nu$ centred around $\nu = 0$. In this manuscript, we used a particular functional form
\begin{align}
   \label{eq:BandPass}
   \lambda^\text{bp}\left(\nu,\Delta \nu\right) = \frac{1}{4} \left( 1 + \tanh \left[ \frac{20}{\Delta \nu} \left(\nu + \frac{\Delta \nu}{2} \right) \right] \right) \left( 1 - \tanh \left[ \frac{20}{\Delta \nu} \left(\nu - \frac{\Delta \nu}{2} \right) \right] \right) ,
\end{align}
that has smooth frequency cut-offs at $\pm \Delta \nu/2$ in order to prevent introducing long lasting ripples to the pulse waveform $a(t)$. Accordingly, we implement $\Xi^\text{bp}$ as 
\begin{align}
   \label{eq:bandPassTransferFn}
   \Xi^\text{bp}(N,\Delta T,\Delta \nu) = \Xi(N,\Delta T,\lambda^\text{bp}, 0) ,
\end{align}
where $\Xi(N,\Delta T,\lambda^\text{bp}, 0)$ is given by Equation \eqref{eq:linearTransferFn}. 

The optimization transfer function $\Xi^\text{opt}$ that we used for the two examples in Section~\ref{sec:examples} combined the action of $\Xi^\text{0}$ and $\Xi^\text{bp}$ and is calculated as
\begin{align}
   \label{eq:optTransferFn}
   \Xi^\text{opt}(N,N_0,\Delta T,\Delta \nu) = \Xi^\text{bp}(N,\Delta T,\Delta \nu) \Xi^{0}(N,N_0) .
\end{align}
The elements of $\alpha$ are consequently determined as
\begin{equation}
\begin{aligned}
   \label{eq:optimizationTransferFnControls}
   \alpha_{1j} &= \text{Re} \left[ \Xi^\text{opt}(N,N_0,\Delta T,\Delta \nu)(\alpha^\text{opt}_1 - i\alpha^\text{opt}_2) \right]_j \\
   \alpha_{2j} &= -\text{Im} \left[ \Xi^\text{opt}(N,N_0,\Delta T,\Delta \nu)(\alpha^\text{opt}_1 - i\alpha^\text{opt}_2) \right]_j ,
\end{aligned}
\end{equation}
for all $1 \leq j \leq N$ , whereas the elements of the Jacobian $\lbrace \frac{\partial}{\partial \alpha^\text{opt}_{i,t}} \alpha_{j,s} \rbrace$ evaluate to
\begin{equation}
\begin{aligned}
   \label{eq:optimizationTransferFnJacobian}
   \frac{\partial}{\partial \alpha^\text{opt}_{1,t}} \alpha_{1,s} &= \text{Re} \left[\Xi^\text{opt}(N,N_0,\Delta T,\Delta \nu) \right]_{s,t} \\
   \frac{\partial}{\partial \alpha^\text{opt}_{2,t}} \alpha_{1,s} &= \text{Im} \left[ \Xi^\text{opt}(N,N_0,\Delta T,\Delta \nu) \right]_{s,t} \\   
   \frac{\partial}{\partial \alpha^\text{opt}_{1,t}} \alpha_{2,s} &= -\text{Im} \left[ \Xi^\text{opt}(N,N_0,\Delta T,\Delta \nu) \right]_{s,t} \\
   \frac{\partial}{\partial \alpha^\text{opt}_{2,t}} \alpha_{2,s} &= \text{Re} \left[ \Xi^\text{opt}(N,N_0,\Delta T,\Delta \nu) \right]_{s,t},
\end{aligned}
\end{equation}
for all $1 \leq s \leq N$ and $1 \leq t \leq N^\text{opt}$.

\section{Symbolic Computation Methods} \label{app:mathematica}

\newcommand{\code}[1]{\texttt{#1}}

In this appendix, we describe the design and usage of code in \mathematica{} for symbolically simplifying the structure of the $C_{i,j}(t)$ matrices of Theorem \ref{theorem:HP_theorem}, with the main goal being to verify Conjecture \ref{conjecture:polynomial}. More generally, automated simplification of the integral expressions arising in Theorem \ref{theorem:HP_theorem} may be useful, as many of the generators naturally constructed in the quantum control context are sparse with repetitive entries, and as such the integral expressions will be amenable to significant simplifications.

The code described here is implemented using \emph{Patterns}, \emph{Pattern Matching}, and \emph{Replacements} in \mathematica{}. In particular, we use symbolic expressions in \mathematica{} to represent integrals arising from Theorem \ref{theorem:HP_theorem}. Expressions that can be simplified are identified using Patterns, which are used in \mathematica{} to identify expressions with a particular structure. If a Pattern matches an expression, then we apply Replacements, which transform one expression into another. For example, if one of the matrices in the symbolic expression for an integral is $0$, we use a Pattern to identify this, and a Replacement to replace the expression with $0$. See \cite{mathematica_patterns} for an introduction to these concepts in the \mathematica{} programming language.

This appendix is organized as follows:
\begin{itemize}
	\item In Section \ref{appendix:data_rep}, we outline a representation for the integrals in Theorem \ref{theorem:HP_theorem} using symbolic expressions in \mathematica{}.
	\item In Section \ref{app:rep_rules}, we define replacement rules to be used for automatically simplifying the integral expressions.
	\item In Section \ref{appendix:computing_general}, we describe the implementation of a function that implements Theorem \ref{theorem:HP_theorem}; i.e., given a list of the $B_{i,j}(t)$ matrices which are the input to Theorem \ref{theorem:HP_theorem}, the function constructs and simplifies the $C_{i,j}(t)$ matrices, which are the outputs of Theorem \ref{theorem:HP_theorem}.
	\item Finally, in Section \ref{app:polynomial_conjecture}, we apply the functionality built in the preceding sections to verify Conjecture \ref{conjecture:polynomial} for $0 \leq s_1, s_2 \leq 15$.
\end{itemize}
The \mathematica{} notebook containing the code developed in this appendix may be found in the online repository \cite{utb_symbolic}.

\subsection{Reserved Symbols and Primitive Expressions} \label{appendix:data_rep}

The first step is specifying a representation for expressions appearing in Theorem \ref{theorem:HP_theorem} in \mathematica{}. We reserve the symbol ``$t$'' to represent integration variables, and the symbol $\I$ to represent the identity matrix, which can be produced in \mathematica{} by typing \texttt{Esc d s 1 Esc}. 

There are two types of expressions which can appear in simplifications of Theorem \ref{theorem:HP_theorem} that we consider. In this appendix, we call these \emph{primitive expressions}, and use the term to refer to both the mathematical objects, as well as their representation in the code. The first kind of primitive expression is the nested integral (note the additional appearance of the pre-factor $t^m$):
\begin{equation}
	t^mU_1(t) \int_0^t dt_1 \dots \int_0^{t_{n-1}} dt_n t_1^{s_1} \dots t_n^{s_n} U^{-1}_1(t_1) A_1(t_1) W_1(t_1) \dots U^{-1}_n(t_n) A_n(t_n) W_n(t_n), \label{equation:nested_integral}
\end{equation}
which is represented in the code using the head ``$\Inte$'':
\begin{equation}
	\Inte\Big[t^m, \{U_1, t^{s_1}*A_1, W_1\}, \dots, \{U_n, t^{s_n}*A_n, W_n\} \Big].
\end{equation}
The second primitive is simply an expression not appearing in an integral, for example $t^mU(t)$, which is represented in the code using the head ''Ex'':
\begin{equation}
	\Ex[t^m,U].
\end{equation}
We remark that the programmatic representation of nested integrals is slightly redundant for representing expressions resulting from Theorem \ref{theorem:HP_theorem}, as primitives from this theorem will always have $W_{i-1} = U_i$, but there is no harm in this redundancy. 

A final technical detail is a special representation for time-independent scalar multiples of matrices, i.e. $c A(t)$, for $c \in \complex$ and $A$ a matrix valued function. These are represented in the code in the following way:
\begin{equation}
	\SM[c, A],
\end{equation}
which eases the discrimination between symbols representing (potentially time-dependent) matrices and time-independent scalars.

\subsection{Simplification of Primitives via Replacement Rules} \label{app:rep_rules}

The simplification of primitive expressions is carried out by defining \emph{replacement rules}, then applying those rules to primitives using the internal functions of \mathematica{}. The replacement rules we specify correspond to basic simplifications of the nested integrals arising in Theorem \ref{theorem:HP_theorem}. The rules are broken into groups.
\begin{itemize}
	\item \emph{Zeroes}
	
	The detection of zero matrices:
	\begin{equation}
		\text{Int}[ x\_\_\_, \{U\_, 0, W\_\}, y\_\_\_] \rightarrow 0.
	\end{equation}
	
	\item \emph{Linearity of integration}
	
	Factoring scalars:
	\begin{equation}
		\text{Int}[ x\_\_\_, \{U\_, \SM[c\_, A\_], W\_\}, y\_\_\_] \rightarrow c~\text{Int}[ x, \{U, A, W\}, y],
	\end{equation}
	and linear combinations:
	\begin{equation}
		\text{Int}[ x\_\_\_, \{U\_, A\_+B\_, W\_\}, y\_\_\_] \rightarrow \text{Int}[ x, \{U, A, W\}, y] + \text{Int}[ x, \{U, B, W\}, y].
	\end{equation}
	
	\item \emph{Performing integrals}
	
	In some cases, an integral can be explicitly performed. The case that we handle is when $U_i = W_i$ and $A_i(t) = t^m\I$ for some index $i$ in Equation~\eqref{equation:nested_integral}. As an explicit example, it holds that
	\begin{align}
		\int_0^t dt_1  \int_0^{t_1} dt_2 \int_0^{t_2}&dt_3 t_2^m U^{-1}_1(t_1) A_1(t_1)W_1(t_1)U^{-1}_3(t_3)A_3(t_3)W_3(t_3)  \\
		\nonumber = \frac{1}{m+1}\bigg(\int_0^t & dt_1 \int_0^{t_1} dt_3 t_1^{m+1}U^{-1}_1(t_1) A_1(t_1)W_1(t_1)U^{-1}_3(t_3)A_3(t_3)W_3(t_3) \\ 
		\nonumber - \int_0^t & dt_1 \int_0^{t_1} dt_3 t_3^{m+1}U^{-1}_1(t_1) A_1(t_1)W_1(t_1)U^{-1}_3(t_3)A_3(t_3)W_3(t_3)\bigg). 
	\end{align}
	 Handling all possible versions of this simplification with replacement rules must be broken into four cases based on the order of the nested integral, as well as where the simplification appears in the nest. Some care must also be taken with powers of $t$; the pattern $t^{m\_}$ will detect powers of $t$ when $m \geq 2$, but not the expressions ``$1$'' and ``$t$'' as, symbolically, they are not powers of $t$. Thus, one must create replacement rules for these cases separately.
	\begin{itemize}
		\item Case 1: A first order integral that can be performed.
		\begin{equation}
		\begin{aligned}
			\text{Int}[x\_,\{U\_, \I, U\_\}] &\rightarrow \text{Ex}[x*t, U]\\
			\text{Int}[x\_,\{U\_,t \I, U\_\}] &\rightarrow \frac{1}{2} \text{Ex}[x*t^2, U]\\
			\text{Int}[x\_,\{U\_,t^{m\_} \I, U\_\}] &\rightarrow \frac{1}{m+1} \text{Ex}[x*t^{m+1}, U]
		\end{aligned}
		\end{equation}
		\item Case 2: An integral of order $\geq 2$ where the last integral can be performed.
		\begin{equation}
		\begin{aligned}
			\text{Int}[x\_\_\_, \{U1\_, A\_, W1\_\}, \{U\_, \I, U\_\}] &\rightarrow \text{Int}[x, \{U1, tA, W\}] \\
			\text{Int}[x\_\_\_, \{U1\_, A\_, W1\_\}, \{U\_, t \I, U\_\}] &\rightarrow \frac{1}{2} \text{Int}[x, \{U1, t^2A, W\}] \\
			\text{Int}[x\_\_\_, \{U1\_, A\_, W1\_\}, \{U\_, t^{m\_} \I, U\_\}] &\rightarrow \frac{1}{m+1} \text{Int}[x, \{U1, t^{m+1}A, W\}]
		\end{aligned}
		\end{equation}
		\item Case 3: An integral of order $\geq 2$ where the first integral can be performed.
		\begin{equation}
		\begin{aligned}
			\text{Int}[&y\_,\{U\_, \I, U\_\}, \{U1\_, B\_, W1\_\}, x\_\_\_] \\  &\rightarrow \text{Int}[y*t,\{U1, B,W1\}, x] - \text{Int}[y,\{U1, tB,W1\}, x] \\
			\text{Int}[&y\_,\{U\_, t \I, U\_\}, \{U1\_, B\_, W1\_\}, x\_\_\_] \\  &\rightarrow \frac{1}{2} \big(\text{Int}[y*t^2,\{U1, B,W1\}, x] - \text{Int}[y,\{U1, t^2B,W1\}, x] \big) \\
			\text{Int}[&y\_,\{U\_, t^{m\_} \I, U\_\}, \{U1\_, B\_, W1\_\}, x\_\_\_] \\  &\rightarrow \frac{1}{m+1} \big(\text{Int}[y*t^{m+1},\{U1, B,W1\}, x]- \text{Int}[y,\{U1, t^{m+1}B,W1\}, x] \big)
		\end{aligned}
		\end{equation}
		\item Case 4: An integral of order $\geq 3$ where an ``internal'' integral can be performed.
		\begin{equation}
		\begin{aligned}
			\text{Int}[&x\_\_\_, \{U1\_, A1\_, W1\_\}, \{U\_, \I, U\_\}, \{U2\_, A2\_, W2\_\}, y\_\_\_] \\ & \rightarrow\text{Int}[x, \{U1, tA1, W1\}, \{U2, A2, W2\}, y]\\
			&\qquad\qquad\qquad\qquad - \text{Int}[x, \{U1,A1, W1\}, \{U2,  tA2, W2\}, y] \\
			\text{Int}[&x\_\_\_, \{U1\_, A1\_, W1\_\}, \{U\_, t\I, U\_\}, \{U2\_, A2\_, W2\_\}, y\_\_\_]  \\ & \rightarrow\frac{1}{2} \Big(\text{Int}[x, \{U1, t^2A1, W1\}, \{U2, A2, W2\}, y]\\ 
			&\qquad\qquad\qquad\qquad- \text{Int}[x, \{U1,A1, W1\}, \{U2,  t^2A2, W2\}, y]\Big) \\
			\text{Int}[&x\_\_\_, \{U1\_, A1\_, W1\_\}, \{U\_, t^{m\_} \I, U\_\}, \{U2\_, A2\_, W2\_\}, y\_\_\_]  \\ & \rightarrow\frac{1}{m+1} \Big(\text{Int}[x, \{U1, t^{m+1}A1, W1\}, \{U2, A2, W2\}, y] \\ 
			&\qquad\qquad\qquad\qquad- \text{Int}[x, \{U1,A1, W1\}, \{U2,  t^{m+1}A2, W2\}, y]\Big)
		\end{aligned}
		\end{equation}
	\end{itemize}
\end{itemize}

\subsection{Computing General Expressions from Theorem \ref{theorem:HP_theorem}} \label{appendix:computing_general}

Computing general expressions arising from Theorem \ref{theorem:HP_theorem} consists of programmatically implementing the mapping
\begin{equation}
 \left(\begin{array}{cccc}
    			B_{1,1}(t) & B_{1,2}(t) & \dots & B_{1,n}(t) \\
			0 & B_{2,2}(t) & \dots & B_{2,n}(t) \\
			\vdots & \ddots & \ddots & \vdots \\
			0 & 0 & \dots & B_{n,n}(t)
			\end{array}\right)
			\rightarrow
			\left(\begin{array}{cccc}
    			C_{1,1}(t) & C_{1,2}(t) & \dots & C_{1,n}(t) \\
			0 & C_{2,2}(t) & \dots & C_{2,n}(t) \\
			\vdots & \ddots & \ddots & \vdots \\
			0 & 0 & \dots & C_{n,n}(t)
			\end{array}\right), \label{equation:exponential_function_mapping}
\end{equation}
where the right-hand side is the time-ordered exponential of the left-hand side. The function \texttt{TOExponential} implements this mapping. As 
\begin{equation}
	C_{i,i}(t) = U_i(t)= \mathcal{T}\exp\Big\{\int_0^t dt_1 B_{i,i}(t_1)\Big\}
\end{equation}
(i.e. $C_{i,i}(t)$ depends only on $B_{i,i}(t)$) rather than specifying $B_{i,i}(t)$, the user specifies the symbols for the $U_i(t)$ matrices as an input. That is, \texttt{TOExponential} takes in two lists of symbols:
\begin{equation}
\begin{aligned}
	\{U_1, \dots, U_n\}\textnormal{, and }
	\{ \{B_{1,2}, \dots, B_{1,n}\}&, \\
	  \{B_{2,3}, \dots, B_{2,n}\}&, \\
	   \vdots \;&, \\
	   \{B_{n-1,n}\}&\},
\end{aligned}
\end{equation}
where the first list of symbols represents the diagonal blocks of the time-ordered exponential, and the second list represents the off-diagonal blocks of the matrix to be time-ordered exponentiated. The output is a list of symbols
\begin{equation}
\begin{aligned}
	C = \{ \{C_{1,1}, \dots, C_{1,n}\}&, \\
	  \{C_{2,2}, \dots, C_{2,n}\}&, \\
	   \vdots \;&, \\
	   \{C_{n,n}\}&\},
\end{aligned}
\end{equation}
representing the upper triangle of matrices in the right-hand side of Equation~\eqref{equation:exponential_function_mapping}. In the notation of Theorem \ref{theorem:HP_theorem}, we have that the expression for $C_{k,k+j}(t)$ is contained in $C[[k,j+1]]$.

The \mathematica{} notebook has several example applications of this code. One basic use-case is to show that:
\begin{equation}
\begin{aligned}
	&\left(\begin{array}{ccc}
    			U(t) & U(t)\int_0^t dt_1 \tilde{A}(t_1) & U(t)\int_0^t dt_1 \int_0^{t_1} dt_2 \tilde{A}(t_1)\tilde{B}(t_2) \\
			0 & U(t) & U(t)\int_0^t dt_1 \tilde{B}(t_1) \\
			0 & 0 & U(t)
			\end{array}\right)\\&\qquad\qquad\qquad\qquad\qquad\qquad\qquad =  \mathcal{T}\exp \left\{ \int_0^t dt_1 \left(\begin{array}{ccc}
    			G(t_1) & A(t_1) & 0 \\
			0 & G(t_1) & B(t_1) \\
			0 & 0 & G(t_1)
			\end{array}\right)\right\},
\end{aligned}
\end{equation}
for $\tilde{A}(t) = U^{-1}(t)A(t)U(t)$ and $\tilde{B}(t) = U^{-1}(t)B(t)U(t)$, where $U(t) = \mathcal{T}\exp\{\int_0^t dt_1 G(t_1)\}$. In this case, the inputs to \texttt{TOExponential} are the lists
\begin{equation}
	\{U, U, U\}\textnormal{, and } \{\{A, 0\}, \{B\} \},
\end{equation}
and the output is the list
\begin{equation}
\begin{aligned}
	\{\{\Ex[1,U], \Inte[1, \{U, A, U\}], \Inte[1, \{U, A, U\}, \{U, B, U\}] \}&, \\
			 \{\Ex[1,U], \Inte[1, \{U, B, U\}] \}&, \\ 
			 	\{\Ex[1,U]\}& \}.
\end{aligned}
\end{equation}
By interpreting the structure of this output according to the data representation for primitives in Appendix~\ref{appendix:data_rep}, we see that this output is correct.

\subsubsection{Implementation of \texttt{TOExponential}}

The implementation of \texttt{TOExponential} requires two pieces: the symbolic construction of expressions for the $C_{s,s+j}(t)$ matrices given in Theorem~\ref{theorem:HP_theorem}, and the simplification of these expressions via the replacement rules of Appendix~\ref{app:rep_rules}. Due to the recursive structure of the matrices in Theorem \ref{theorem:HP_theorem}, it is both conceptually natural and computationally more efficient to perform this computation in a recursive fashion. The recursion proceeds by computing successive off-diagonals of the $C_{s,s+j}(t)$ matrices (i.e. successive values of $j$ for all $s$). As a reminder, the recursion relation for the $C_{s,s+j}(t)$ matrices for $j \geq 1$ is:
\begin{equation}
	C_{s,s+j}(t) = \Inte_{(s, s+j)}(t) + \sum_{i=1}^{j-1}U_{s}(t) \int_0^t dt_1U^{-1}_{s}(t_1) A_{s, s+i}(t_1) C_{s+i, s+j}(t_1). \label{equation:recursive_exp}
\end{equation}

Roughly, the computation goes as follows:
\begin{enumerate}
	\item Base case: Initialize the case $j=1$ using Equation~\eqref{equation:recursive_exp} and apply replacement rules to simplify any expressions.
	\item Recursive step: Construct expressions for $C_{s,s+j}(t)$ using Equation~\eqref{equation:recursive_exp} and the already computed $C_{s,s+i}(t)$ matrices for $1 \leq i < j$. Apply replacement rules to simplify resulting expressions.
\end{enumerate}

The most basic recursive construction step is to produce a single term in the sum in Equation~\eqref{equation:recursive_exp}. That is, given $C(t)$ (a linear combination of primitives) and two symbols $U$ and $A$, we must programmatically produce the mapping
\begin{align}
	C(t) \rightarrow U(t) \int_0^t dt_1 U^{-1}(t_1) A(t_1) C(t_1).
\end{align}
This mapping is performed by two functions both named \texttt{RecursiveRule}, where the second handles the case when the $A$ symbol is given as some scalar multiple $\SM[c,A]$. The replacement rules are
\begin{align}
	\Ex[q\_, z\_] \rightarrow \Inte[1, \{U, q*A, z\}]
\end{align}
and
\begin{align}
	\Inte[q\_, \{U1\_, A1\_, W1\_\}, z\_\_\_] \rightarrow \Inte[1, \{U, q*A, U1\}, \{U1, A1, W1\}, z],
\end{align}
and for the version which handles scalar multiples they are
\begin{align}
	\Ex[q\_, z\_] \rightarrow c*\Inte[1, \{U, q*A, z\}]
\end{align}
and
\begin{align}
	\Inte[q\_, \{U1\_, A1\_, W1\_\}, z\_\_\_] \rightarrow c*\Inte[1, \{U, q*A, U1\}, \{U1, A1, W1\}, z].
\end{align}

A full recursion step (for computing the next off-diagonal of elements from the previous) is implemented by \texttt{RecursiveConstruct}, which simply uses the \texttt{RecursiveRule} functions to construct the whole of the right-hand-side of Equation~\eqref{equation:recursive_exp}.

Lastly, the implementation of \texttt{TOExponential} is to do the initialization step of constructing and simplifying the $C_{s,s+1}(t)$ matrices, then recursively calling \texttt{RecursiveConstruct} to populate the output consisting of all $C_{s,s+j}(t)$ matrices.

\subsection{Bivariate Polynomials and Conjecture \ref{conjecture:polynomial}} \label{app:polynomial_conjecture}

Here, we describe code for working with and analyzing the generators $L_{s_1,s_2}(t)$ for the purposes of verifying Conjecture \ref{conjecture:polynomial}. A description of $L_{s_1,s_2}(t)$ is in Section~\ref{section:polynomial_system}, in the lead up to the statement of Conjecture~\ref{conjecture:polynomial}.

The first step is simply to produce the generator $L_{s_1, s_2}(t)$ and its time-ordered exponential. The functions \texttt{BPODGenerator} and \texttt{BPGenerator} both serve the function of specifying $L_{s_1,s_2}(t)$, with the first producing the upper-off-diagonal pieces of the generator, and the second including the diagonal. The function \texttt{BPTOExponential} simply applies the function \texttt{TOExponential} of Appendix~\ref{appendix:computing_general} to the generator given in \texttt{BPGenerator} and \texttt{BPODGenerator}.

The problems of verifying Conjecture \ref{conjecture:polynomial}, and providing explicit linear combinations of the blocks of $\mathcal{T}\exp\left(\int_0^t dt_1 L_{s_1,s_2}(t_1)\right)$ for computing integrals of the form
\begin{equation}\label{equation:bivariate_int}
	U(t)\int_0^t dt_1 \int_0^{t_1}dt_2 p(t_1,t_2)U^{-1}(t_1)A_1(t_1)U(t_1)U^{-1}(t_2)A_2(t_2)U(t_2) 
\end{equation}
for polynomials of degree at most $(s_1,s_2$), are fundamentally problems of linear algebra, and hence it is necessary to represent the upper right $(s_1 + 1) \times (s_2 + 1)$ blocks of $\mathcal{T}\exp\{\int_0^t dt_1 L_{s_1,s_2}(t)\}$ in a way that they may be analyzed using the linear algebra functions in \mathematica{}.

\subsubsection{Linear Algebraic Representation}

Due to the linearity of integration, for a polynomial $p$ of degree $(s_1,s_2)$, the expression in Equation~\eqref{equation:bivariate_int} may be viewed as member of the vector space of expressions
\begin{equation}
	\textnormal{span}\left\{ \Dyson_U(t^iA_1,t^jA_2) : 0 \leq i \leq s_1, 0 \leq j \leq s_2\right\},
\end{equation}
where, again
\begin{equation}
	\Dyson_U(t^iA_1,t^jA_2) = U(t) \int_0^t dt_1 \int_0^{t_1}dt_2 t_1^i t_2^jU^{-1}(t_1)A_1(t_1)U(t_1)U^{-1}(t_2)A_2(t_2)U(t_2),
\end{equation}
and each $\Dyson(t^iA_1,t^jA_2)$ is considered to be linearly independent from all others. Denoting this vector space of expressions as $P(s_1,s_2)$, we call the $\Dyson(t^iA_1,t^jA_2)$ the \emph{elementary basis} for this vector space.

We may represent $\poly{s_1}{s_2}$ as column vectors by defining a linear mapping
\begin{equation}
	Z : \poly{s_1}{s_2} \rightarrow \complex^{(s_1 + 1)(s_2 + 1)},
\end{equation} 
which acts on the standard basis as
\begin{equation}
	Z \Dyson_U(t^iA_1,t^jA_2) = e_{(s_2+1)i + j + 1}, \label{equation:vector_iso}
\end{equation}
where $e_n$ is the column vector with a $1$ in the $n^{th}$ position and a $0$ everywhere else (i.e. $Z$ sends the standard basis of $\poly{s_1}{s_2}$ to the standard basis of $\complex^{(s_1+1)(s_2+1)}$). Note that ordering of the images of the standard basis elements of $\poly{s_1}{s_2}$ according to this mapping is the \emph{lexicographic ordering} of the basis elements according to the powers $(i,j)$ of $t_1$ and $t_2$ appearing in the integral. 

The function \texttt{BPVectorRep} constructs a list of replacement rules that corresponding to this mapping. The function \texttt{BPTopRightBlockVectors} produces a matrix whose column vectors correspond to the top-right $(s_1+1) \times (s_2 + 1)$ blocks of $\mathcal{T}\exp\left(\int_0^t dt_1 L_{s_1,s_2}(t)\right)$ under the above mapping, with the blocks ordered in terms of the lexicographic ordering of their indices (which is automatically implemented by the \texttt{Flatten} function in \mathematica{}).

\subsubsection{Testing Conjecture \ref{conjecture:polynomial}}

With the terminology of the previous section, the content of Conjecture \ref{conjecture:polynomial} is simply that the top right $(s_1 + 1) \times (s_2 + 1)$ blocks of the time-ordered exponential of $\mathcal{T}\exp\left(\int_0^t dt_1 L_{s_1,s_2}(t)\right)$ are a basis for the formal vector space $\poly{s_1}{s_2}$. This is equivalent to the columns of the matrix $Q$ output by \texttt{BPTopRightBlockVectors} being a basis for $\complex^{(s_1+1)(s_2+1)}$. The function \texttt{TestClaim1} checks if this is the case by computing the rank of the matrix $Q$, with the claim being true if and only if the rank is $(s_1 + 1)(s_2+ 1)$. Note that before checking this, the code checks if any primitive expressions are present (i.e. whether the symbols ``Ex'' or ``Int'' remain, which would mean that there are expressions in the top right $(s_1 + 1) \times (s_2 + 1)$ blocks appearing that are not expected). If any such expressions are found, the function outputs the value $\infty$ indicating that Conjecture \ref{conjecture:polynomial} has failed catastrophically. If no such expressions are found, the rank is checked, with an output of $1$ meaning Conjecture \ref{conjecture:polynomial} is verified, and $0$ meaning it is invalidated. For example, the matrix $Q$ in the $s_1 = s_2 = 1$ case is
\begin{align}
	\left(\begin{array}{cccc}
    			0 & 0 & 1 & 0 \\
			1 & 0 & 0 & 1 \\
			1 & 0 & 0 & 0 \\
			0 & 1 & 0 & 0
		\end{array}\right),
\end{align}
which is clearly full rank.

The function \texttt{TestClaim1Range} applies \texttt{TestClaim1} to a range of values of $s_1$ and $s_2$. We have tested this claim for $0 \leq s_1, s_2 \leq 15$ and have found it to be true in all cases. While this does not constitute a proof, it does provide confidence that it is true in general, and in practical settings one can simply check its validity for a specific $s_1$ and $s_2$ of interest.

\subsubsection{Computing Integrals with Bivariate Polynomials}
Given an arbitrary polynomial $p(t_1, t_2) = \sum_{i=0}^{s_1} \sum_{j=0}^{s_2} c_{i,j}t_1^i t_2^j$ we wish to write the integral
\begin{equation}
\begin{aligned}
	U(t)&\int_0^T dt_1 \int_0^{t_1} dt_2 p(t_1, t_2) U^{-1}(t_1) A_1(t_1) U(t_1)U^{-1}(t_2) A_2(t_2)U(t_2)\\  &= \sum_{i=0}^{s_1} \sum_{j=0}^{s_2} c_{i,j} \Dyson_U(t^i A_1,t^j A_2)
\end{aligned}
\end{equation}
as a linear combination of the top-right $(s_1 + 1) \times (s_2 + 1)$ blocks of $\mathcal{T}\exp\left(\int_0^t dt_1 L_{s_1,s_2}(t)\right)$. Assuming the truth of Conjecture \ref{conjecture:polynomial}, the top-right blocks of this time-ordered exponential are a basis for $\poly{s_1}{s_2}$, which we will call the \emph{exponential basis}. Let $S : \poly{s_1}{s_2} \rightarrow \complex^{(s_1 + 1)(s_2+1)}$ be the mapping which takes an element of $\poly{s_1}{s_2}$ and returns the column vector of coefficients corresponding to expanding it according to the exponential basis (in lexicographic ordering). 

The matrix $Q$ output from \texttt{BPTopRightBlockVectors} then simply represents a change of basis from the image of the exponential basis under $S$ to the image of the standard basis under $Z$. That is, for every $r \in \poly{s_1}{s_2}$, it holds that
\begin{align}
	Zr = QSr.
\end{align}
Hence, the vector of coefficients of $r$ representing its expansion in the exponential basis is
\begin{align}
	Sr = Q^{-1}Zr.
\end{align}
As an example, in the $s_1 = s_2 = 1$ case, for an arbitrary $r \in \poly{s_1}{s_2}$ corresponding to the polynomial
\begin{align}
	p(t_1, t_2) = c_{0,0} + c_{0,1}t_2 + c_{1,0}t_1 + c_{1,1}t_1t_2,
\end{align}
we have
\begin{align}
Zr = \left(\begin{array}{c}
    			c_{0,0} \\
			c_{0,1} \\
			c_{1,0} \\
			c_{1,1}
		\end{array}\right)\textnormal{, and }
Q^{-1} = \left(\begin{array}{cccc}
    			0 & 0 & 1 & 0 \\
			0 & 0 & 0 & 1 \\
			1 & 0 & 0 & 0 \\
			0 & 1 & -1 & 0
		\end{array}\right),
\end{align}
and hence
\begin{align}
	Q^{-1}Zr = \left(\begin{array}{c}
    			c_{1,0} \\
			c_{1,1} \\
			c_{0,0} \\
			c_{0,1} - c_{1,0}
		\end{array}\right).
\end{align}
Denoting the blocks of $\mathcal{T}\exp\left(\int_0^t dt_1 L_{1,1}(t)\right)$ by $C_{i,j}(t)$, we may write the desired integral explicitly as a linear combination of $C_{i,j}(t)$ matrices by taking the following dot product
\begin{equation}
\begin{aligned}
 \left(\begin{array}{c}
    			C_{1,6}(t) \\
			C_{1,7}(t) \\
			C_{2,6}(t) \\
			C_{2,7}(t)
		\end{array}\right) 
		\cdotp
		&\left(\begin{array}{c}
    			c_{1,0} \\
			c_{1,1} \\
			c_{0,0} \\
			c_{0,1} - c_{1,0}
		\end{array}\right)
		\\&\qquad\qquad=c_{1,0}C_{1,6}(t) + c_{1,1} C_{1,7}(t) + c_{0,0}C_{2,6}(t) + (c_{0,1} - c_{1,0}) C_{2,7}(t), \label{equation:blockdecomp_output}
\end{aligned}
\end{equation}
where the vector of $C_{i,j}(t)$ matrices is just the top $2 \times 2$ blocks of $\mathcal{T}\exp\left(\int_0^t dt_1 L_{1,1}(t)\right)$ in lexicographic ordering.

The function \texttt{PolynomialBlockDecomposition} outputs the right hand side of Equation~\eqref{equation:blockdecomp_output} for an arbitrary $s_1$ and $s_2$ by performing the above computation in generality. An example when $s_1=s_2=2$ is included in the code.

\section{Proof of Theorem 1} \label{app:proof}
Recall, the goal is to prove that the $C_{i,j}(t)$ matrices, defined implicitly by the equation
\begin{equation}
\begin{aligned} \label{equation:appendix_exponential}
	&\left(\begin{array}{cccc}
    			C_{1,1}(t) & C_{1,2}(t) & \dots & C_{1,m}(t) \\
			0 & C_{2,2}(t) & \dots & C_{2,m}(t) \\
			\vdots & \ddots & \ddots & \vdots \\
			0 & 0 & \dots & C_{m,m}(t)
			\end{array}\right)\\
	&\qquad\qquad\qquad\qquad\qquad=
	\mathcal{T}\exp \left[ \int_0^t dt_1 \left(\begin{array}{cccc}
    			B_{1,1}(t_1) & B_{1,2}(t_1) & \dots & B_{1,m}(t_1) \\
			0 & B_{2,2}(t_1) & \dots & B_{2,m}(t_1) \\
			\vdots & \ddots & \ddots & \vdots \\
			0 & 0 & \dots & B_{m,m}(t_1)
			\end{array}\right)  \right],
\end{aligned}
\end{equation}
satisfy 
\begin{equation}
	C_{s,s}(t) = U_s(t) = \mathcal{T}\exp\left(\int_0^t dt_1 B_{s,s}(t_1) \right), \label{equation:unitary_explicit_reproduced}
\end{equation}
and
\begin{equation}
	C_{s,s+j}(t) = \Inte_{(s, s+j)}(t) +  \sum_{r=1}^{j-1} \sum_{s < i_1 < \dots < i_r < s+j} \Inte_{(s, i_1, \dots, i_r, s+j)}(t) \label{equation:explicit_reproduced}
\end{equation}
for all $1 \leq s \leq m$ and $1 \leq j \leq m-s$. We also show that alternatively, for $1 \leq j \leq m-s$, they can be given recursively:
\begin{equation}
	C_{s,s+j}(t) = \sum_{i=1}^{j}U_{s}(t) \int_0^t dt_1U^{-1}_{s}(t_1) B_{s, s+i}(t_1) C_{s+i, s+j}(t_1). \label{equation:recursive_reproduced}
\end{equation}
The approach of the proof is to first show that Equations~\eqref{equation:unitary_explicit_reproduced} and  \eqref{equation:recursive_reproduced} hold through an application of the uniqueness of solutions to differential equations. Afterwards, we show that the expression in Equation~\eqref{equation:explicit_reproduced} satisfies the same recursion relation as is given in Equation~\eqref{equation:recursive_reproduced}, and therefore Equation~\eqref{equation:explicit_reproduced} is also correct.

To prove that Equations~\eqref{equation:unitary_explicit_reproduced} and  \eqref{equation:recursive_reproduced} hold, we first use the definition of the time-ordered exponential to obtain the explicit form of the differential equation that the $C_{s,s+j}$ satisfy. Differentiating both sides of Equation~(\ref{equation:appendix_exponential}), we see that:
\begin{equation}
\begin{aligned}
	&\left(\begin{array}{cccc}
    			\dot{C}_{1,1}(t) & \dot{C}_{1,2}(t) & \dots & \dot{C}_{1,m}(t) \\
			0 & \dot{C}_{2,2}(t) & \dots & \dot{C}_{2,m}(t) \\
			\vdots & \ddots & \ddots & \vdots \\
			0 & 0 & \dots & \dot{C}_{m,m}(t)
			\end{array}\right) \\	
	 &\qquad=\left(\begin{array}{cccc}
    			B_{1,1}(t) & B_{1,2}(t) & \dots & B_{1,m}(t) \\
			0 & B_{2,2}(t) & \dots & B_{2,m}(t) \\
			\vdots & \ddots & \ddots & \vdots \\
			0 & 0 & \dots & B_{m,m}(t)
			\end{array}\right) \left(\begin{array}{cccc}
    			C_{1,1}(t) & C_{1,2}(t) & \dots & C_{1,m}(t) \\
			0 & C_{2,2}(t) & \dots & C_{2,m}(t) \\
			\vdots & \ddots & \ddots & \vdots \\
			0 & 0 & \dots & C_{m,m}(t)
			\end{array}\right) .
\end{aligned}
\end{equation}

Hence, for $1 \leq s \leq m$ and $0 \leq j \leq m-s$, the $C_{i,j}(t)$ matrices satisfy the differential equation
\begin{align}
	\dot{C}_{s,s+j}(t) = \sum_{i=0}^j B_{s,s+i}(t) C_{s+i,s+j}(t), \label{equation:de_system}
\end{align}
with initial conditions
\begin{align}
	C_{s,s+j}(0) = \begin{cases} 
      \I_n & \textnormal{if } j=0 \\
      0 &\textnormal{else}
   \end{cases}, \label{equation:de_initial_values}
\end{align}
where we are using a dot to denote differentiation with respect to $t$. From here, the validity of Equation~\eqref{equation:unitary_explicit_reproduced} follows from the $j=0$ case of Equations~\eqref{equation:de_system} and \eqref{equation:de_initial_values}. To establish the recursion relation in Equation~\eqref{equation:recursive_reproduced}, we differentiate it:
\begin{equation}
\begin{aligned}
&\dot{C}_{s,s+j}(t) =
	\frac{d}{dt}\sum_{i=1}^{j-1}U_{s}(t) \int_0^t dt_1U^{-1}_{s}(t_1) B_{s, s+i}(t_1) C_{s+i, s+j}(t_1)\\
	&\quad=  \sum_{i=1}^{j}\left( B_{s,s}(t)U_s(t) \int_0^t dt_1U^{-1}_{s}(t_1) B_{s, s+i}(t_1) C_{s+i, s+j}(t_1)  +  B_{s, s+i}(t) C_{s+i, s+j}(t)\right) \\
	&\quad= B_{s,s}(t)C_{s,s+j}(t) + \sum_{i=1}^j B_{s,s+i}(t)C_{s+i,s+j}(t) \\
	&\quad= \sum_{i=0}^j B_{s,s+i}(t)C_{s+i,s+j}(t),
\end{aligned}
\end{equation}
where in the last equality we have applied the recursion relation. Hence, the expressions in the recursion relation satisfy the same differential equations as the $j \geq 1$ case of Equations~\eqref{equation:de_system} and \eqref{equation:de_initial_values}, and hence they are correct under the assumption that this differential equation has a unique solution.

To show that Equation~\eqref{equation:explicit_reproduced} is also correct, we show that it satisfies the same recursion relations given by Equation~\eqref{equation:recursive_reproduced}. As the first term in Equation~\eqref{equation:explicit_reproduced} is immediately equal to the $i=j$ term in the sum of Equation~\eqref{equation:recursive_reproduced}, and the remaining terms in both are only non-zero when $j \geq 2$, it is sufficient to establish that 
\begin{equation} \label{equation:to_prove}
	\sum_{r=1}^{j-1} \sum_{s < i_1 < \dots < i_r < s+j} \Inte_{(s, i_1, \dots, i_r, s+j)}(t) = \sum_{i=1}^{j-1}U_{s}(t) \int_0^t dt_1U^{-1}_{s}(t_1) B_{s, s+i}(t_1) C_{s+i, s+j}(t_1)
\end{equation}
when $j \geq 2$. First, expand the left-hand side of the above equation:
\begin{equation} \label{equation:r1_and_r2}
\begin{aligned}
	\sum_{r=1}^{j-1} \sum_{s < i_1 < \dots < i_r < s+j} &\Inte_{(s, i_1, \dots, i_r, s+j)}(t)\\
			&= \sum_{s < i_1 < s+j} \Inte_{(s, i_1, s+j)}(t) + \sum_{r=2}^{j-1} \sum_{s < i_1 < \dots < i_r < s+j} \Inte_{(s, i_1, \dots, i_r, s+j)}(t),
\end{aligned}
\end{equation}
where we have broken the sum into the $r=1$ and $r \geq 2$ case. We may rewrite the first summation as:
\begin{equation}\label{equation:r1_term}
\begin{aligned}
	 \sum_{s < i_1 < s+j} \Inte_{(s, i_1, s+j)}(t)&= \sum_{i=1}^{j-1}\Inte_{(s, s+i, s+j)}(t)\\
	 &= \sum_{i=1}^{j-1} U_s(t) \int_0^t dt_1 U_s^{-1}(t_1)B_{s,s+i}(t_1) \Inte_{(s+i,s+j)}(t_1), 
\end{aligned}
\end{equation}
where in the second equality we have applied the recursion relation for $\Inte$ in Equation~\eqref{equation:int_recursive}. We may similarly rewrite the $r \geq 2$ terms in Equation~\eqref{equation:r1_and_r2} using a series of steps given below, with a description of each step given after the chain of equalities:
\begin{align}
	\nonumber \sum_{r=2}^{j-1} &\sum_{s < i_1 < \dots < i_r < s+j} \Inte_{(s, i_1, \dots, i_r, s+j)}(t)\\ \nonumber &= \sum_{r=2}^{j-1} \sum_{i=1}^{j-r} \sum_{s+i < i_2 < \dots < i_r < s+j} \Inte_{(s,s+i, i_2, \dots, i_r, s+j)}(t)\\ 
	\nonumber &=  \sum_{i=1}^{j-2}\sum_{r=2}^{j-i} \sum_{s+i < i_2 < \dots < i_r < s+j} \Inte_{(s,s+i, i_2, \dots, i_r, s+j)}(t)\\ 
	\nonumber &=  \sum_{i=1}^{j-2}\sum_{r=2}^{j-i} \sum_{s+i < i_2 < \dots < i_r < s+j} U_{s}(t) \int_0^t dt_1U^{-1}_{s}(t_1) B_{s, s+i}(t_1) \Inte_{(s+i, i_2, \dots, i_{r}, s+j)}(t_1)\\ 
	\nonumber &= \sum_{i=1}^{j-2}U_{s}(t) \int_0^t dt_1U^{-1}_{s}(t_1) B_{s, s+i}(t_1) \bigg(\sum_{r=2}^{j-i}\sum_{s+i < i_2 < \dots < i_r < s+j} \Inte_{(s+i, i_2, \dots, i_{r}, s+j)}(t_1) \bigg)\\
	\nonumber &= \sum_{i=1}^{j-2}U_{s}(t) \int_0^t dt_1U^{-1}_{s}(t_1) B_{s, s+i}(t_1) \bigg(\sum_{r=1}^{(j-i)-1}\sum_{s+i < i_1 < \dots < i_r < s+j} \Inte_{(s+i, i_1, \dots, i_{r}, s+j)}(t_1) \bigg)\\
	\nonumber &= \sum_{i=1}^{j-2}U_{s}(t) \int_0^t dt_1U^{-1}_{s}(t_1) B_{s, s+i}(t_1) \bigg(C_{s+i,s+j}(t_1) - \Inte_{s+i,s+j}(t_1) \bigg) \\
	&= \sum_{i=1}^{j-1}U_{s}(t) \int_0^t dt_1U^{-1}_{s}(t_1) B_{s, s+i}(t_1) \bigg(C_{s+i,s+j}(t_1) - \Inte_{s+i,s+j}(t_1)\bigg). \label{equation:r2_term}
\end{align}
The operation in each equality given by:
\begin{enumerate}[(1)]
	\item Break up the inner sum, with the new index $i$ having the correspondence $i_1 = s+i$.
	\item Swap the order of summation over $r$ and $i$.
	\item Apply the recursion relation for $\Inte$ given in Equation~\eqref{equation:int_recursive}.
	\item Move the summation for $r$ and $i_2, \dots, i_r$ past all terms that have no dependence on these indices.
	\item Change the limits for summation over $r$ to start at $1$.
	\item Substitute the expression in the brackets using Equation~\eqref{equation:explicit_reproduced}.
	\item Increase the upper limit of summation over $i$ to include $j-1$, which does not change the sum as $C_{s+j-1,s+j}(t) = \Inte_{(s+j-1,s+j)}(t)$.
\end{enumerate}

To complete the proof, add the new forms for the $r=1$ and $r \geq 2$ terms in Equation~\eqref{equation:r1_and_r2}, given respectively in Equations~\eqref{equation:r1_term} and \eqref{equation:r2_term}, to conclude that Equation~\eqref{equation:to_prove} holds.

\bibliographystyle{unsrt}

\end{document}